\documentclass[3p]{elsarticle}

\usepackage{amssymb}
\usepackage{amsmath}
\usepackage{units}

\usepackage{lineno}

\journal{Astroparticle Physics}

\begin{document}

\begin{frontmatter}

\title{Radio detection of air showers with the ARIANNA experiment on the Ross Ice Shelf}



\author[uci]{S.~W.~Barwick}
\author[kansas,rus]{D.~Z. Besson}
\author[upp]{A.~Burgman}
\author[uci_e]{E.~Chiem}
\author[upp]{A.~Hallgren}
\author[ccapp]{J.~C.~Hanson}
\author[berk1,berk2]{S.~R.~Klein}
\author[uci_e]{S.~A.~Kleinfelder}
\author[uci]{A.~Nelles\corref{cor1}}
\author[uci]{C.~Persichilli}
\author[uci]{S.~Phillips}
\author[uci_e]{T.~Prakash}
\author[uci]{C.~Reed}
\author[uci]{S.~R.~Shively}
\author[uci,oit]{J.~Tatar}
\author[upp]{E.~Unger}
\author[uci]{J.~Walker}
\author[uci]{G.~Yodh}

\address[uci]{Dept.\ of Physics and Astronomy, University of California, Irvine, USA}
\address[kansas]{Dept. of Physics and Astronomy, University of Kansas, USA}
\address[rus]{National Research Nuclear University MEPhI (Moscow Engineering Physics Institute) 115409, Moscow, Russia}
\address[upp]{Dept.\ of Physics and Astronomy, Uppsala University, Sweden}
\address[uci_e]{Dept. of Electrical Engineering and Computer Science, University of California, Irvine, USA}
\address[ccapp]{Center for Cosmology and Astro-Particle Physics (CCAPP), Dept.\ of Physics, Ohio State University, USA}
\address[berk1]{Lawrence Berkeley National Laboratory, USA}
\address[berk2]{Dept.\ of Physics, University of California, Berkeley, USA}
\address[oit]{Office of Information Technology, Research Computing, University of California, Irvine, USA}

 \cortext[cor1]{Corresponding author: anelles@uci.edu}

\begin{abstract}
The ARIANNA hexagonal radio array (HRA) is an experiment in its pilot phase designed to detect cosmogenic neutrinos of energies above \unit[$10^{16}$]{eV}. The most neutrino-like background stems from the radio emission of air showers. This article reports on dedicated efforts of simulating and detecting the signals of cosmic rays. A description of the fully radio self-triggered data-set, the properties of the detected air shower signals in the frequency range of \unit[100-500]{MHz} and the consequences for neutrino detection are given. 38 air shower signals are identified by their distinct waveform characteristics, are in good agreement with simulations and their signals provide evidence that neutrino-induced radio signals will be distinguishable with high efficiency in ARIANNA. The cosmic ray flux at a mean energy of \unit[$6.5^{+1.2}_{-1.0}\times10^{17}$]{eV} is measured to be \unit[$1.1^{+1.0}_{-0.7}\times10^{-16}$]{eV$^{-1}$km$^{-2}$sr$^{-1}$yr$^{-1}$} and one five-fold coincident event is used to illustrate the capabilities of the ARIANNA detector to reconstruct arrival direction and energy of air showers. 
\end{abstract}


\end{frontmatter}


\section{Introduction}
\label{sec:Intro}
The ARIANNA experiment (Antarctic Ross Ice shelf ANtenna Neutrino Array) is a surface array of radio antennas designed to detect cosmogenic neutrinos \cite{2015APh70}. Currently in its pilot phase the hexagonal radio array (HRA) \cite{2014IEEE62}, the experiment has been taking data since 2014. 

ARIANNA is aimed at detecting the radio emission of neutrino induced showers in ice. The radio emission is caused by the changing charge imbalance that is created in the shower front as it accumulates electrons from the medium and shower positrons are annihilated \cite{1965JETPAskaryan}. While test-beam experiments have shown that radio emission is indeed created in a shower \cite{2007PhRvL..99q1101G}, the first measurement of the radio emission due to a neutrino is still being awaited. 

Considerable progress has been made in the last years in radio detection of air showers (see \cite{2016PhR...620....1H} for a review). Radio emissions of air showers are now routinely measured at various experiments, such as AERA \cite{2016PhRvL.116x1101A}, LOFAR \cite{2013A&A...560A..98S} and Tunka-Rex \cite{2015NIMPA.802...89B} in coincidence with particle arrays or optical methods such as Fluorescence or Cherenkov detectors. Detections have also been reported from the balloon-based experiment ANITA \cite{2016APh....77...32S}.

The radio emission of air showers is created analogously to the emission of showers in dense media (such as ice), with differences due to differing density and index of refraction of the two media. In the atmosphere the shower develops over several kilometers, and its development is affected by its propagation through the geomagnetic field. The magnetic field leads to a charge separation of electrons and positrons, which creates a changing transverse current that is responsible for the dominant \emph{geomagnetic emission} in air showers \cite{1966RSPSA.289..206K, 2008APh....29...94S}. The charge imbalance, or \emph{Askaryan effect} \cite{1965JETPAskaryan}, is only of secondary importance in air, and accounts for about 5-20\% of the total emission depending on zenith angle and observer distance to the shower axis \cite{2014PhRvD..89e2002A, 2014JCAP...10..014S}. In ice, the increased density enhances the charge excess effect and causes showers to develop over a shorter length-scale, which diminishes the geomagnetic effect. Both contributions can be identified by their polarization signature. The polarization of the geomagnetic emission is determined for all observer positions through the Lorentz-force and so set by the shower axis and the geomagnetic field. Since the charge excess is symmetric around the shower axis, the polarization of the electric-field vector induced by the Askaryan effect always points toward the shower axis for all observer positions. The signal power radiated from each effect scales with the number of particles in the shower, which itself scales with the energy of the primary particle. Experimental results support the theoretical findings that the pulse amplitude scales linearly with the energy of the shower \cite{2016PhRvL.116x1101A}.

The frequencies at which the radio emission of a shower is observable are determined by the length-scale (in space and time) of the shower front, because it is a coherence effect. The size of the shower front determines the minimum wavelength over which the radiation is coherent. The shower front in air is larger than the one in ice, so air showers generally show the strongest power at frequencies below \unit[100]{MHz}, while the power emitted from neutrino showers peaks above \unit[100]{MHz} \cite{1991PhLB..257..432H}. Whether coherence is obtained at a given observer location is also subject to the index of refraction of the medium. At the Cherenkov angle the emitted radiation of all frequencies will arrive at the same time, increasing the signal amplitude. In air, the typical Cherenkov angle is about $1^{\circ}$. For a vertical air shower observed at sea-level, $1^{\circ}$ corresponds to a distance of less than \unit[100]{m} from the shower axis. At this radius, the signals undergo Cherenkov-like compression and a ring of strong emission is visible \cite{2015APh....65...11N}. Longer wavelengths remain coherent further off the Cherenkov cone than shorter wavelengths, so most of the air shower detectors operating at \unit[30-80]{MHz} observe a non-uniformly illuminated ellipse in the order of hundred meters in radius, showing a clear increase in size with zenith angle \cite{2015APh....60...13N}. In cold ice, the index of refraction is about 1.7, corresponding to a Cherenkov angle of about $53^{\circ}$. For all frequencies the emission from neutrino showers in ice will therefore be on a ring distant from the shower axis, with its width determined by the observing frequency band \cite{2000PhRvD..62f3001A,2016arXiv160504975H}. 

The radio emissions of both types of showers will only be detectable once their emitted power exceeds the ambient background, determined by the thermal noise of antenna systems as well as man-made and astronomical radio emission. The pulse power at a fixed distance from the shower scales quadratically with the energy of the primary particle \cite{2016PhRvL.116x1101A}. The exact threshold energy of detectability is also determined by the observation frequency band, as well as the gain of the antenna system. In addition, real trigger effects such as spacing of the antenna array, the threshold of an external particle array or the efficiency of the self-trigger have to be considered. 

Short, bipolar, nanosecond-scale pulses are predicted for showers from both neutrinos and cosmic rays, where the frequency content of the pulse is determined by the dimensions of the shower and the observer location. Consequently, air showers, as they are more abundant than neutrinos interactions at energies of \unit[$10^{16}$]{eV}, are a significant confusion background for neutrino observations. They will therefore have to be identified and removed for any neutrino analysis. However, measuring air showers with neutrino telescopes will provide a unique calibration source and a training set for methods and algorithms designed to identify neutrino signals. Both aspects of air shower detection will be covered in this article. 


\section{The ARIANNA experiment}
The design concept of ARIANNA is to deploy high-gain, directional, and wide bandwidth (\unit[100]{MHz} - \unit[1]{GHz}) antennas just under the surface of the Ross ice shelf, facing downward into the ice. This will allow for the detection of the radio emission created by particle showers induced by neutrinos of energy above \unit[$10^{16}$]{eV} \cite{2015APh70}. At these energies, the Earth is no longer transparent to neutrinos, with the exception of $\nu_{\tau}$-regeneration. So ARIANNA will mostly detect neutrinos arriving from near, or above, the horizon. Additional sensitivity to downward going neutrinos is gained by using the reflective properties of the water-ice interface at the bottom of the ice shelf. The ARIANNA-site is located in an area of very smooth ice, where the reflectivity at the bottom interface is high \cite{2015JGlac..61..438H}.

\begin{figure}
\includegraphics[width=0.47\textwidth]{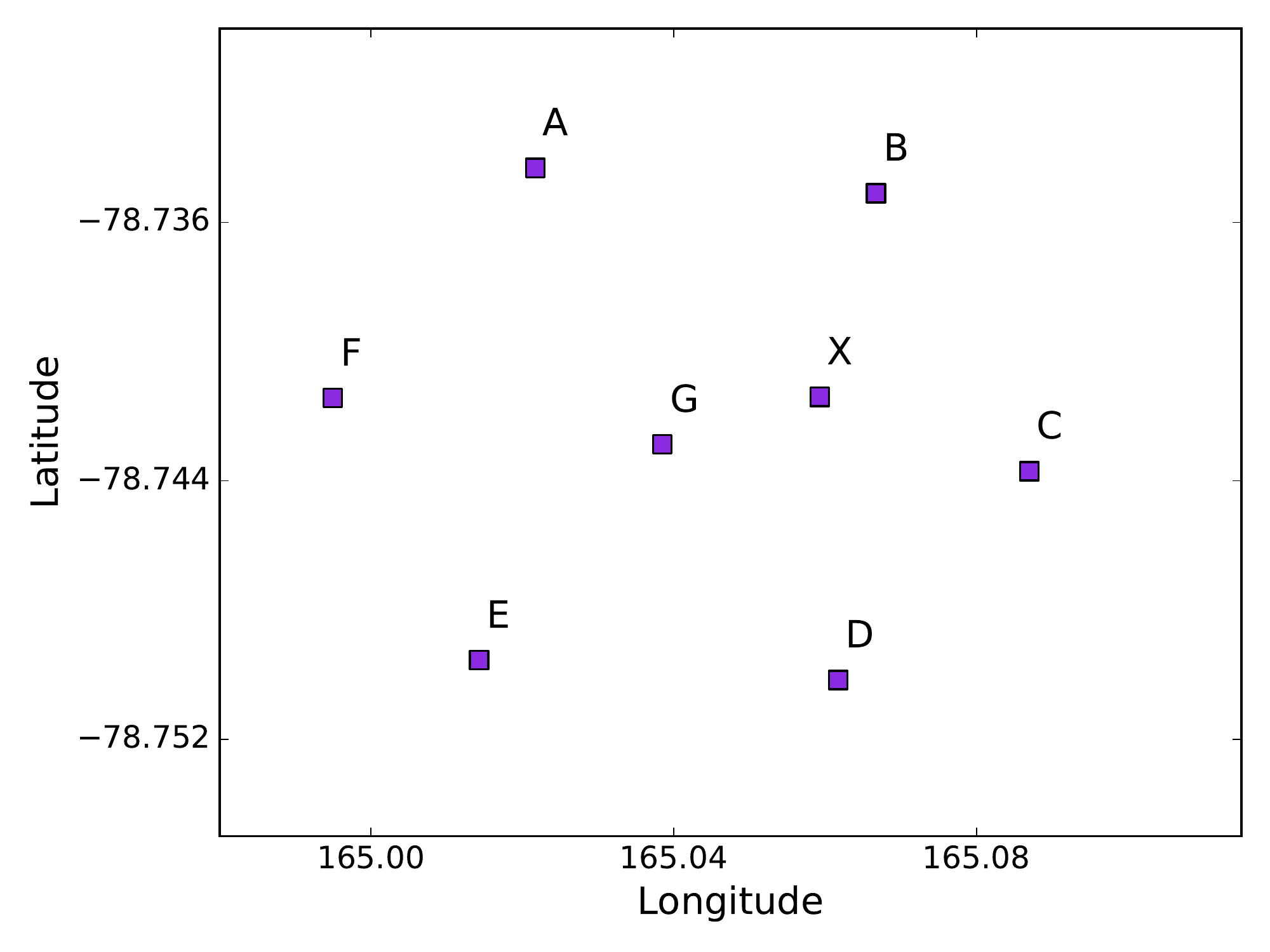}
\includegraphics[width=0.52\textwidth]{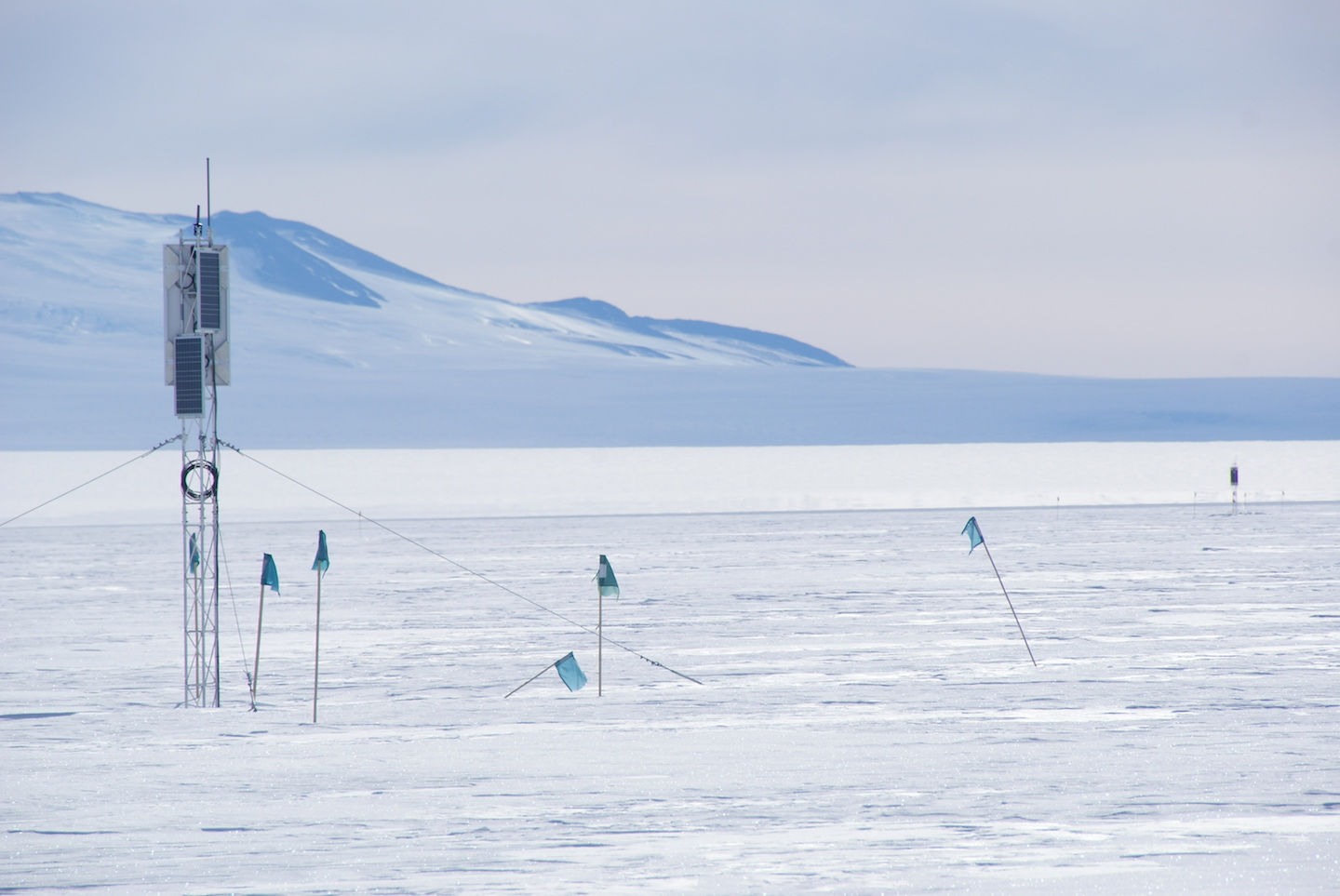}
\caption{The layout of the ARIANNA HRA. The left panel shows the locations of the seven stations, while the photo on the right shows station X with station B in the background. The image has been taken in the 2015/16 polar season.}
\label{fig:arianna}
\end{figure}

ARIANNA targets the neutrino flux that is proposed to be caused by cosmic rays above an energy of \unit[$10^{19}$]{eV} and their interactions with the cosmic microwave background \cite{1973Ap&SS..20...47S,1969PhLB...28..423B,1975Ap&SS..32..461B}. The best current limits on these fluxes have been found by IceCube \cite{2016PhRvL.117x1101A} and the Pierre Auger Observatory \cite{2015PhRvD..91i2008A}. This article will show that, in addition to improving on the sensitivity of current neutrino limits by more than an order of magnitude at \unit[$10^{18}$]{eV}, ARIANNA will also be sensitive to cosmic rays themselves. 

\subsection{Experimental set-up of the HRA}
\label{chap:exp_setup}

The ARIANNA baseline design comprises 1296 autonomous stations on a grid of $\unit[36\times36]{km^2}$. Each station will be equipped with at least four downward and two upward facing antennas (Create CLP5130-2N). The antennas are log-periodic dipole antennas (LPDA) and are sensitive from 100 to \unit[1000]{MHz} in air, and down to \unit[80]{MHz} when immersed in firn \cite{2010NIMPA.624...85G}. The directionality of the LPDAs allows the downward facing antennas to be sensitive to the neutrino signal, while the upward facing antennas provide a unique tag for cosmic ray induced showers, which will be used to calibrate the array as discussed later in this article. 

A pilot program of the ARIANNA observatory, the Hexagonal Radio Array (HRA), consists of seven stations (A-G) with four downward facing antennas each. The HRA was completed in 2014, and was augmented in 2015 by the addition of one station, marked as X in Figure \ref{fig:arianna}, targeted at the detection of air showers, with two upward and two downward facing antennas. The upward facing antennas are tilted by $45^{\circ}$ and point North and West, respectively. The layout of the array is shown in Figure \ref{fig:arianna}, together with a photo from the site taken in 2015.

ARIANNA stations are designed for an extremely low power consumption of \unit[4]{W} per station. This allows each station to operate continuously from sunrise in mid-September until sunset at the end of April \cite{NellesICRC2015}, powered by solar panels and cold-resistant batteries (LiFePO$_4$). Direct communications with the stations is enabled through either long-range wifi via a relay station on a nearby mountain, or short burst data (SBD) messages using the Iridium network. Both modes allow for real-time data transfer and remote configuration of the stations in-situ, if needed \cite{2014IEEE62}.

\subsection{Data acquisition}
\label{sec:daq}
The data-acquisition (DAQ) boards for the HRA contain \emph{SST}-chips which were custom designed for the HRA. The SST samples data on four  channels at \unit[2]{Gsamples/s} (HRA) or  \unit[1]{Gsamples/s} (station X) into ring-buffers of 256 samples, with a dynamic range of 12 bits RMS and at a bandwidth of \unit[1.5]{GHz} \cite{SST}.  Each channel is connected to one LPDA after amplification. The boards also house an embedded CPU, which can perform simple online analysis and can be remotely programmed. Data is stored on the local 32 GB solid-state disk and is transmitted in real-time to the server in the United States as soon as the stations enter a communication window. The trigger is formed with minimal buffering. The current trigger requires the waveform voltage to exceed a high and a low threshold within \unit[5]{ns} on one channel, and again on another channel within \unit[32]{ns} for a two of four majority logic requirement. The relative timing of the chips is extremely stable with a trigger-jitter in the order of picoseconds. The absolute timing is based on the Iridium network and accurate to microseconds. As the HRA stations are independent, there is no high-precision time-synchronization between them.

The DAQ system is designed to be easily reconfigured during its normal operation.  A configuration file is added to a queue on the communication server, which is transmitted and applied to the station when it communicates with the server at the next regular interval. The configuration file contains parameters to control, among other things, trigger thresholds and logic, application of the level one (L1) trigger, frequency of forced triggers to collect background data, if/how to transmit data, and the time between communication windows. HRA stations pause data-taking while they are communicating, which is usually true for a couple minutes every half-hour.

The trigger thresholds are tuned to obtain a target trigger rate. Given the extreme quietness of the ARIANNA site, the stations generally run a two out of four majority logic coincidence at $4\sigma$ of the thermal noise level, and still obtain trigger rates of less than \unit[$10^{-3}$]{Hz}. There are very few narrow-band transmitters detectable at the ARIANNA site. The only relevant signals are a radio communication frequency (VHF band) at \unit[140]{MHz}, air-traffic control at \unit[220]{MHz} and a search and rescue channel at \unit[400]{MHz}. Once communication is on-going on these channels, the transmitted power increases and the signal triggers the ARIANNA DAQ. This is typically the case only a few minutes per day. Since these signals are sinusoidal, they maximize the trigger rate when they are active. To suppress these narrow-band transmitters a L1 trigger has been designed that vetoes events on station level. Whenever the power in one frequency bin is higher than a set-able fraction (typically 0.3) of the remaining spectrum, the event will be vetoed. This L1 veto has an efficiency to retain more than 99.99\% of simulated neutrino signals. The stations are also equipped with a pre-scale that records a fraction of the otherwise vetoed signals. 


\section{Simulation framework}
Identifying cosmic rays -- and ultimately neutrinos -- in ARIANNA relies solely on methods developed with Monte Carlo simulations, as the radio emission is measured without coincidence with another method such as a particle detector. Several studies have shown that the radio signal of air showers is well understood. Especially in the frequency range of \unit[30-80]{MHz}, which is used by dedicated radio air shower experiments, details are well studied. The dependence on energy or the absolute scale of the signal \cite{2016APh....75...72A,2016PhRvL.116x1101A}, the height of the shower maximum \cite{2014PhRvD..90f2001A,2014PhRvD..90h2003B,2016JCAP...01..052B}, and the signal distribution on ground \cite{2015APh....60...13N,2016APh....74...79K} are reproduced with deviations of less than 10\% by air shower codes such as CoREAS \cite{2013AIPC.1535..128H} and ZHAireS \cite{2012APh....35..287A}. In combination with a full detector-simulation, these codes are used to precisely predict the characteristics of the signals of air showers in ARIANNA.  

A thorough understanding of the detector response is prerequisite for a signal prediction. For radio detectors, the most important component remains the antenna \cite{2012JInst...7P0011A}. The measured ARIANNA antenna characteristics have been discussed in previous work \cite{2015APh....62..139B}. A simulation of the antenna response has the advantage of being able to study systematic effects and to have a uniform parameter space for arrival directions and frequencies. In this chapter, all steps of the simulation chain and its uncertainties will be discussed. 

\subsection{Antenna modelling}
\label{sec:ant-mod}
The software WIPL-D \cite{Kolundzija2011}, which has been used earlier by the LOFAR Collaboration \cite{2013A&A...560A..98S}, has been chosen to model the frequency and phase dependence of the ARIANNA antennas. A detailed model of the antenna  has been constructed in WIPL-D, including all plastic parts and connector wires. The program can simulate antennas immersed in any dielectric medium, as well as close to a boundary to study reflection and refraction effects. For the purpose of this analysis the antenna was placed in an infinite firn medium, $n =1.3, \epsilon=1.7$, which is an approximation for the depth of a couple of meters below the snow surface \cite{2010NIMPA.624...85G}. The software returns the response of the antenna to an incoming electric field of a certain polarization as a function of frequency. The response is a complex number, so it contains the gain and phase response. 

The antenna response is mostly influenced by the detailed antenna dimensions. Immersing the antenna in firn shifts its sensitivity to lower frequencies, which is explained by the change in the propagation speed from air to firn.
While the antenna response in firn will be used throughout the cosmic ray analysis, the antenna simulation itself was checked with data taken in air \cite{2015APh....62..139B}, and augmented by measurements of gain and group delay with different antennas from the same manufacturer to help assess systematic variations of the LPDA response. More details can be found in \ref{sec:app_sim}.

Systematic offsets in gain and group delay have been found between simulations and data of order 10\%, while the spread between data and simulation is on average 30\%. For a single frequency, the systematic difference as a function of incident angle has been found to be as low as 3\% and as high as 35\%. 

It has been determined that the detailed features of the frequency dependent response are very sensitive to the precise antenna tine length, where high frequencies are subject to larger changes. Small changes, such as bending or shortening by a few millimeters, induce changes as function of frequency, while the overall gain remains similar. For small deformations, the changes in the angular response are small and therefore negligible. It should, however, be noted that small changes affect the back-lobe sensitivity more than the front-lobe, which makes the accurate reconstruction of signal measurements with the back-lobe more challenging. 
Significant differences have been found between two measurement set-ups. It is likely that these are due to systematics of the measurement set-ups, such as size of the Anechoic chamber, other local interference, or other external influences, as the same measurement of two antennas in the same set-up does not reveal such large differences. 

Small changes in the frequency-dependent structure of the gain or group delay of the antenna are likely to affect the pulse form. It will be discussed later in this analysis, in how far such uncertainties are problematic. However, the overall sensitivity is mostly determined by the average again, which has been found to agree to 10\% between simulations and measurements.

\subsection{Simulating air shower signals in the HRA}
\label{chap:emiss}
Predicting the radio emission of air showers with an accuracy that can match modern experiments is best done using a full air shower simulation like CORSIKA \cite{1998cmcc.book.....H} or Aires \cite{1999astro.ph.11331S} with their radio extensions CoREAS \cite{2013AIPC.1535..128H} or ZHAireS \cite{2012APh....35..287A}. These codes take into account not only the different emission mechanisms on a particle by particle basis, but also allow one to change atmospheric and local magnetic field settings. However, they are highly CPU intensive. While parameterizations have been shown to be useful to reconstruct events \cite{2015APh....60...13N, 2015JCAP...05..018N,2016PhRvL.116x1101A}, so far, they are based on total power and provide insufficient information about the precise characteristics of the electric field, which are needed for a full detailed simulation of the signal. Therefore, we chose to simulate \emph{generic} CoREAS showers on a star-shaped pattern of antennas \cite{2014PhRvD..90h2003B}, so that showers can be reused during the analysis as they are independent of the actual detector configuration.  

A set of 1300 air showers has been simulated with CoREAS (CORSIKA 7.5000, QGSJet-II-04, FLUKA 2011.2c) for the conditions at the ARIANNA site (magnetic field of \unit[62.3]{nT}, inclination angle of $-80.4^{\circ}$, \unit[30]{m} above sea level). As the simulations will be used to both explore the parameter space of air shower detections and for flux calculations, it has been chosen to use only proton primaries as they cover the widest range of heights of shower maximum. The simulation parameters were randomly chosen from the energy interval of \unit[$10^{16}-10^{20}$]{eV} and the arrival direction of $\theta=[0^{\circ},90^{\circ}]$ and $\phi=[0^{\circ},360^{\circ}]$, while ensuring that all bins of $\Delta\log(E) = 0.25$ and five bins in zenith angle of equal steradian were populated. 

Every simulation prescribes 160 electric field vectors at different positions on the ground surrounding the shower axis. The CoREAS simulation stops at the observer position in air and delivers a time-dependent electric field in three components. The time-sampling was chosen at \unit[0.1]{ns} which over-samples the fastest HRA sampling by a factor of 10 to provide sufficient frequency resolution.

For every one of the simulated electric field vectors a prediction of the signal as it would be measured in ARIANNA can be obtained. In order to be able to account for the Fresnel reflection that takes place when entering the firn and to be able to fold the electric field with the antenna response, the electric field is converted to on-sky coordinates $\vec{e}_{r}$, $\vec{e}_{\phi}$ and $\vec{e}_{\theta}$ that correspond to the propagation direction and s-polarized and p-polarized radiation, respectively. Since the electric field in the propagation direction is zero, the $\vec{e}_r$ component is commonly ignored for the calculation of the antenna response. 

Using Snell's law with $\theta_i$, the incident angle, $\theta_t$ the refracted angle, and the indices of refraction $n_{\mathrm{air}}$ and $n_{\mathrm{ice}}$ leads to the following standard transmitted power fractions $T_s$ and $T_p$: 
\begin{equation}
T_s = 1 - \left| \frac{n_{\mathrm{air}} \cos(\theta_i) - n_{\mathrm{firn}} \cos(\theta_t)}{n_{\mathrm{air}} \cos(\theta_i) + n_{\mathrm{firn}} \cos(\theta_t)} \right|^2, \:
T_p = 1 - \left| \frac{n_{\mathrm{air}} \cos(\theta_t) - n_{\mathrm{firn}} \cos(\theta_i)}{n_{\mathrm{air}} \cos(\theta_t) + n_{\mathrm{firn}} \cos(\theta_i)} \right|^2.
\end{equation}
In the field calculation the square root of the values is applied. This refraction is predicted to not only weaken the signal, but also to have an effect on the overall direction of the electric field vector and thereby the polarization. 

After refracting the electric field it is folded with the modeled antenna response, which converts the electric field into a voltage $\mathcal{E}$ in the respective antenna for an arrival direction $(\theta,\phi)$. This operation is performed in the on-sky coordinate system $(\vec{e}_{\phi},\vec{e}_{\theta})$ and in the frequency domain, where the convolution is then a multiplication of the complex electric field $\vec{E}(f, \theta,\phi)$ with the complex antenna effective height $\vec{h}_{\text{eff}}(f,\theta,\phi)$:
\begin{equation}
\mathcal{E} = \begin{pmatrix}h_{\phi},h_{\theta}\end{pmatrix} \cdot \begin{pmatrix}E_{\phi}\\ E_{\theta}\end{pmatrix}
\end{equation}
The simulation software WIPL-D provides the antenna effective height in terms of complex voltages $\vec{I}(f,\theta,\phi)$ that are the reaction of the simulated antenna to a generator current of $V_0=1$ Volt: 
\begin{equation}
\mathcal{E}= \frac{2~\lambda  ~  Z(f)}{i ~Z_0 ~ V_0}\cdot  \vec{I}(f,\theta,\phi) \cdot \vec{E}(f,\theta,\phi),
\end{equation}
where $\lambda$ is the wavelength and $Z_0$ the vacuum impedance. An example of the antenna impedance $ Z(f)$ and the complex voltages $\vec{I}(f,\theta,\phi)$ resulting from the WIPL-D simulations are shown in Figure \ref{fig:wipl-d}.

In processing the simulations, the voltage $\mathcal{E}$ is corrected for the frequency dependent amplifier gain (see \cite{2014IEEE62}) and cable losses, as well as additional filtering, which then delivers the voltage that is measured by the digitizer. The conversion ADC units to Volts is applied to the data, so that simulations and data can be compared in physical units. 

\begin{figure}
\includegraphics[width=0.5\textwidth]{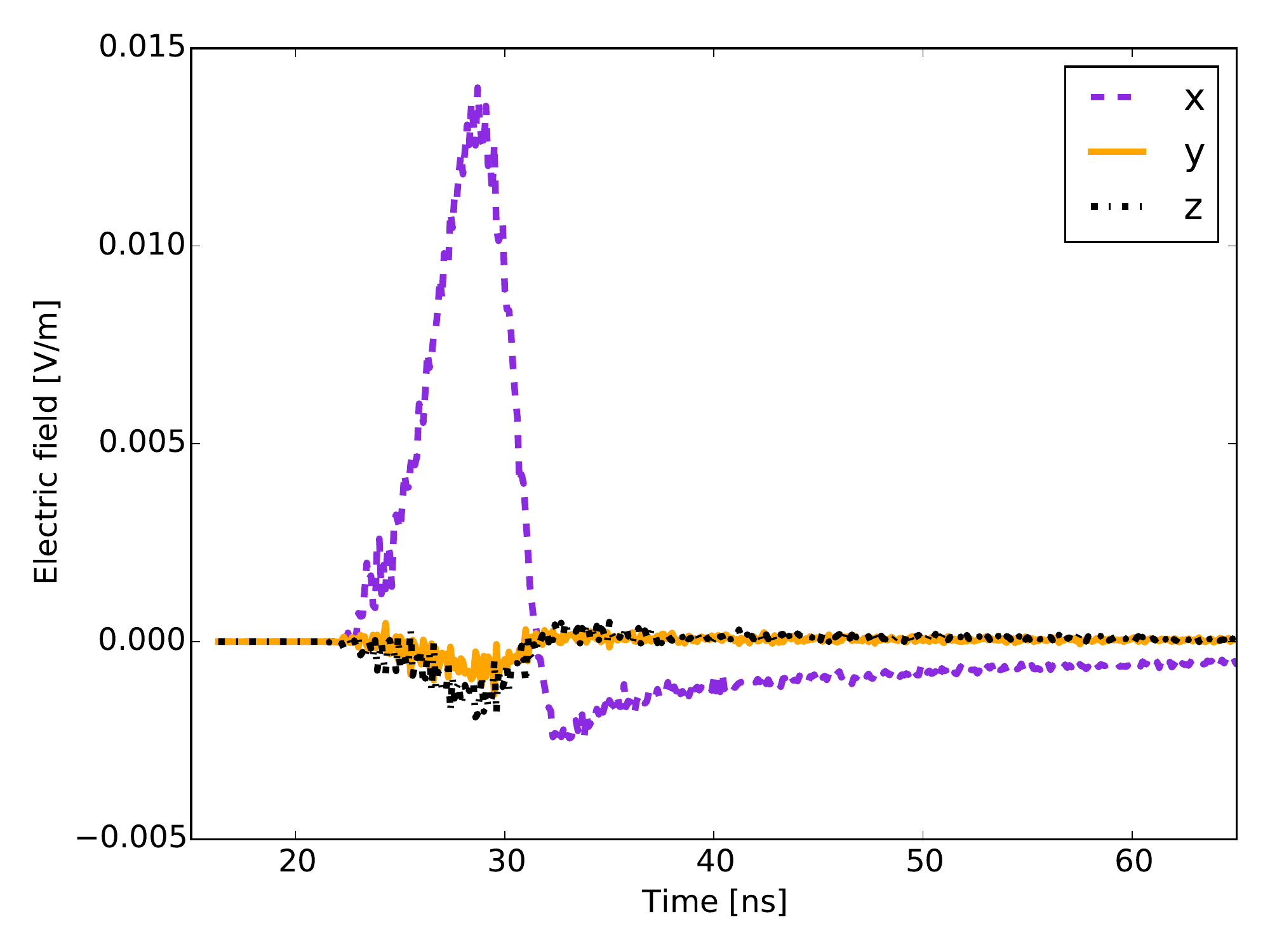}
\includegraphics[width=0.5\textwidth]{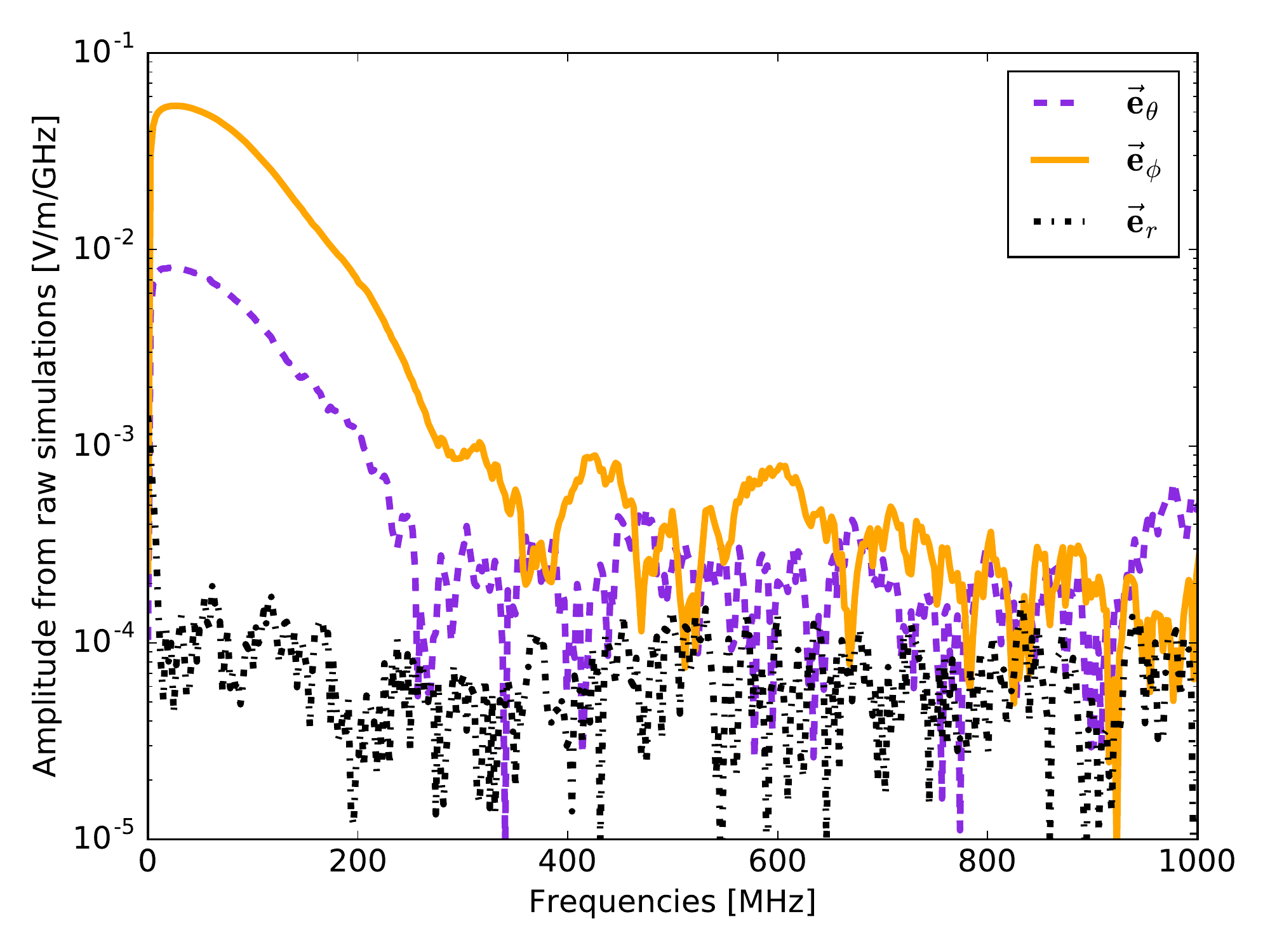}
\includegraphics[width=0.5\textwidth]{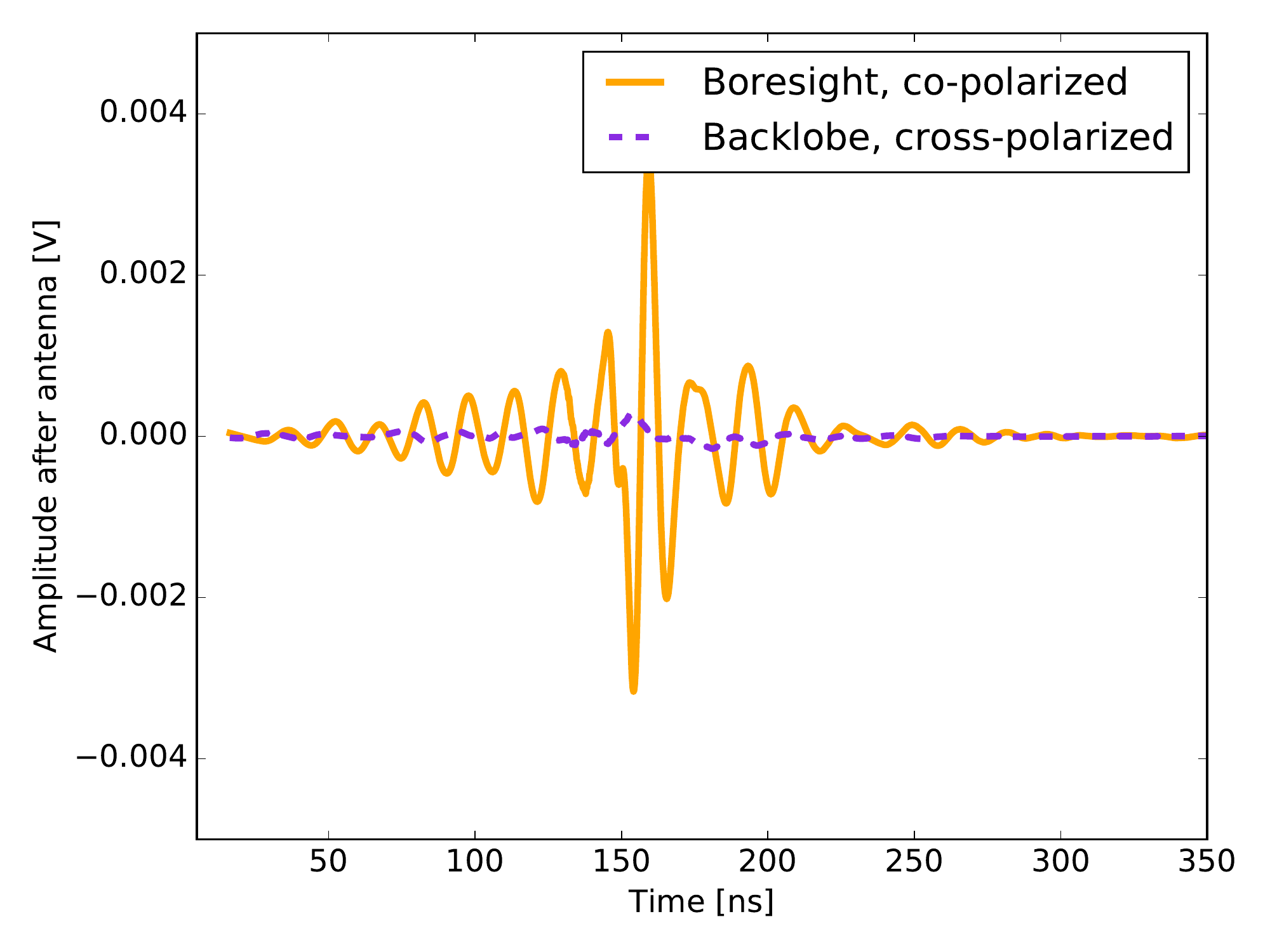}
\includegraphics[width=0.5\textwidth]{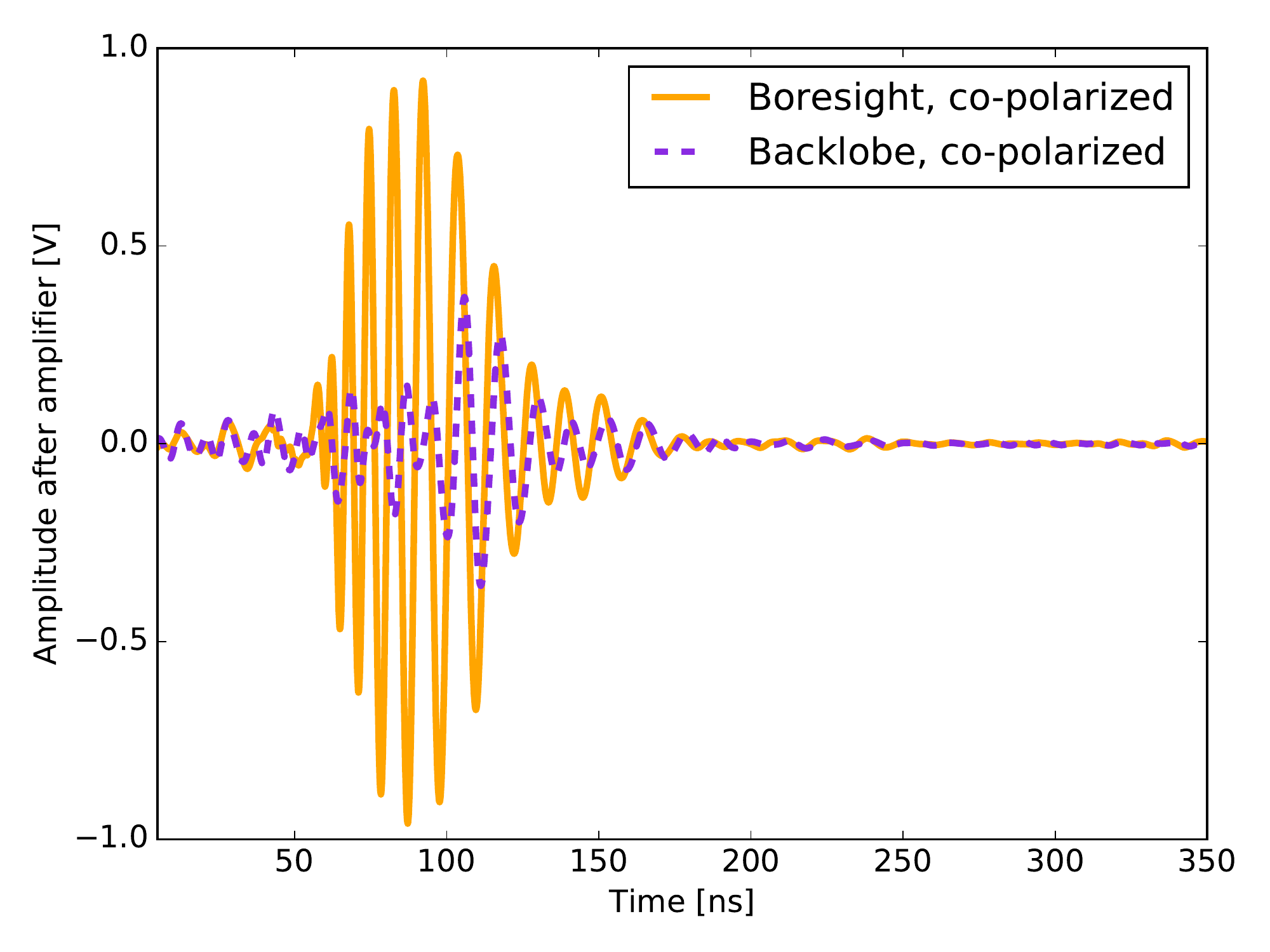}
\caption{Illustration of the simulation chain using an air shower of an energy of \unit[$5\times10^{18}$]{eV}. Top left: Raw electric field amplitude in air as function of time from CoREAS simulations. Top right: Frequency spectrum of the raw data, rotated to on-sky-coordinates. There is no signal in the propagation direction $\vec{e}_r$. Bottom left: Simulated pulse after applying the antenna model. Bottom right: Simulated pulse after amplification and filtering as it would be measured in the HRA stations. The position of the pulse is shifted due to an absolute phase offset in antenna model and measured amplifier response.}
\label{fig:sim_pulses}
\end{figure}

An example of an electric field simulation is shown in Figure \ref{fig:sim_pulses}. The signal is generated for an air shower from a specific arrival direction and distance to the shower axis. No noise from thermal fluctuations is present. By down-sampling to the instrumental sampling rate and truncating the signals \emph{search templates} are obtained that can be used to compare to measured data. The simulation set provides more than 200,000 templates. If, in addition, real noise as recorded by forced triggers is added, mock-data is obtained that mimics recorded events. After measured noise is added to the simulations, a software trigger is applied that mimics the hardware trigger in an HRA station. Only events that pass the trigger are used for later analyses. 

\subsection{Absolute scale of signal strength}
\label{sec:abs}
As elaborated above, the radio emission amplitude is directly proportional to the energy of the shower. For any cosmic ray study, consequently, the absolute calibration of the detector is essential to determine the threshold energy for air shower detection.
The simulation codes used to predict the emission from air showers have been reported to be accurate up to the limit of current experimental uncertainties \cite{2016PhRvL.116x1101A}. Therefore, very minimal uncertainties on the absolute scale are expected from the predicted emission of air showers. It is more difficult to obtain a calibration of the detector to such a precise scale. A detailed measurement of every single component, as well as the monitoring of temperature dependency of the components and other environmental effects is needed to obtain a precision at the 10\% level. The ARIANNA HRA was not designed as a precision instrument and therefore many of the following uncertainties are reducible for a future detector. 

\textbf{Antenna gain:} It is not practical to measure the antenna response directly in the ice to the precision needed for this analysis. This analysis therefore relies on the simulated antenna response. As discussed in Section \ref{sec:ant-mod}, it has been shown for measurements in air and at room temperature that the modeled and measured gain and group delay agree over all frequencies with a standard deviation of 30\% with a systematic offset of 10\%. While a fraction of these uncertainties can be explained by the measurement uncertainties, the in-ice environment adds additional uncertainties such as the index of refraction and boundary effects. Therefore, a systematic uncertainty of $\pm15$\% is added for the antenna model.  

\textbf{Antenna alignment:} Station X has antennas tilted towards the horizon. While it is relatively easy to ensure a perfectly vertical antenna alignment, a tilted angle is more complicated to deploy. An uncertainty of $10^{\circ}$ in zenith angle is assumed, while the azimuthal alignment is accurate to $5^{\circ}$. These relatively large uncertainties also cover the fact that antennas might shift while the snow is settling, even though excavating one antenna has provided no evidence for significant movement. As these alignment uncertainties translate only indirectly into an absolute scale uncertainty,  they are included as an uncertainty in the arrival direction of the air showers. 

\textbf{Amplifier gain:} The amplifiers are custom made and show systematic gain variations of up to $\pm20$\%. Especially with respect to the low frequencies, where high gain is achieved, the amplifiers are not uniform. In a future set-up, either the amplifiers will have to be more similar so that one calibration measurement suffices, or records of every individual component will have to be kept to reduce this uncertainty.

In addition, the gain of the amplifiers decreases with temperature. Reference measurements have been taken at room temperature. For the recorded temperatures in ice, this accounts to an uncertainty of up to $+15$\% of the amplitude scale. Once a model for the temperature dependence of the amplification has been developed, corrections can be applied.

\textbf{Losses in cables and connector:} All HRA stations are equipped with industrially manufactured cables and connectors. The losses have been found to increase with frequency and are about \unit[0.5]{dB} at room temperature, with additional group delay. The variations between cables are small and add an uncertainty of $\pm2$\%. 

\textbf{ADC calibration:} The conversion from ADC counts to mV has been performed for all boards in the lab. They show a spread of 3\% in the slope of the linear conversion between ADC count and voltage and a $\pm15$\% uncertainty on the absolute scale. Unfortunately, the individual calibration of station X has been faulty, and therefore unusable. It was therefore decided to use an average calibration and use the spread on the measurements as the corresponding uncertainty. 

\textbf{Ice parameters:} The properties of the ice at the ARIANNA-site are reasonably well-studied. The index of refraction of the firn determines the Fresnel angles and antenna properties and has been measured to be $n=1.29\pm0.02$ at the surface \cite{2015JGlac..61..438H}. The index of refraction increases as a function of depth, however, the antennas remain close to the surface so that no additional correction is needed. The resulting consequences of the absolute scale are small. The same holds true for the assumption of a flat surface. The ARIANNA site shows negligible surface roughness, at least in the length scales relevant for the observing band. In addition, inhomogeneities of the firn could induced additional scattering and weaken the signal. This will have to be investigated with additional measurements. For this analysis, we add uncertainties of $-10$\% in signal amplitude to account for these effects. 

The uncertainties are included in every relevant step of the simulation framework directly when the uncertain components are used. For all analysis a best and a worst case scenario is calculated and final parameters are given quoting the resulting scale uncertainty.

\section{Overview of data}
\label{chap:ov}
In this article we concentrate on data taken between December 6\textsuperscript{th} 2015 and the end of data-taking at the end of April 2016. In this period the HRA was in a uniform configuration and station X was operational. All data is available for analysis with the exception of a fraction from station C, where data still has to be retrieved in the next polar season. The live-time fractions range between 70\% and 95\%. Station X has been designed with two upward and two downward facing antennas, which gives its data unique characteristics for the detection of cosmic rays. Its data will be discussed in more detail. 

\subsection{Data taken with station X}

\begin{figure}
\includegraphics[width=\textwidth]{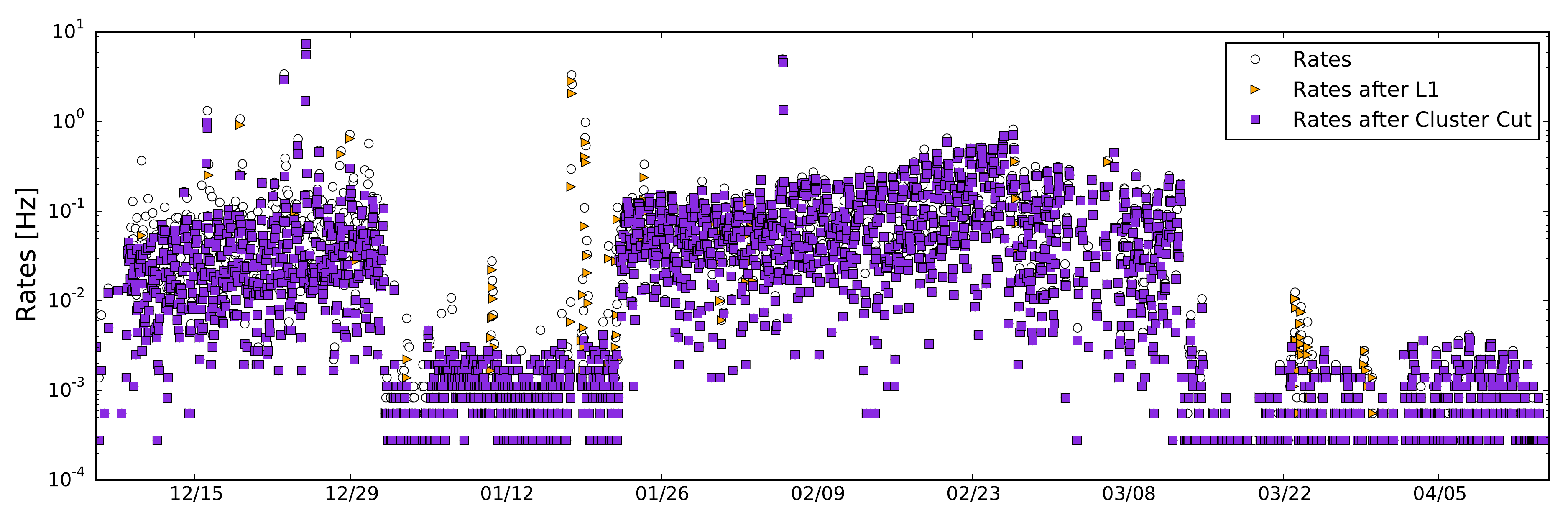}
\includegraphics[width=\textwidth]{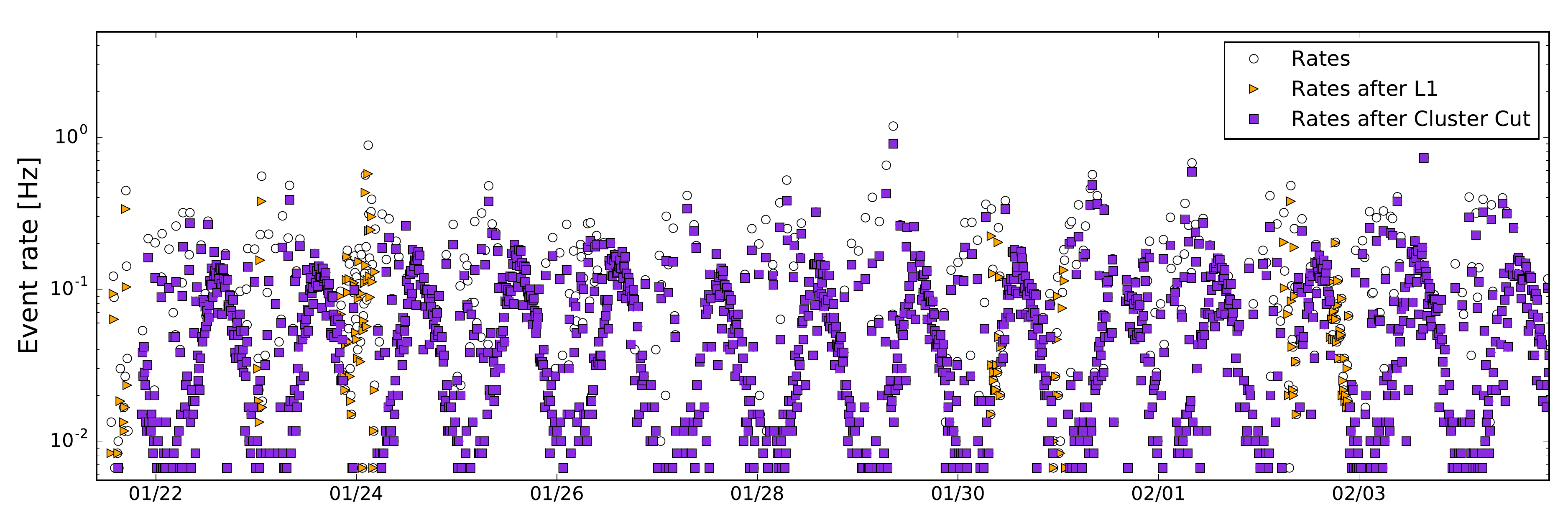}
\caption{Top panel: Event rates as function of time from data recorded with the cosmic ray station X in the Antarctic season of 2015/16. The open circles show the recorded event rate. The triangles depict the rate after the level one trigger (on station or offline) and the filled squares indicate the cleaned event rate after L1 and cluster cut (see Section \ref{chap:ov_sel}). Five periods of different trigger settings are visible as described in Table \ref{table:data}. Discrete event rate levels of less than $10^{-3}$ Hz are caused by small numbers of events per time-bin. Bottom panel: Zoom on period in January, illustrating the sidereal variation.}
\label{fig:Rates}
\end{figure} 

\begin{table}
\begin{tabular}{ll}
\hline
Period&Settings\\
\hline
December 6\textsuperscript{th} -- January 4\textsuperscript{th}& Trigger 2/2 upward channels, no L1, threshold: \unit[70]{mV}\\
January 4\textsuperscript{th} -- January 22\textsuperscript{nd}& Trigger 2/2 downward channels, no L1, threshold: \unit[70]{mV}\\
January 22\textsuperscript{nd} -- February 26\textsuperscript{th}& Trigger 2/2 upward channels, L1 on, threshold: \unit[70]{mV}\\
February 26\textsuperscript{th} -- March 2\textsuperscript{nd}& Trigger 2/2 upward channels, L1 on, threshold: \unit[72]{mV}\\
March 2\textsuperscript{nd} -- March 12\textsuperscript{th}& Trigger 2/2 upward channels, L1 on, threshold: \unit[74]{mV}\\
March 12\textsuperscript{th} -- March 14\textsuperscript{th}& Trigger 2/2 upward channels, L1 on, thresholds: \unit[82]{mV}\\
March 12\textsuperscript{th} -- April 23\textsuperscript{rd}& Trigger 2/2 upward channels, L1 on, thresholds: \unit[84]{mV}\\
\hline
\end{tabular}
\caption{Periods of data taking in the polar season 2015/16 with the cosmic ray station X. The trigger settings are described including the majority logic, the L1 setting and the threshold. Periods with different settings of less than one day (such as to tune thresholds) are not included. Excluding periods for data-transfer and accounting for dead-time, the total live-time for the cosmic ray station amounts to 85 days in this data-set.}
\label{table:data}
\end{table}

Figure \ref{fig:Rates} and Table \ref{table:data} give an overview of the data recorded with station X in the polar season of 2015/16. Several periods of distinct event rates can be identified, which correspond to different trigger conditions. During the first period, the station was triggering on a coincidence of the two upward facing antennas with no level one trigger (see Section \ref{sec:daq}). The latter was also true for the second period of data-taking in which the station triggered on a coincidence of the two downward facing antennas. While the thresholds were similar in both cases, the rates were significantly lower in the second period. This indicates that the directional LPDAs observed more down-coming noise. The trigger rates oscillate between \unit[$10^{-2}$]{Hz} and \unit[$10^{-1}$]{Hz}, with a periodicity of slightly less than 24 hours, as shown in the bottom panel of Figure \ref{fig:Rates}. As will be discussed in Chapter \ref{chap:galaxy}, this periodicity is caused by Galactic emission, which dominates the signal in the upward facing antennas. On January 22\textsuperscript{nd} the station was reconfigured to trigger on the upward facing antennas, and the L1 trigger to eliminate narrowband contributions was implemented. Later in the season, the thresholds were adapted to account for the increase in amplification caused by the drop in temperature. In March, the thresholds were raised to allow all data to be transmitted through limited-bandwidth Iridium communication.

\subsection{Measuring the Galactic emission}
\label{chap:galaxy}
The measurement of diffuse Galactic radio emission provides an important tool to assess the system sensitivity and quality. The brightness of the emission can be approximated by a steeply falling power-law as a function of frequency. In a well-designed system the Galactic emission should be larger than the thermal emission generated by the amplifiers and the antenna itself at frequencies below \unit[100-150]{MHz}.

In station X, the periodic variation of the trigger rates (see Figure \ref{fig:Rates}) is caused by a continuous variation of the detected noise power. The variation is periodic as function of local sidereal time (LST), which indicates that it is caused by an astronomical phenomenon. The power variation is strongest in the band of \unit[80-120]{MHz} and the rise in power towards smaller frequencies is compatible with the expectation from the Galaxy. The lower bound on this frequency band is determined by the cutoff in frequency caused by filters and amplifiers, while the upper bound is given by the frequency where the Galactic emission is no longer detectable above the system noise \cite{LFmap}. The Galactic emission is used to test whether the simulation chain is an accurate representation of the data and to check the absolute scale of the predictions \cite{2015JInst..10P1005N}. 

All background data (forced triggers) from station X have been used for this analysis. The measured waveforms of \unit[256]{ns} are Fourier transformed into frequency space, squared to obtain the power and integrated between \unit[82-121]{MHz}. The frequency bins are determined by the resolution of the Fourier transform (FFT), while the boundaries are chosen to maximize the observed power-variation as function of LST derived from the Galaxy. At higher frequencies the signal-to-noise ratio decreases, so only bins with a measurable power variation as function of LST are included. Finally, the average power in bins of one sidereal hour is calculated for all four channels of station X.

\begin{figure}
\includegraphics[width=0.49\textwidth]{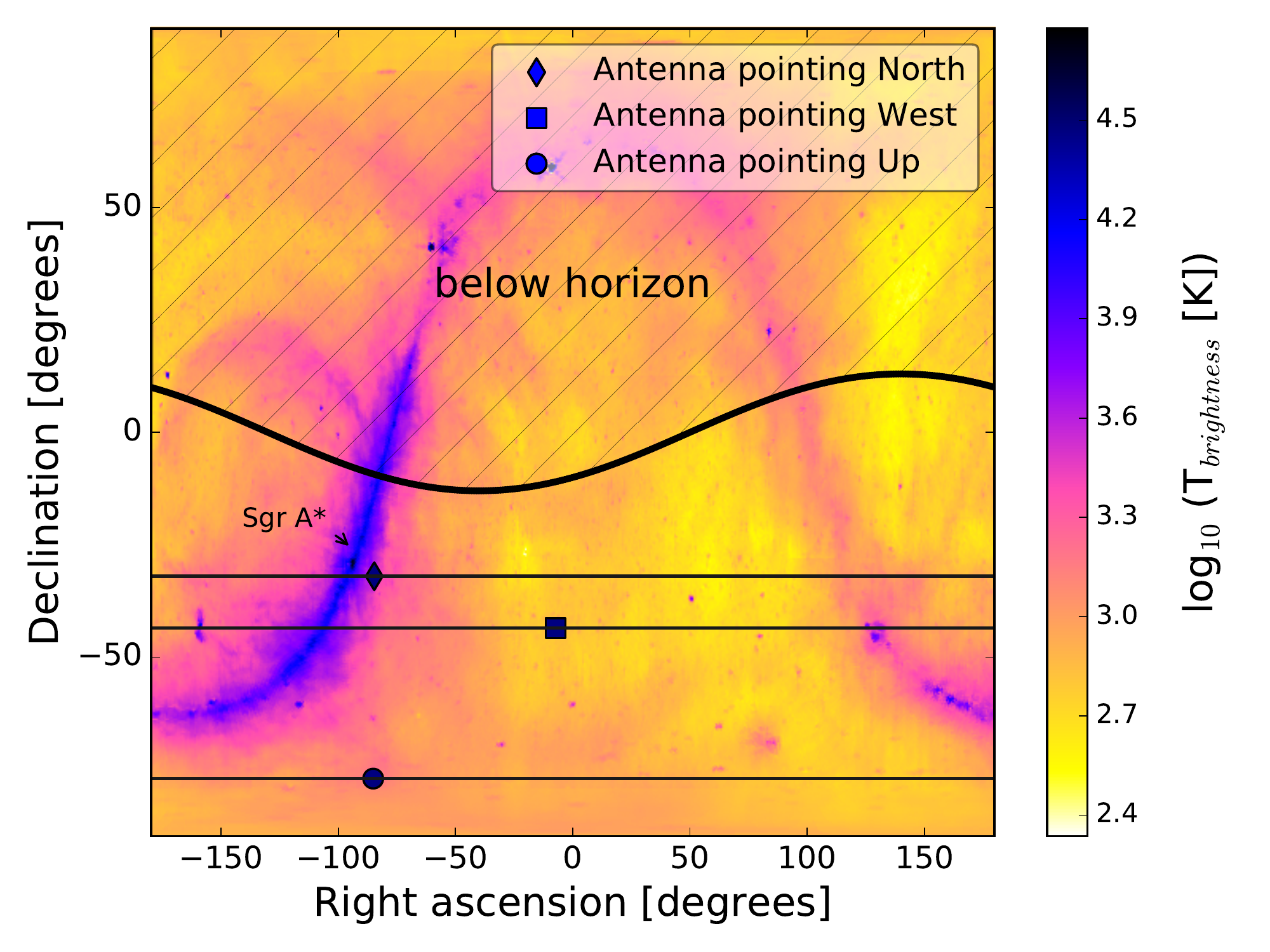}
\includegraphics[width=0.49\textwidth]{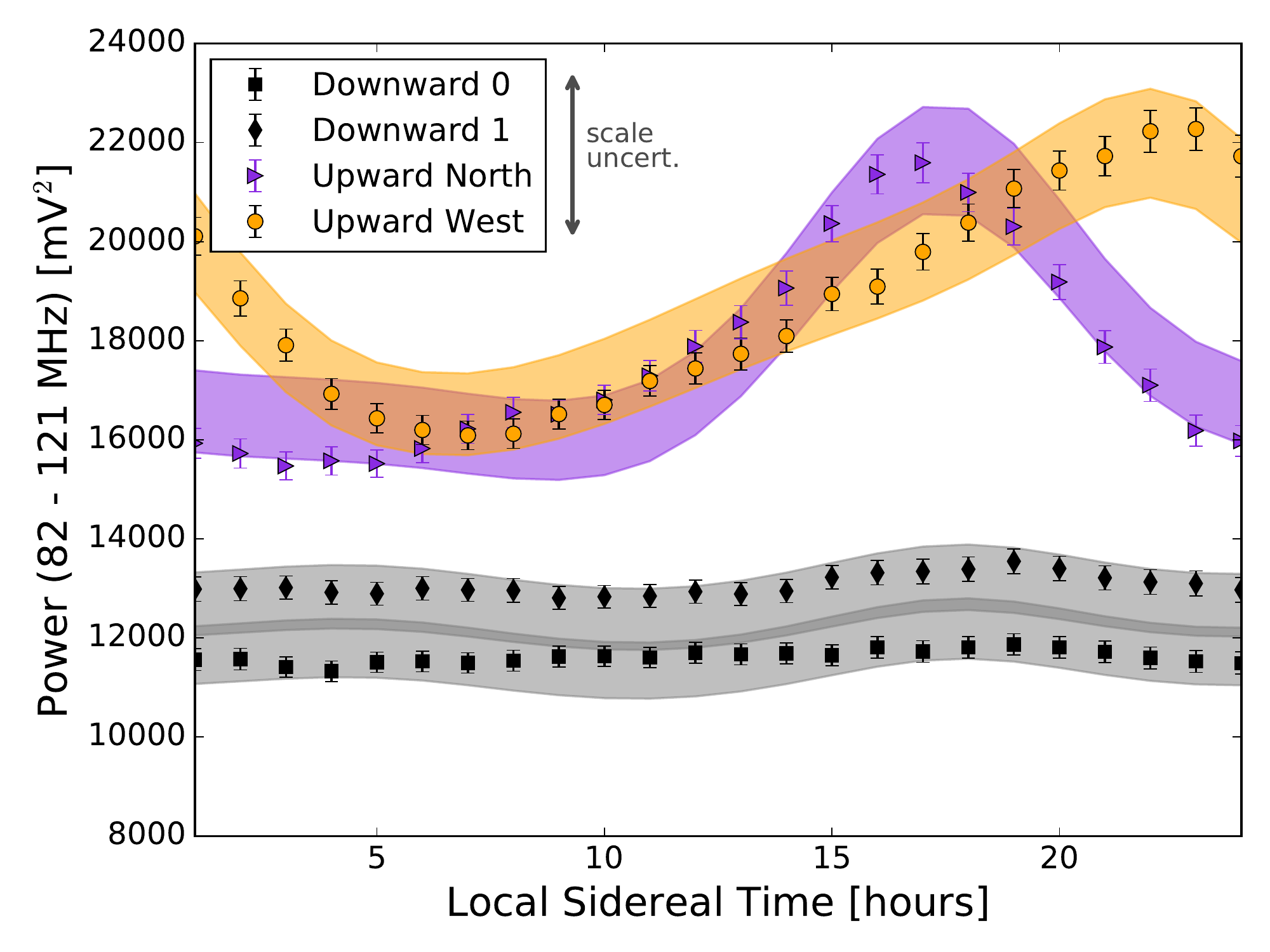}
\caption{Left: Brightness temperature of the diffuse Galactic emission as a function of right ascension and declination at an observing frequency of 101 MHz, as given by LFmap. The broad line and hashed area indicate the horizon at the ARIANNA site at a given time. The markers indicate the pointing of the three different HRA antennas at that time. The thin lines indicate how the pointing changes throughout the day. Right: Recorded power in \unit[256]{ns} as function of local sidereal time. The markers indicate the measured average power in each antenna of station X from data taken over a period of four months. The bands of the corresponding color depict the prediction derived from LFmap and the detector simulation, including systematic effects. The uncertainty on the absolute scale is illustrated by the grey arrow. This uncertainty may shift the curves vertically, but does not influence the shape.}
\label{fig:Galaxy}
\end{figure}

A model of the brightness temperature of the Galactic environment, here LFmap \cite{LFmap}, is used to predict the power that is present in a specific frequency range, $P_{\mathrm{Galactic}}$. As an example, the Galactic emission $T_{\mathrm{brightness}}$ at \unit[101]{MHz} is shown on the left in Figure \ref{fig:Galaxy}. The uncertainties of LFmap are quoted to better than 10\% in absolute scale and to about 2\% in angular variation \cite{LFmap}.

The instantaneous Rayleigh-Jeans power spectral density $S_{\lambda}$ is related to a brightness temperature $T(\theta,\phi,\lambda)$ by:
\begin{equation}
S_{\lambda} = \frac{2ck_B}{\lambda^4}T(\phi,\theta,\lambda)
\end{equation}
The received power $P_r$ is then:
\begin{equation}
P_{r,\lambda} = S_{\lambda} \cdot A_{\text{eff}}
\end{equation}
with $A_{\text{eff}}$ the effective area of the antenna. Using the relation\footnote{Note that $A_{\text{eff}}$ is the absolute part of the effective antenna height $\vec{h}_{\text{eff}}$. Using this relation, the results from WIPL-D have been checked for consistency, see \ref{sec:app_sim}.} between $A_{\text{eff}}$ and the power gain $G$ 
\begin{equation}
A_{\text{eff}}(\phi,\theta,\lambda) = \frac{G(\phi,\theta,\lambda)  \cdot \lambda^2}{4\pi}, 
\end{equation}
and taking into account a factor $\frac12$, because the linearly polarized antenna is only sensitive to one half of the unpolarized emission, gives:
\begin{equation}
P_{r,\lambda} = \frac{ck_B}{4\pi\lambda^2}T(\phi,\theta,\lambda)~G(\phi,\theta,\lambda)
\end{equation}
Integrating over the bandwidth of the LPDA that contains a detectable fraction of the Galactic power (\unit[$\lambda$ = 2.50-3.75]{m}, \unit[$f$= 82-121]{MHz}) and solid angle (all angles above the horizon at a given time), gives:
\begin{equation}
P_{\text{Galactic}} = \frac{ck_B}{4\pi}\int_{\lambda} \int_{\Omega}  \frac{1}{\lambda^2} T(\phi,\theta,\lambda)~G(\phi,\theta,\lambda)~ d\Omega\ d\lambda.
\end{equation}
The model of the Galactic emission is given in a binned map for the whole sky, which turns the integration over all angles into a sum over all steradian bins. The brightness temperature falls monotonically over the frequency band. For simplification, the emission is assumed to be constant over the frequency band and the emission at the central frequency $f_b = 101\ \text{MHz}$ is used. The error induced by this assumption is small with respect to the absolute scale uncertainties discussed in Section \ref{sec:abs}.
\begin{equation}
P_{\text{Galactic}} = \frac{k_B}{4\pi} \cdot (121-83)\text{MHz} \cdot \sum^{\Omega}_i T(\phi_i,\theta_i, f_{b})\ G(\phi_i,\theta_i, f_b) \cdot \delta\omega_i.
\end{equation}
To obtain the voltage $V$measured by the HRA data-acquisition, the following equation holds:
\begin{equation}
V = \sqrt{P_{\text{Galactic}}\cdot Z_{\text{ant}} \cdot G_{\text{sys}}},
\end{equation}
with $Z_{\text{ant}}$ the antenna impedance and $G_{\text{sys}}$ the gain of the system at the central frequency of \unit[101]{MHz}. The latter consists of half the gain of the amplifier (for an impedance matched system \cite{2002aaa..book.....K}) and losses in cables and connectors. For direct comparison to the data, the finite data length of \unit[256]{ns} has to be normalized. 

The model of the Galactic emission is given in right ascension and declination, so at every time of the day the emission pattern will be shifted with respect to the local coordinates $(\theta, \phi)$. As the signal is periodic in LST, the predicted signal power for every hour in LST for all channels is calculated. The combination of the antenna sensitivity pattern and the Galactic emission is different for all antenna orientations, as illustrated on the left of Figure \ref{fig:Galaxy}, which leads to a different shape of the LST dependence for each antenna orientation.

To compare model results to data, the predicted Galactic power $P_{i, \text{Galactic}}$ per channel $i$ is co-added with thermal noise $P_{\text{thermal}}$ from amplifiers and antenna. Even at Galactic minimum the Galactic emission detected in an antenna is still considerable, however not sufficient to explain all power observed in the system. Unfortunately, the precise fraction of thermal noise is difficult to estimate as it would require a measurement at cold temperatures, isolated from any emission from the Galactic sky. Instead, $P_{\text{thermal}}$ was chosen as the best theoretical estimate for the noise of an amplifier at cold temperatures and is also constant in time. As this is an underestimate of the noise contribution and there are differences between channels, we allow for a free parameter $\alpha_i$ when fitting the total power variation $P_{i, \text{total}}$ to the data per channel with
\begin{equation}
P_{i, \text{total}}(t) = P_{\text{i,Galactic}}(t) + \alpha_i \cdot P_{\text{thermal}}.
\end{equation}
The obtained values are $\alpha_i = [1.18 , 1.24,  1.23 , 1.30]$ and the observed spread is consistent with the uncertainties discussed in Section \ref{sec:abs}.   

The best fitting prediction of LFmap (bands) and the measured power (markers) are compared in Figure \ref{fig:Galaxy}. The shape of the prediction is determined by the response of the antenna and the distribution of the Galactic emission. The thermal noise contribution is a constant offset and does not affect the shape of the oscillation in time. The bands include the uncertainties induced by the uncertainty on the antenna orientation, the uncertainty of the antenna model itself and the angular variation of LFmap. They account to between 5\% and 10\% of the prediction. The additional uncertainties discussed in Section \ref{sec:abs} as well as the uncertainty on the absolute scaling of LFmap lead to an absolute scale uncertainty that is indicated with the vertical arrow in Figure \ref{fig:Galaxy}. The scale uncertainty allows to shift the complete model up and down within the arrow which corresponds to $1\sigma$.

Agreement is found between the shape of the predicted variation and the measured power variation. For both upward-facing antennas the maxima, minima and phase have been accurately predicted and all data points are compatible with the prediction within $1 \sigma$.  Given the fitted thermal noise, the contribution of the Galactic power to the total noise budget $P_{\text{i,Galactic}}/P_{i, \text{total}}$ accounts to between 40\% and 50\% in this frequency range for the upward facing antennas and between 10\% and 15\% for the downward facing antennas. This is in line with what is expected from a well-designed antenna system \cite{2002aaa..book.....K}. The two downward channels are, in principle, identical. The observed difference between the two is an additional indication of the scale uncertainty. The agreement of data and prediction indicates that the antenna response is adequately modeled and that the absolute scale is in agreement with the predicted emission strength. 


\section{Identification of air shower signals}
Several ground-based radio air shower experiments have attempted to detect air showers solely based on their radio emission (e.g. \cite{2012JInst...7P1023A,2008ICRC....5.1081A,2013NIMPA.725..133K}). The challenge usually is to select air shower signals from a background of highly abundant anthropogenic noise pulses. While it is possible to record all pulses and separate signal from background by a search for coincidences with other particle detectors, most experiments deal with the high data-rates by using sophisticated algorithms and trigger criteria at the cost of a loss in efficiency. There are several observables, such as polarization, that can be used to confirm the cosmic origin without relying on a coincidence with a particle detector \cite{2016PhRvD..93l2005A}. However, all dedicated surface air shower experiments ultimately rely on a particle array to uniquely flag air showers for their analyses. 

In this section it will be shown that the ARIANNA site is a remarkable location and that the HRA stations do not require external confirmation to detect the radio emission of air showers. The background rates are so low that the experiment can be run at full efficiency at a low energy threshold. Also, the site has virtually no anthropogenic impulsive background that can be mistaken for signal. 

The identification of air shower signals initially proceeds by selecting events from station X with upward facing antennas (see Chapter \ref{chap:exp_setup}). The data from the other stations in the HRA will be used to support and augment the results. In particular, we determine the energy and arrival direction for an interesting event that was detected in five stations in the array. 

\begin{figure}
\includegraphics[width=0.5\textwidth]{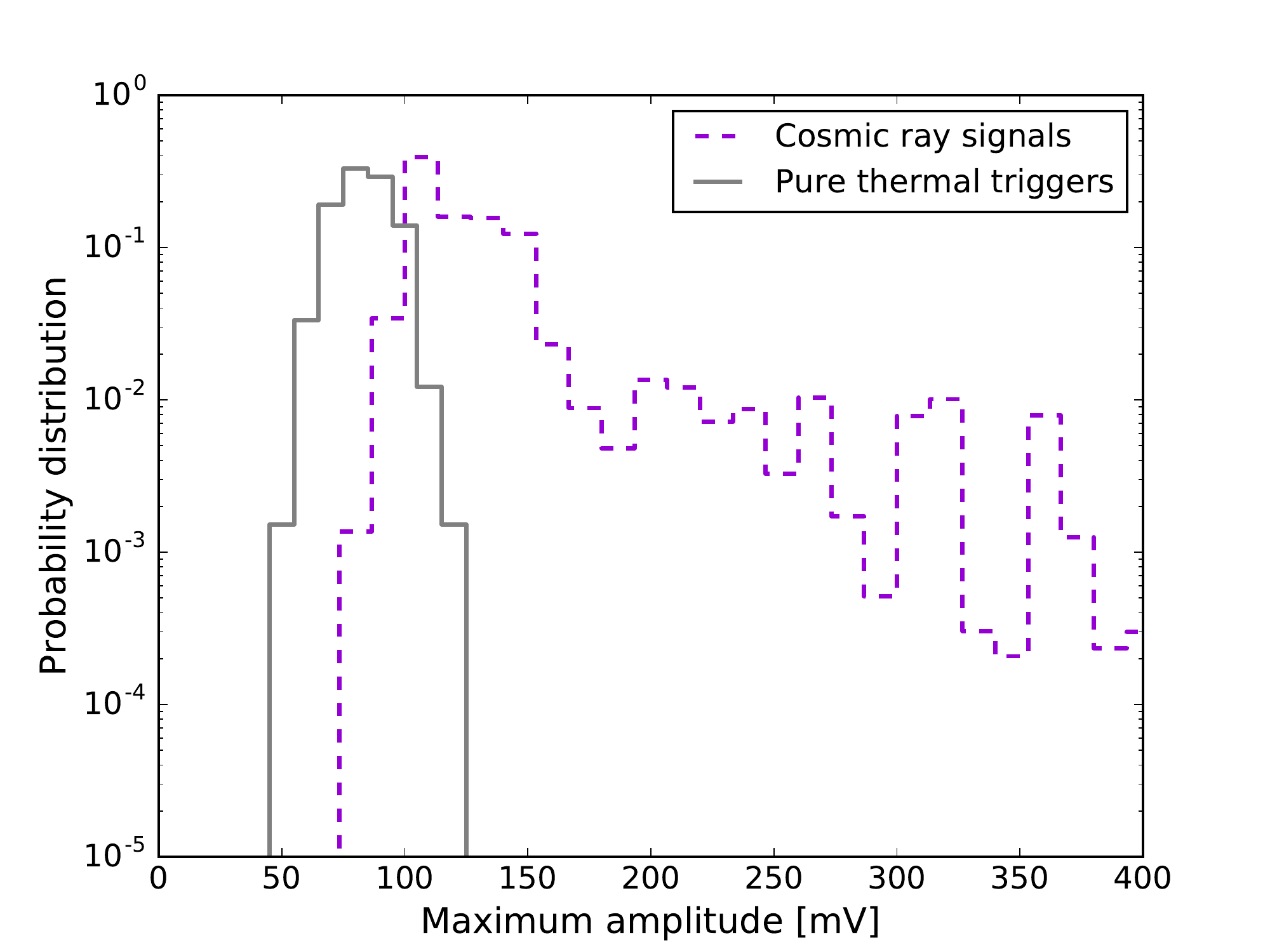}
\caption{Probability distributions of simulated cosmic ray amplitudes as detected in station X of the HRA in comparison to the expected distribution of pure thermal noise. Isotropic arrival directions and a flux of $E^{-3.3}$ are assumed. It should be noted that both distributions shift slightly, when the trigger thresholds are adapted. The depicted distributions were created for an average threshold of \unit[75]{mV}.}
\label{fig:max_amp}
\end{figure}

First, we describe a very basic search for cosmic ray candidates that relies purely on two assumptions: Air shower signals have higher amplitudes than the thermal background (see Figure \ref{fig:max_amp}) and air showers are unlikely to be clustered in time. This first search is neither optimized for efficiency (finding all cosmic rays) nor for purity (finding exclusively cosmic rays), but more directly illustrates the analysis methods and the quality of the HRA data, compared to the more sophisticated analysis methods described in Section \ref{sec:correl}.

\subsection{Initial candidate selection}
\label{chap:ov_sel}
To obtain a homogenous data-set the L1 trigger was implemented in software and applied again as a filter offline to all the data in which the L1 trigger had not been applied on station-level. As is visible in Figure \ref{fig:Rates}, the stabilizing effect of the L1 trigger is obvious in December, while the small differences in January can be explained by data that had failed the L1 criterion at the station level, but was still kept as an event due to the pre-scale trigger. Those events are tagged specifically by the DAQ.

\begin{figure}
\includegraphics[width=\textwidth]{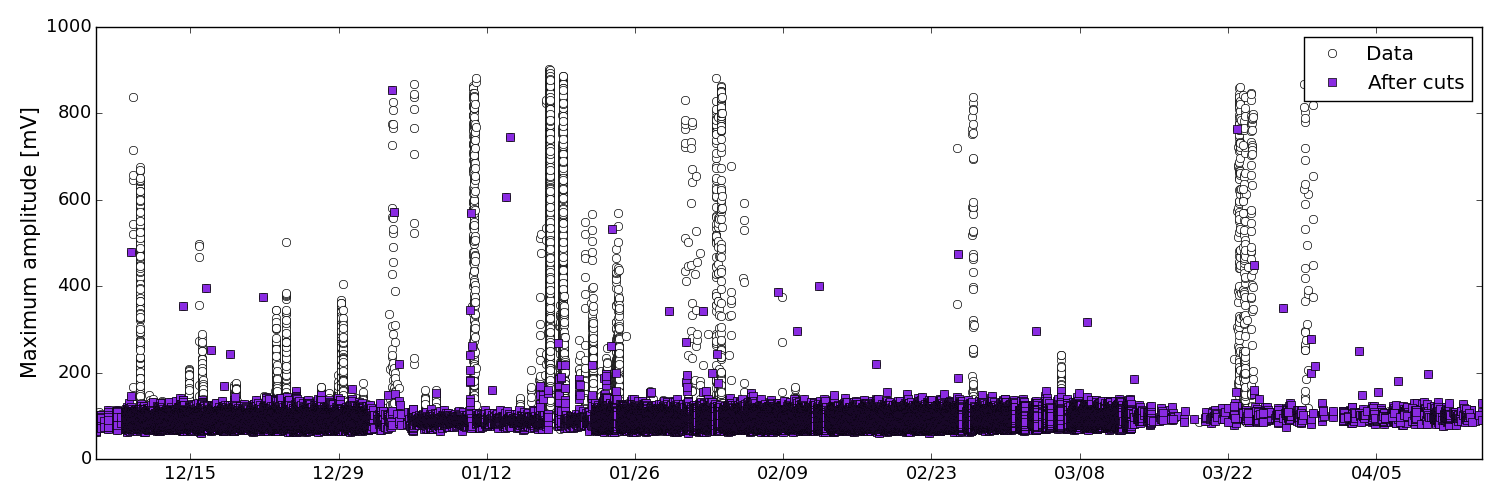}
\caption{Maximum amplitude in any waveform recorded per event as function of time. Shown are all events in the data set (open circles) and the events remaining after L1 cut and cluster cut (filled squares).}
\label{fig:Threshold}
\end{figure}

Figure \ref{fig:Rates} also shows that there are periods with higher trigger rates that are not significantly affected by the L1 criterion. As reported earlier, these are correlated to periods of high winds \cite{2015APh70} or increased solar activity \cite{NellesICRC2015}. A cluster cut was implemented to remove these events from the data-set. The data is divided into sequential periods of time: $T_c$. Whenever two events above a certain threshold $A_c$ occur within a short time period $T_c$, all events are removed within this period. The cut is implemented in two ways:  to remove long periods of very high signals, such as storms: $A_c=300~\text{mV}$ and $T_c=3600~\text{seconds}$, and to remove short periods of smaller signals from e.g. local transmitters:  $A_c=150~\text{mV}$ and $T_c=60~\text{seconds}$. The effect of the cluster cut is illustrated in Figure \ref{fig:Threshold}, where periods of increased signal amplitude are vetoed. 

The accumulated dead-time due to the two cluster cuts is 6211 and  693 minutes, respectively, with the first being dominated by three multi-day storms.
They represent a conservative estimate of the induced dead-time corresponding to a loss in live-time of 6.6\% of the total season. So, applying this cut removes about 7\% of all cosmic rays or neutrinos randomly distributed in time, as expected from non-explosive sources.

After the L1 and cluster cut, more than 99\% of the events can be found just above the trigger threshold and originate from random thermal fluctuations. Only 92 events remain above \unit[150]{mV}, which is roughly twice the trigger threshold. Some of these events are likely to have survived the cluster cut, as a non-physically motivated cluster cut can hardly be fully efficient. However, these events are consistent with a random distribution in time and warrant further study.

\subsection{Correlation analysis based on simulations}
\label{sec:correl}
As reported earlier, the ARIANNA collaboration follows the strategy of using simulated waveforms as search templates to identify signals from neutrino interactions \cite{2015APh70}. As discussed above, air shower signals show similar distinct characteristics, so it was chosen to apply a simulation-based correlation analysis to identify air shower signals. If the initial cut indeed selected mostly air shower signals and the correlation method has merit, then a correlation analysis will reveal the same distinct subset. Also, refining the correlation analysis with measured air shower signals, allows us to study its capabilities on real data and draw conclusions for the analysis efficiency for neutrinos of ARIANNA. 

Using the simulations and templates described in Section \ref{chap:emiss}, it is first evaluated, how similar all cosmic ray templates are with respect to each other and then how the data correlates to the templates. For this study, the variable $\chi$ is used to describe the best correlation with one reference template out of four station channels. The variable $\bar{\chi}$ indicates the average of all best correlations $\chi$ over all available reference templates. 

\begin{figure}
\includegraphics[width=0.6\textwidth]{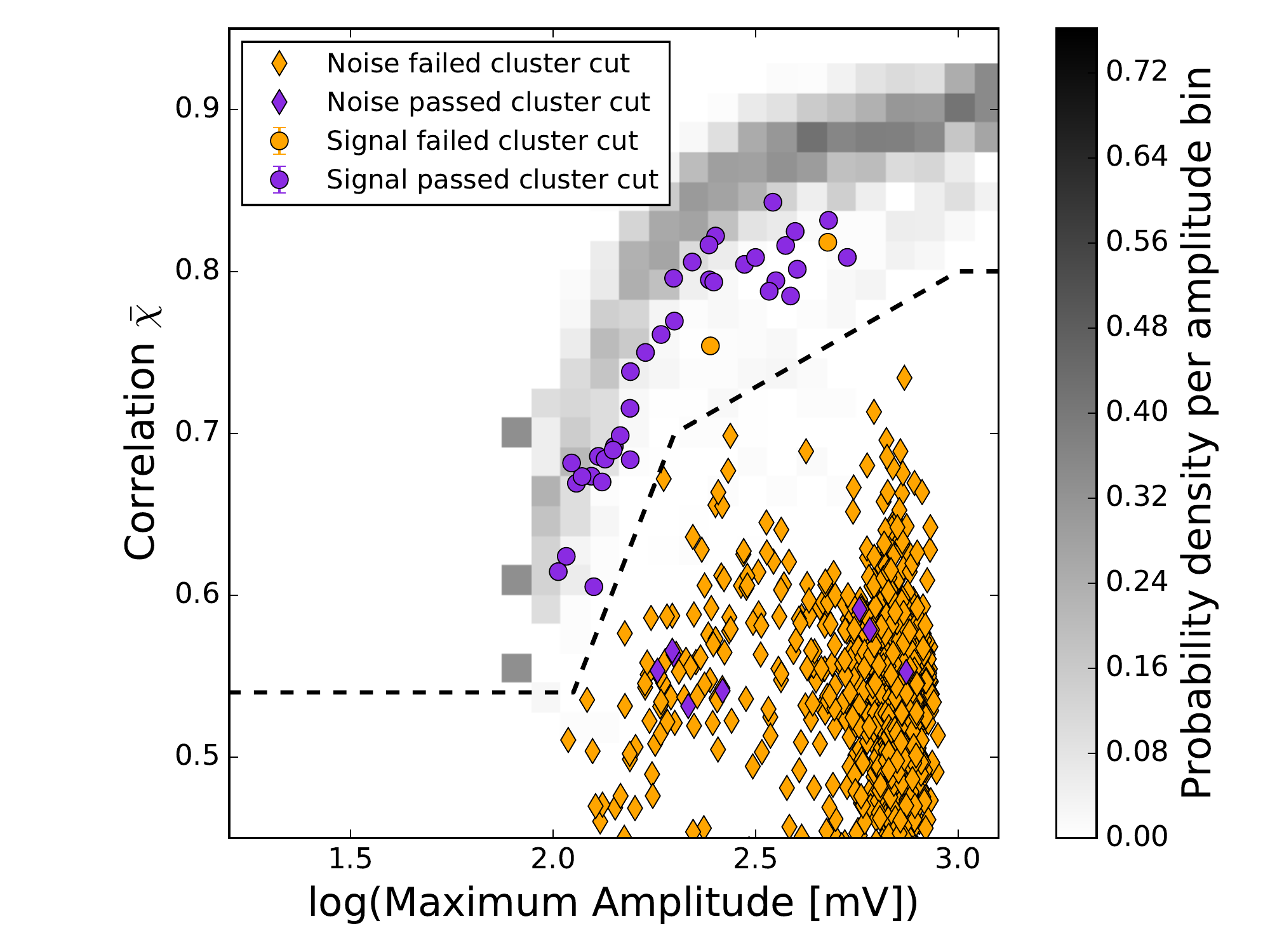}
\caption{Average correlation $\bar{\chi}$ as function of the maximum signal amplitude over all four channels per event. The background density map shows the probability distributions for simulated air showers given an amplitude. The markers show the average correlation value $\bar{\chi}$ of measured events. The line indicates a cut separating the cosmic ray signals from the background. All diamonds are background events, while the signals are indicated by circles. All events corresponding to light markers are vetoed by the cluster cut.}
\label{fig:corr_distr}
\end{figure}

For every simulated event (four channels including noise) the values for $\chi$ are calculated for all other 200,000 noise-free templates. As discussed in Section \ref{chap:emiss} the templates cover the whole expected parameter space of different distances to the shower axis, arrival directions, energies and heights of shower maximum. The resulting distribution of correlation values $\chi$ per simulated event typically is Gaussian centered around $\bar{\chi}=0.8$. The values for $\bar{\chi}$ of these distributions range from 0.6 to 0.9 and are mostly a function of signal amplitude. Signals with high amplitudes correlate better with the templates than those where signal and noise are of similar amplitude, as the correlation is then dominated by noise. The theoretical distribution of $\bar{\chi}$ is used to devise a cut for the data to identify air shower signals, as shown in Figure \ref{fig:corr_distr}.

It should be noted that all distributions have a longer tail towards low correlation values ($<0.6$) and about 2\% of the templates show an overall poor correlation with all other templates. These templates are due to an artifact of the CoREAS simulations. At frequencies above the coherence cut-off, the simulations are dominated by numerical noise, which scales with the energy of the shower \cite{2013AIPC.1535..121H}, allowing cases where numerical noise passes the HRA trigger condition. A perfect filter for these events is challenging without biasing against small amplitude pulses, so a contamination of about 2\% is expected. 

Figure \ref{fig:corr_distr} shows the average correlation $\bar{\chi}$ of measured events with the search templates as a function of the maximum amplitude detected per event. In the background the same distribution is shown as obtained from correlating mock-data from simulations with the templates. The measured events surviving the cluster cut follow the distribution obtained from air shower simulations. At larger amplitudes the correlation obtained is, however, slightly smaller than predicted. This indicates that the simulations are not a perfect description of the detector. The measured pulses tend to be longer than predicted, which could indicate additional effects of the ice such as scattering, which are not modeled yet. Still, the differences in correlation are small. One can also observe that the simulations have no limit on the amplitude like the data, which is assuming a perfect linear amplification in dynamic range of the amplifiers of the HRA. 

It can be seen that the cluster cut vetoes two events that belong to the cosmic ray population and does not exclude eight events that belong to the noise population, illustrating the difficulty of optimizing a cluster cut. Retaining 36 out of 38 signals is compatible with a live-time loss of 7\%. We therefore choose to cut on correlation as indicated in Figure \ref{fig:corr_distr} as final event selection criteria, together with the narrow-band line removal L1 at station level. The cluster cut is not applied. The cut on the correlation (dashed line) contains 98.2\% of the simulated templates, which is about equal to the fraction of simulated pulses that are not affected by the CoREAS artifact. This value is taken as conservative analysis efficiency of retrieving cosmic ray signals that have triggered station X. 

\begin{figure}
\includegraphics[width=0.33\textwidth]{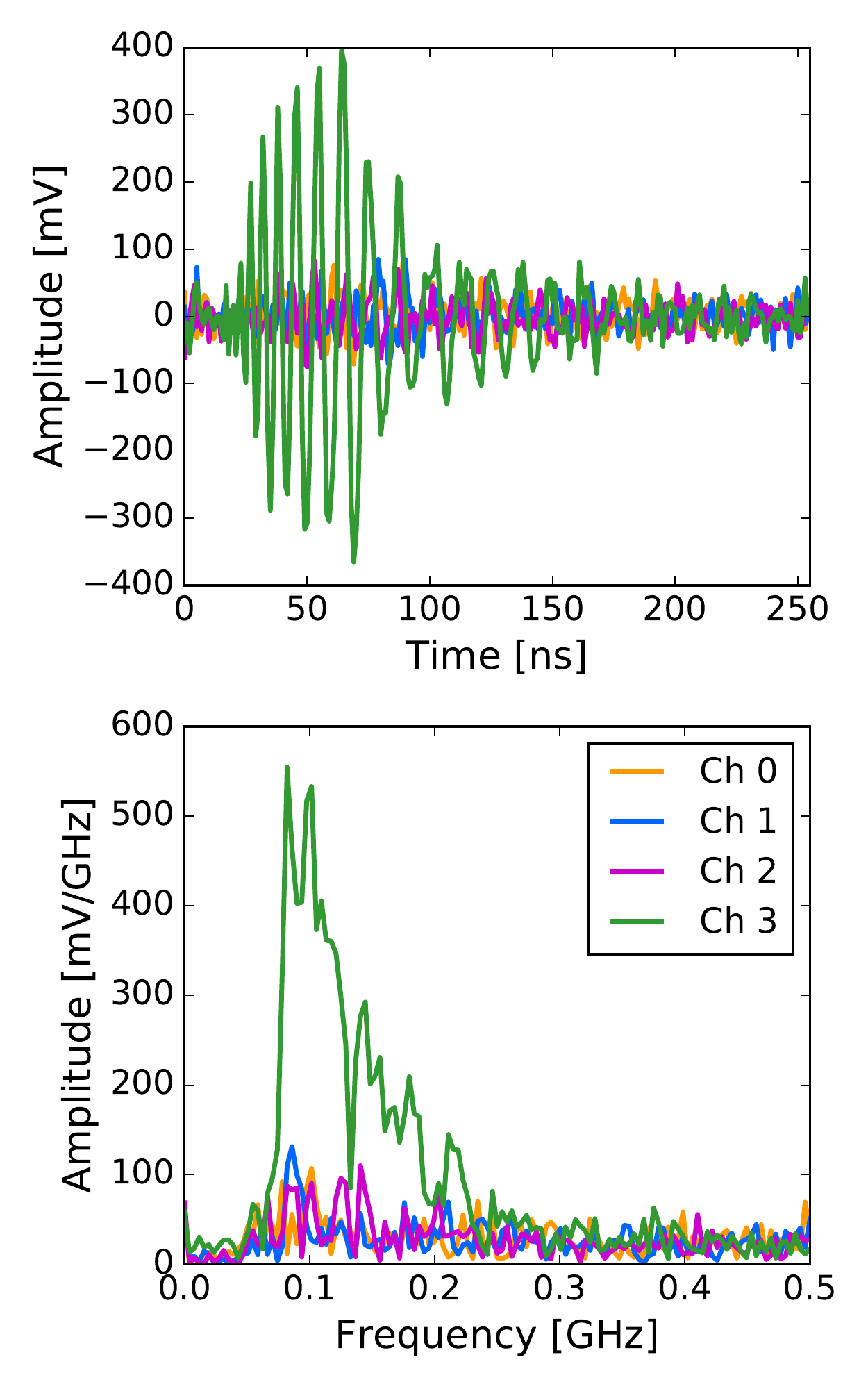}
\includegraphics[width=0.33\textwidth]{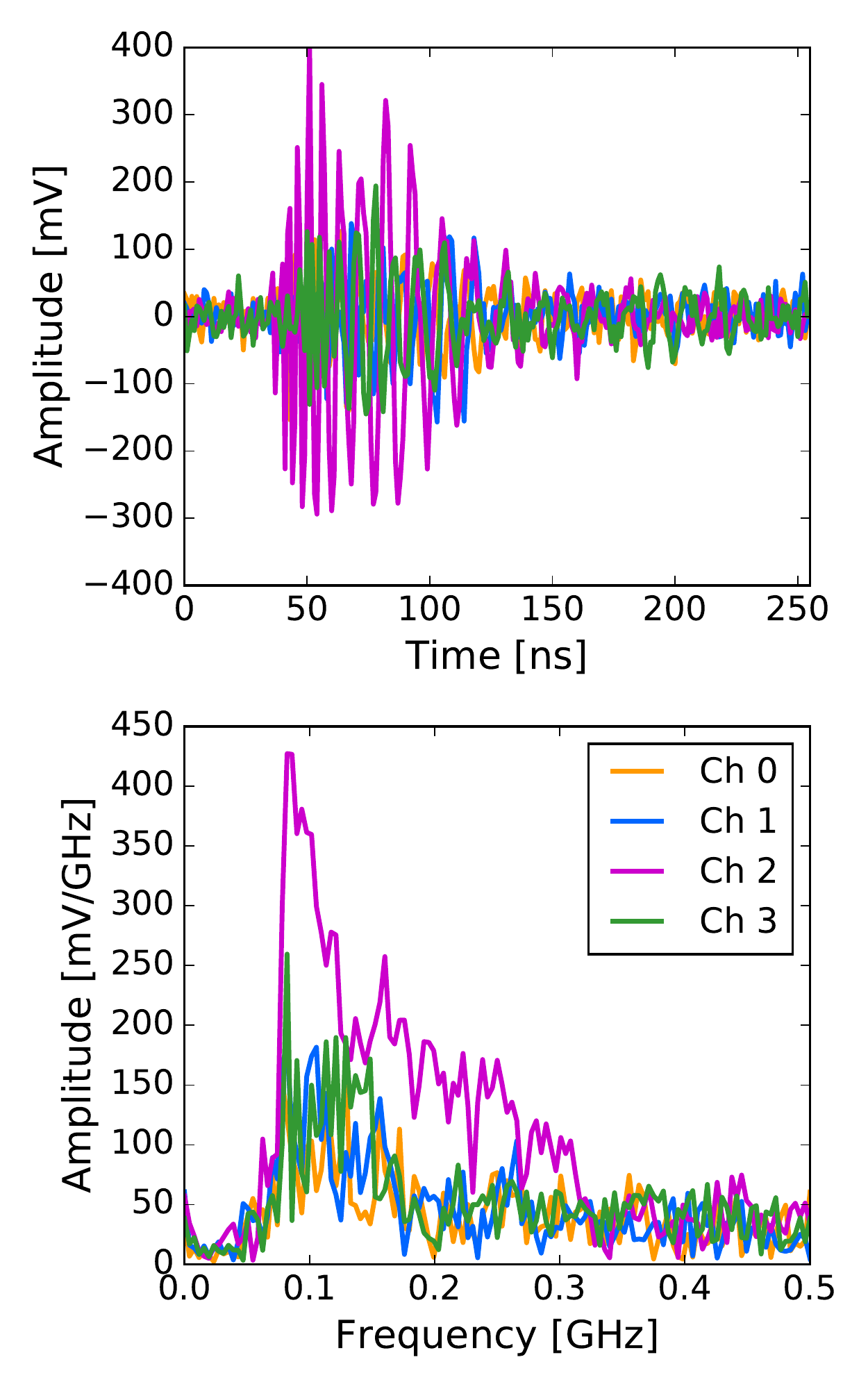}
\includegraphics[width=0.33\textwidth]{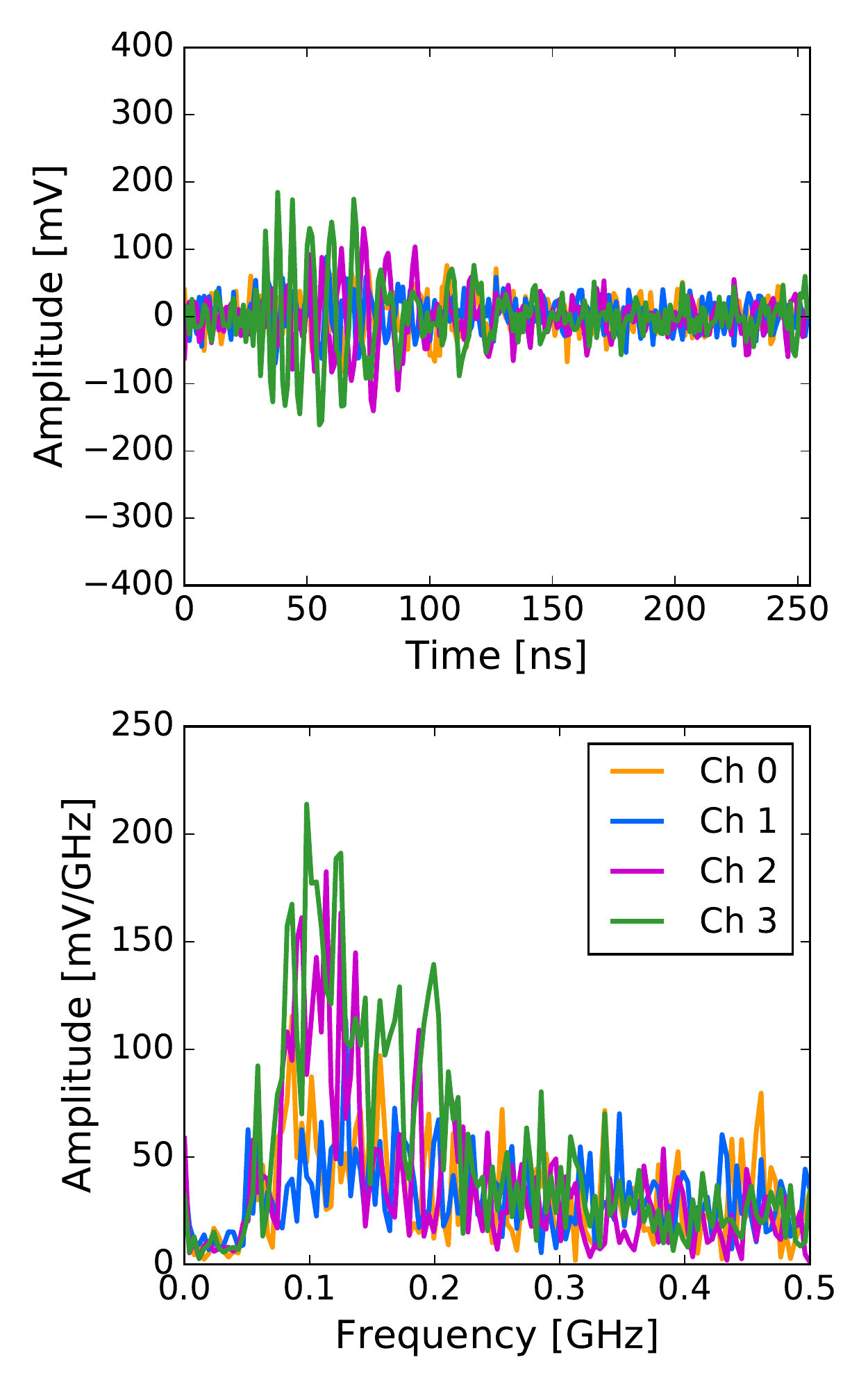}
\caption{Three examples of events selected during the cosmic ray search. The top row shows the recorded waveforms, the bottom row the corresponding frequency spectra. Channels 0 and 1 are the downward facing antennas, antenna 2 points upward north and antenna 3 upward west. }
\label{fig:Examples}
\end{figure}

The final event selection selects 38 cosmic ray events in station X of the HRA in the 2015/16 season. An overview of the cuts is given in Table \ref{table:cuts}. Three example events are shown in Figure \ref{fig:Examples}. All events show the fall-off in frequency characteristic of air shower signals, the typical rise and fall of a pulse, and most are strongly polarized in one of the two orthogonal antenna polarizations.

\begin{table}
\begin{tabular}{llll}
\hline
Description&Number of events&Fraction&Note\\
\hline
All data & 653,447 & 100\%\\
After L1 & 578,745 & 88\% & On station level: 75\%\\
\hline
Option 1: \\
After cluster cut & 538,198 & 82\% & Live-time loss: 6.6\%\\
Events above \unit[150]{mV}& 92 & 0.01\%&Unclear contamination with noise\\
\hline
Option 2:\\
Cut on correlation & 38 &0.005\%& $>$ 98\% analysis efficiency, 100\% live-time\\
\hline
\end{tabular}
\caption{Overview of the different cuts to obtain the event selections as discussed in sections \ref{chap:ov_sel} and \ref{sec:correl}. Two different options are given, of which the latter is used as final cut criterion.}
\label{table:cuts}
\end{table}

\begin{figure}
\includegraphics[width=\textwidth]{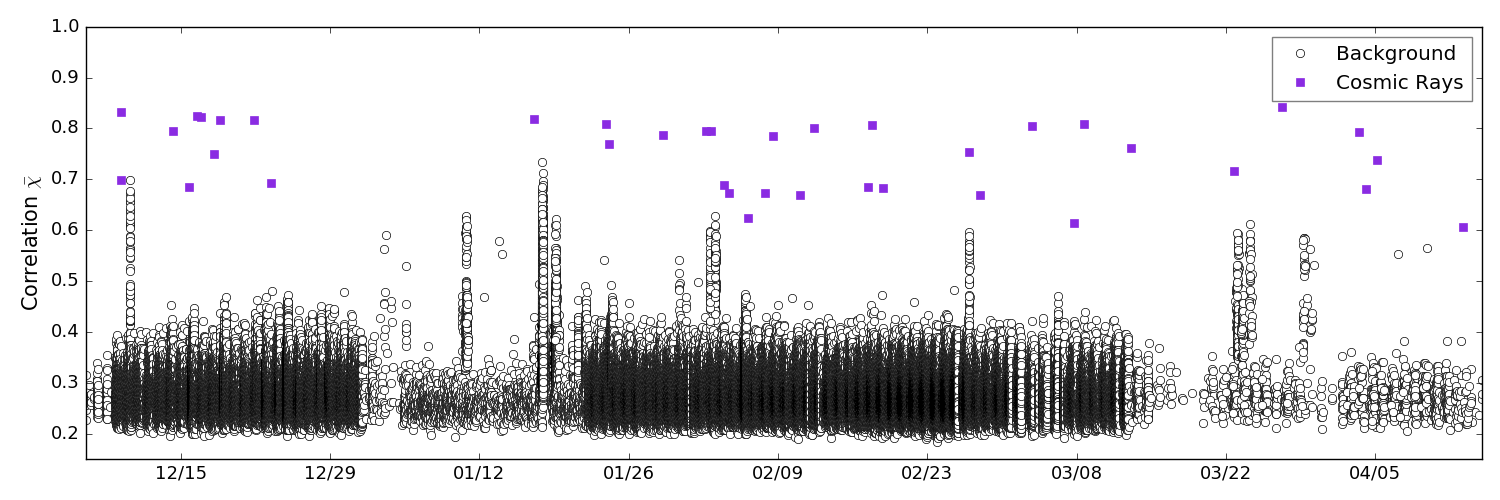}
\caption{Average correlation values $\bar{\chi}$ for all data of station X as function of arrival time. Shown are all events in the data set (open circles) and the data events after using the selection as shown in Figure \ref{fig:corr_distr} based on amplitude and correlation (filled squares).}
\label{fig:corr_overview}
\end{figure}

Next to ensuring that the identified signals indeed show the expected shape, several cross-checks have been conducted to ensure that these events are valid air showers. For example, there is an overlap of 25 events between those 92 events that survived the cluster cut and the amplitude cut of 150 mV and the 38 selected events. Given the unknown noise contribution in the 92 events, it can only be concluded that selecting 25 out of 38 by an amplitude cut is within what has been expected from simulations. In addition, it has been confirmed that the selected events are compatible with stemming from a Poisson distribution in time. There is no evidence for a periodicity in the event times and the distribution of the inter-arrival times is well-described ($\chi^2/\text{ndf}=0.8$, $p=0.6$) by an exponential function. The distribution of arrival times is shown in Figure \ref{fig:corr_overview}. Furthermore, all events show larger amplitudes in the upward facing antennas than in the downward facing antennas, which is consistent with directional antennas and signals arriving from above. The ratio between the amplitudes in the two upward facing antennas is compatible with a uniform distribution. As air shower signals are at the ARIANNA-site to first order polarized horizontally and orthogonal to the arrival direction, a uniform distribution of amplitude ratios is compatible with random arrival directions. A final cross-check to confirm the cosmic origin of these signals is the search for coincidences with other stations. 

\subsection{Coincidences with station X}
Simulations indicate that about 40\% of cosmic ray events measured in station X will trigger also another station in the HRA, even though those stations have antennas that face downwards. However, the L1, which is extremely efficient for neutrino and air shower signals arriving in the antenna front-lobe, is less efficient for back-lobe events. In the back-lobe, the antennas act like a dipole antenna and therefore the signal is likely to be strongest in one frequency bin that corresponds to the longest antenna tine and will trigger the L1 veto. Consequently, the efficiency for back-lobe cosmic ray signals is about 50\%. When also correcting for the live-time of the stations, we expect $7\pm3$ out of 38 air showers to have been measured in coincidence with at least one other station.

The HRA stations have coordinated absolute times to one second accuracy. While every station uses the time set by the Iridium network, which allows for a time-coordination of $0.01$ s, the stations currently only record the second of each event and the time-difference between events in microseconds \cite{2014IEEE62}. In an initial search, multi-station coincidences were identified using a $\pm 1$ second window around the time of the air shower pulse in station X. We find coinciding events for five air showers. On visual inspection, all five waveforms show a pulsed signal, which is in agreement with the predicted $7\pm3$ coinciding events. A search with a broader search window has revealed no additional pulses, only random coincidences of thermal triggers. 
Of the five coincidences, four events are a two-fold coincidence involving only one other station, while one is a coincidence of five stations. The expected number of five-fold coincidences was 0.1. 

\begin{figure}
\includegraphics[width=0.33\textwidth]{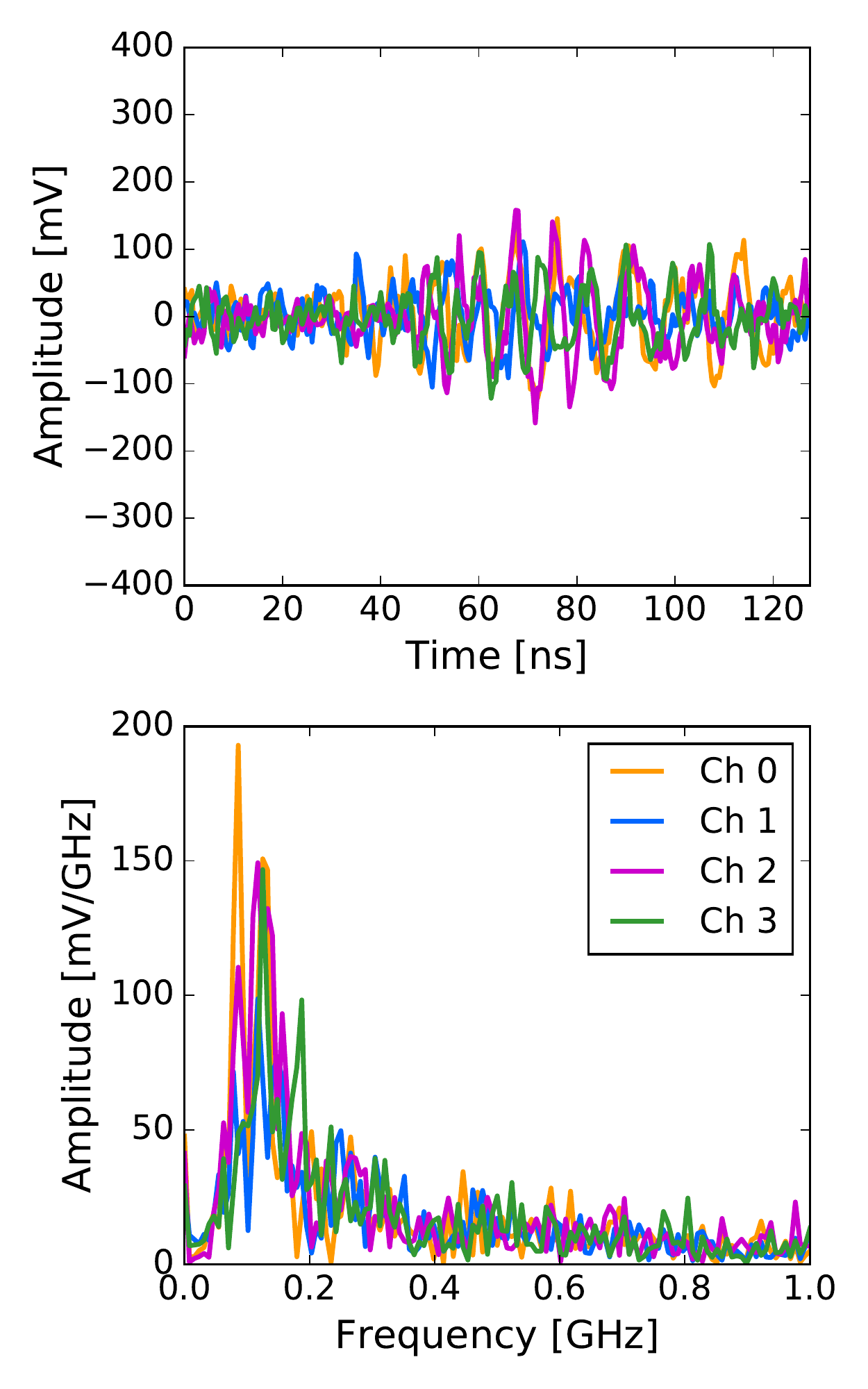}
\includegraphics[width=0.33\textwidth]{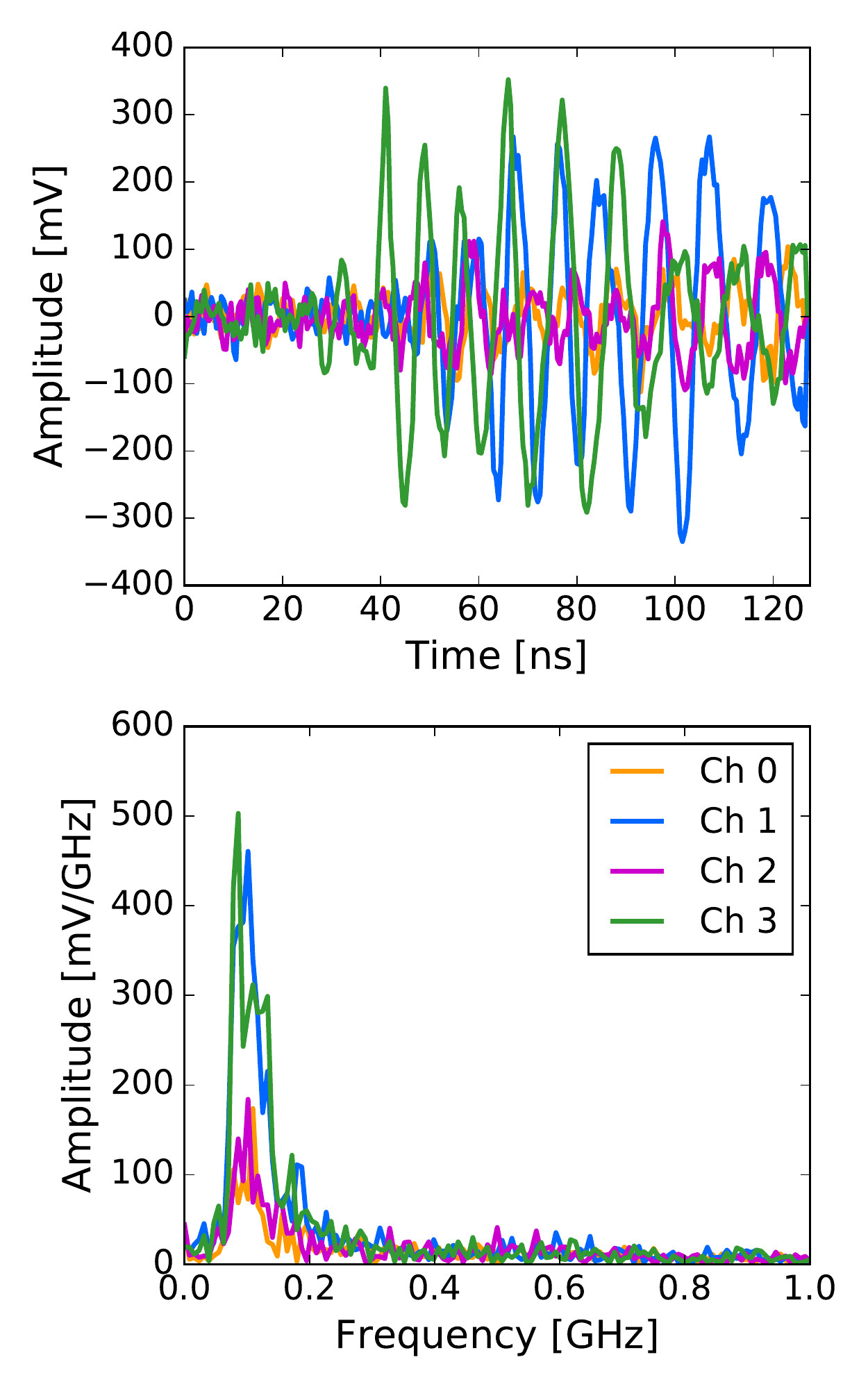}
\includegraphics[width=0.33\textwidth]{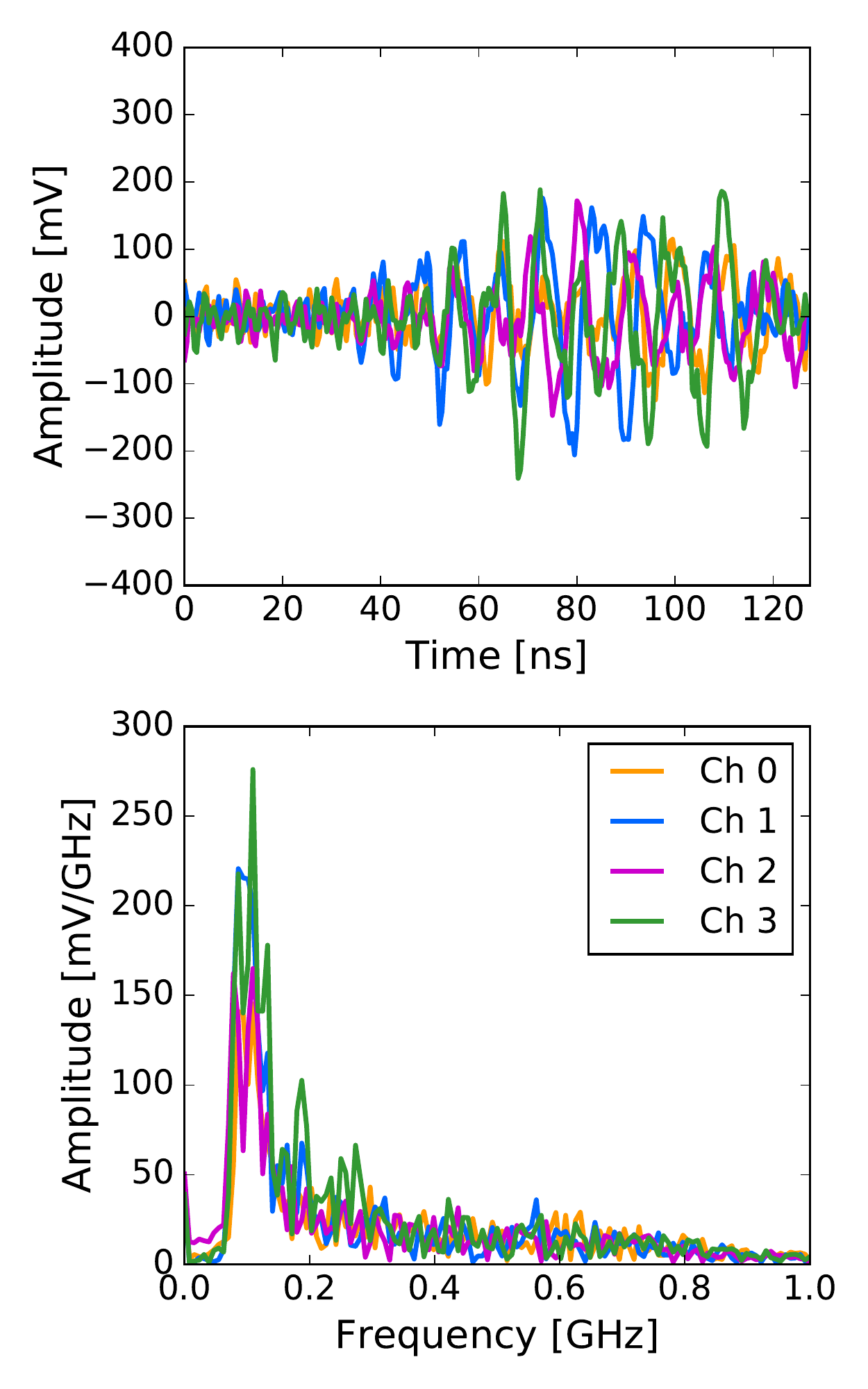}
\caption{Three examples of signals detected in downward-facing antennas in coincidence with air shower signals in station X. Shown are the waveforms as function of time (top row)  and the corresponding amplitude spectra (bottom row). It should be noted that both time and frequency resolution are different in the standard HRA stations compared to station X.}
\label{fig:Examples_down}
\end{figure}

The pulses detected through the back-lobe have different characteristics than those detected in the front-lobe due to the rather different antenna response as discussed in Section \ref{chap:emiss}, especially with respect to high frequencies. The expectations are consistent with the observed waveforms (Figure \ref{fig:Examples_down}), providing evidence for the downward direction of the coincidence events, and the cosmic-ray hypothesis. 

\subsection{Coincidences between standard stations}
The coincident waveforms in standard HRA stations are not as unique as the ones in station X. Since the back-lobe suppresses high frequencies and all pulsed background signals arrive from above, there is insufficient discrimination power between cosmic rays signals and noise to reliably identify cosmic ray events in these stations. Also, due to the faster sampling, the recorded waveforms in the regular HRA stations are a factor of two shorter than the ones of station X, which means that the full pulse of an air shower is not always recorded, again hampering the identification of cosmic ray events. 

If we repeat the correlation study from Section \ref{sec:correl} for simulated events including noise with clean templates for regular HRA stations, the signal space is no longer fully disjunct from the tails and statistical fluctuation of the background noise. As opposed to the front-lobe templates, the back-lobe templates show a broad distribution of correlation with each other, so they cannot be used to uniquely identify air shower signals without additional information such as the arrival direction. That said, the templates show a correlation better than pure thermal noise, so they can still be used to tag likely air shower signals. 

Interestingly, all identified coincident signals with station X show correlations between 0.5 and 0.9 with each other, with high correlations between events measured in the same stations. This provides evidence that the back-lobe might be more affected by physical differences between antennas than the arrival direction and therefore might not be uniform between antennas. Evidence can also be found in Figure \ref{fig:Examples_down}, where parallel channels (0 and 2, 1 and 3) are not identical. It was therefore chosen to apply both the L1 and the cluster cut to all data of the standard stations before a back-lobe template is used for a correlation study. All events that show a correlation of better than 0.5 are selected and a coincidence between two or more stations in a time-window of one second is required. This identifies three two-fold, and two three-fold coincidences. For four out of these five coincidence events, station X was in a communications window during which the trigger was disabled. The non-detection of one of the events in station X is compatible with the prediction of a size-limited signal footprint. 

If considering all ten identified coincidences, the central station G, is a part of a coincidence event in six cases, which can be explained by its closeness to station X and its central location. All but one station have been found in at least one coincidence. All these characteristics provide confidence that air shower signals have been detected in multiple stations. 

Simulations predict that there should be fewer events in the regular HRA stations as opposed to station X due to the suppressed gain. Since ARIANNA is focussed on neutrino detection, the suppression of cosmic ray signals in the back-lobe signal is desirable. However, upward facing antennas are needed to uniquely tag cosmic rays. The baseline design of ARIANNA includes upward facing antennas at all locations to allow for cosmic ray identification with high efficiency and allow for a full reduction of the relatively small possibility to confuse back-lobe cosmic ray signals with neutrino signals. 

\subsection{Event properties derived from five-fold event}
To illustrate the capabilities of ARIANNA, the five-fold multi-station event discussed in the previous section will be investigated in more detail. The following section describes different methods to obtain the energy and arrival direction for this particular event. 

\subsubsection{Reconstruction of arrival direction}
It is possible to reconstruct the direction of an arriving pulse accurately from two pairs of antennas in a single station using cross-correlation, given that the positions of the antennas are well known, and the timing of the system is stable \cite{2015APh70}. The antenna positions within a station of the HRA have been measured to an accuracy of about \unit[10]{cm} with respect to each other, the timing stability of the electronics boards is better than \unit[100]{ps} and the cable lengths have been measured to sub-nanosecond timing \cite{2014IEEE62}. With a baseline of \unit[6]{m}, this should allow for an angular reconstruction for signals arriving in the forward-direction to the theoretical limit of $0.2^{\circ}$. This number assumes fully efficient fitting algorithms, a limited influence of noise and that signals modulo noise are the same in parallel channels. For back-lobe signals, it is therefore unclear whether the same accuracy can be reached. 

\begin{figure}
\includegraphics[width=0.5\textwidth]{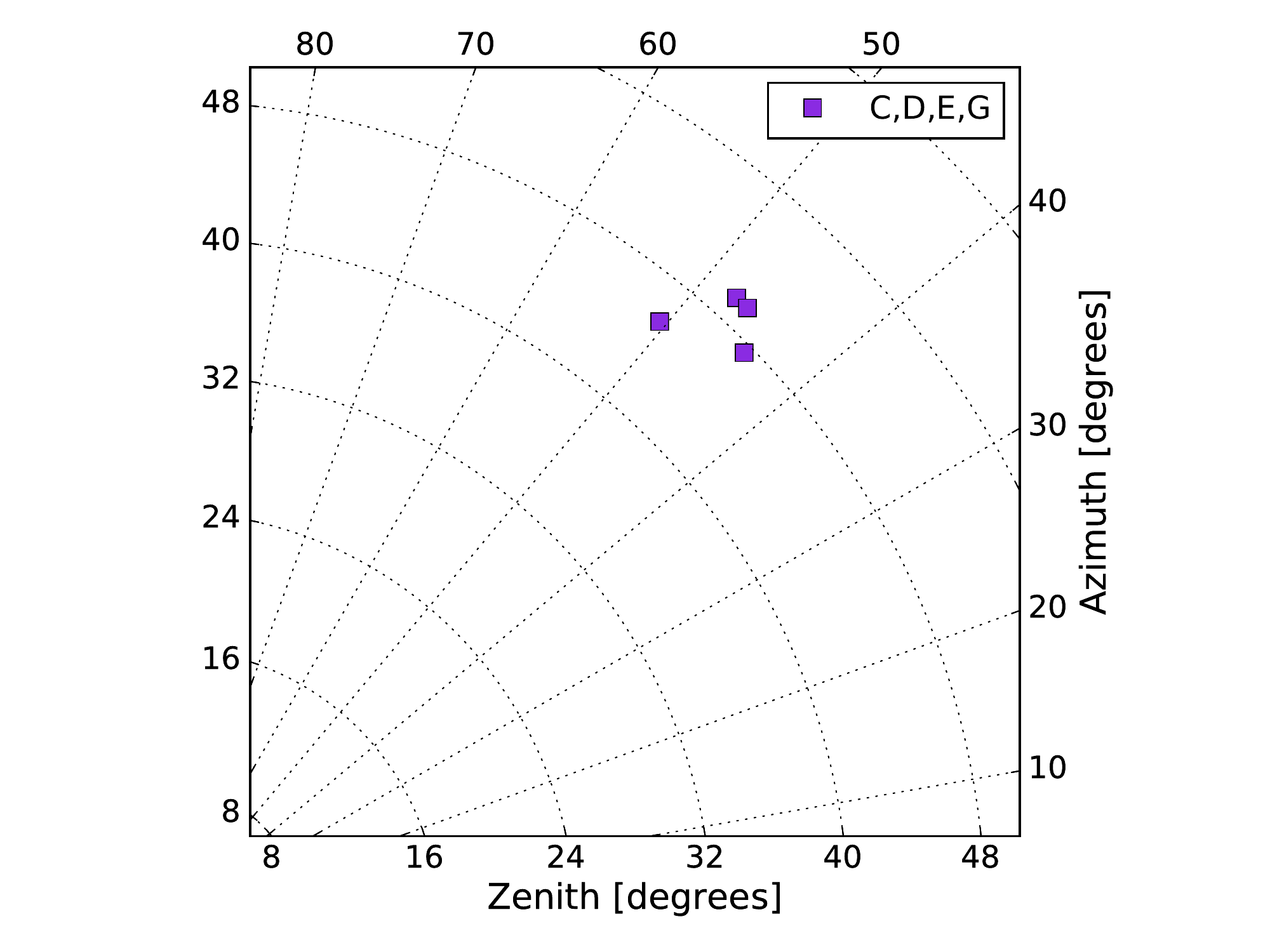}
\caption{The reconstructed signal arrival directions corresponding to the five-fold coincidence event. The event has been detected in the standard stations C, D, E, and G, as well as station X.}
\label{fig:dir}
\end{figure}

Station X, the only station that measures signals in the forward direction of the antenna, has a non-optimal geometry for direction reconstruction. The two upward facing antennas are rotated by 90 degrees to each other and tilted by 45 degrees to vertical. This design was chosen to maximize the detection efficiency for cosmic rays, confirm that the signal is strongly polarized and study the up-to-down ratio in signal amplitude. Consequently, any signal from an air shower will arrive in different sectors of antenna sensitivity and simulations indicate that the signals will look very different due to the variation in gain and group-delay. This complicates using cross-correlations of parallel channels to obtain an arrival direction. An iterative loop of direction fitting and unfolding the antenna response can provide a more complete angular reconstruction, after weighting the two strong signals and two weak signals. For now, the angular analysis excludes station X. The reconstructed arrival directions are shown in Figure \ref{fig:dir}. The clustering is encouraging, however, the space angle between the directions is $15^{\circ}$, which serves as estimator for all events detected by a standard station in the back-lobe. Including Fresnel refraction, the average arrival direction of the detected air shower was found to be $\theta=75.0^{\circ}, \phi=47.3^{\circ}$. 

\subsubsection{Reconstruction of the energy using the slope of the frequency spectrum}
\label{sec:slope}
Typically, the received power on ground is used to reconstruct the energy of an air shower (e.g. \cite{2015JCAP...05..018N}). The arrival direction is determined from timing and the signals are transformed into the shower plane, where a multi-parameter fit is executed, containing a dependence on the distance to the shower axis, the energy and the distance to the shower maximum. This procedure has been optimized for radio measurements in the frequency range of \unit[30-80]{MHz}, with at least 3 independent measurements of the full electric-field of the shower. The energy resolution is at least comparable to particle arrays (25-50\%) \cite{2016PhRvD..93l2005A,2015JCAP...05..018N}. If the methods are adapted to the frequency range of ARIANNA, they should deliver similar results for all air showers measured in at least three stations, which will be less than 20\% of the detected air showers.  

Radio measurements have also shown to be sensitive to the height of shower maximum with accuracies between \unit[15]{g/cm$^2$} for dense (distance between antennas $\sim$\unit[10]{m}) and \unit[40]{g/cm$^2$} for sparse arrays (distance between antennas $\sim$\unit[200]{m}) \cite{2014PhRvD..90f2001A, 2016JCAP...01..052B,2016PhRvD..93l2005A}. A similar sensitivity is likely for ARIANNA, however, we will focus on the energy reconstruction in this analysis.  

The above mentioned reconstruction methods are based on the total power. They do not use the detailed signal shape of the waveforms (their frequency and phase content). It has been suggested (e.g. \cite{2013AIPC.1535..209B,GrebePhD, JansenPhD}) that one can extract shower parameters from just a single measurement of the electric field of the shower. The general form of a signal spectrum has been shown in Figure \ref{fig:sim_pulses}. Air shower signals usually show the strongest power at frequencies \unit[$< 100$]{MHz}, with a characteristically falling spectrum towards higher frequencies. The slope of this falling spectrum, as well as the frequency at which the power is strongest, depend on the distance to the Cherenkov cone, where the slope is the flattest as coherence is obtained for all frequencies. The position of the Cherenkov cone with respect to the shower axis depends on the geometric shower development, mostly the distance to the height of the shower maximum. Given the changing frequency content, the absolute power contained in a pulse is also a function of the distance to the shower axis. However, at the same distance from the Cherenkov cone, the amplitude of the pulse scales on average linear with energy \cite{2015JCAP...05..018N}, so all determining air shower parameters are encoded in the signal shape. In recent air shower experiments, the bandwidth of \unit[30-80]{MHz} limits the power of the spectral slope in comparison to other methods. For a wider frequency band, such as in ARIANNA, this is no longer necessarily true. As the ANITA collaboration has shown \cite{2016APh....77...32S} an estimate of the shower energy with statistical uncertainties of as low as 25\% can be determined from a pulse measurement with a wide frequency-band.

\begin{figure}
\includegraphics[width=0.5\textwidth]{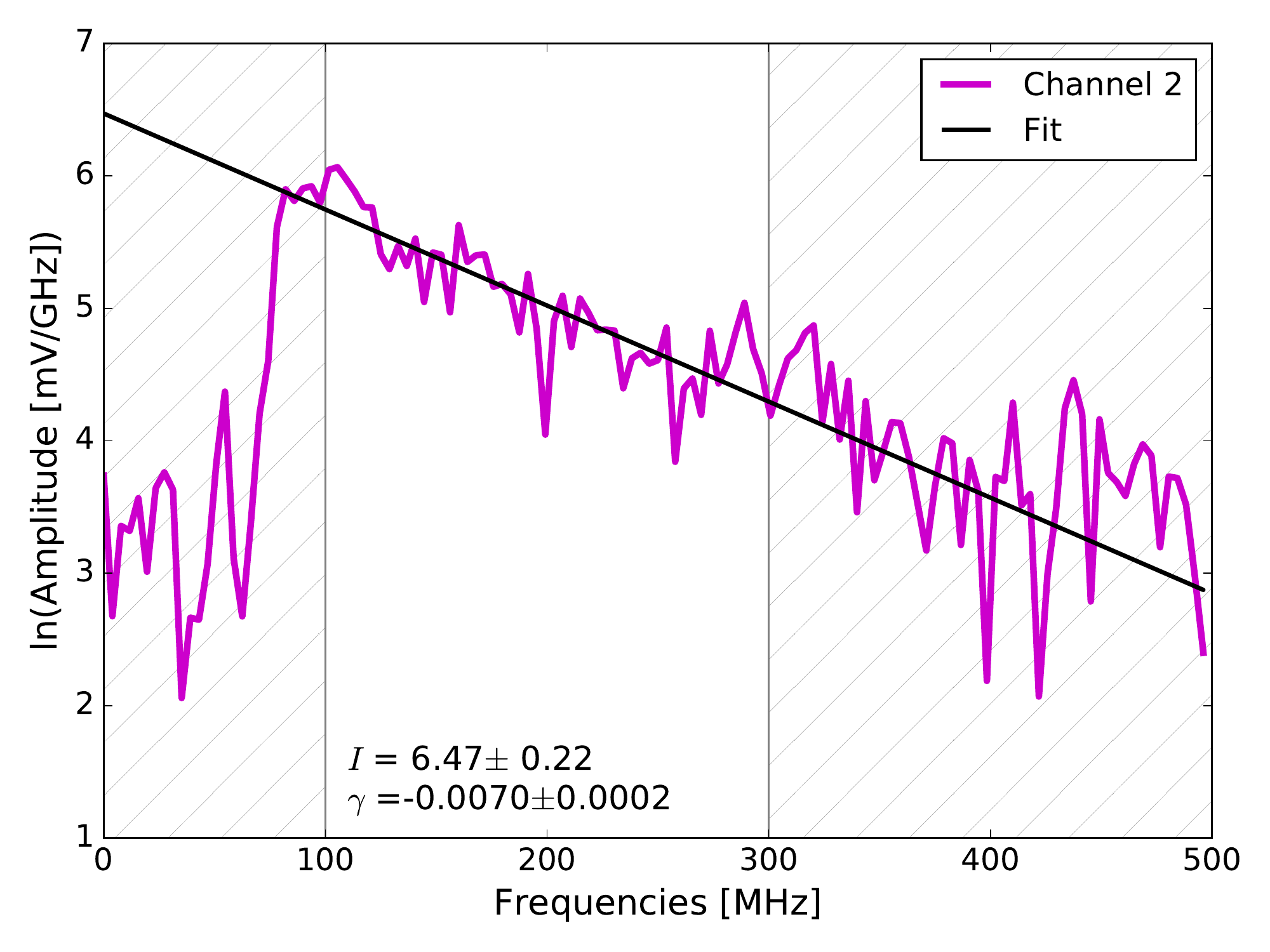}
\includegraphics[width=0.5\textwidth]{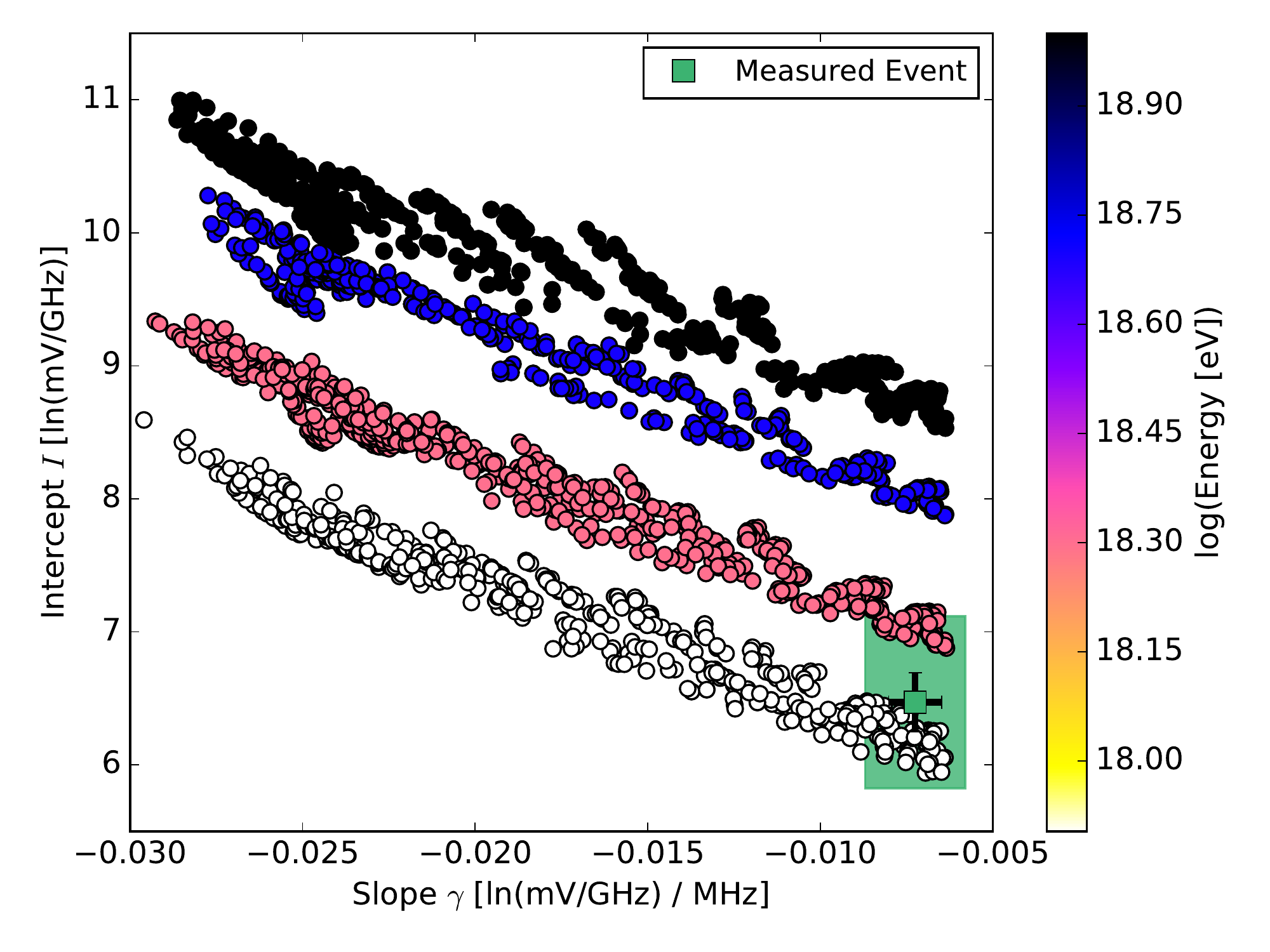}
\caption{Left: Measured frequency spectrum of the five-fold cosmic ray event in station X. The strongest channel is highlighted together with the fit to the data between \unit[100]{MHz} and \unit[300]{MHz}. Right: The two fitted spectral parameters, slope $\gamma$ and intercept $I$, as functions of each other. The colors of the circles indicate the energy of the simulated showers.  Their signals have been processed to include the hardware effects. Shown are four discrete energies ($8\times10^{17},2\times10^{18},5\times10^{18},$ and $ 10^{19}$ eV). The square marker indicates the measured value of the spectral slope in the channel with the strongest amplitude in station X. Systematic uncertainties are denoted with the box. Note that the relation between $I$ and $\gamma$ is different for different arrival directions and for different antenna configurations.}
\label{fig:spec}
\end{figure}

The spectral slope as an energy estimator is tested with the five-fold coincidence event. An exponential function $\ln(A) = I + \gamma \cdot f$ is fit to the amplitude spectrum $A$ of the pulse in the interval $ f = [100 \text{ MHz}, 300\text{ MHz}]$, which is where the combination of the expected pulse power and the ARIANNA sensitivity maximizes the signal-to-noise-ratio. The slope $\gamma$ and the y-axis intercept $I$ are fitted to the data of station X, which is shown on the left in Figure \ref{fig:spec}. The slope measurements with the downward-facing antennas in both station X and in the standard-stations carry very large uncertainties compared to the upward facing antennas and will not be included for the energy reconstruction. In addition to the measured data, $I$ and $\gamma$ are fitted for a dedicated set of simulated showers having the same arrival direction as the measured one (average of stations C, D, E, and G, $\theta=75.0^{\circ}, \phi=47.3^{\circ}$). 30 showers with this arrival direction have been simulated, covering different values of $\text{X}_{\text{max}}$ and several discrete steps of energy. The resulting $I$ and $\gamma$ from a subset of simulations are shown together with the measured values from the cosmic ray event on the right in Figure \ref{fig:spec}.
 
The slope $\gamma$ for one shower varies strongly with the distance to shower axis (e.g. smallest absolute value at the Cherenkov angle, largest at the shower axis). Secondary effects are the interplay of the antenna response and the polarization of the signal, as well as the height of the shower maximum. Together, these three effects explain the distribution of combination of $I$ and $\gamma$ for the same energy. The energy determines the intercept that is found at the Cherenkov angle. An increase in energy, while keeping all other parameters fixed, corresponds to a vertical shift upward in the diagram. Additional information can be found in \ref{sec:app_slope}.

In the absence of noise and for this particular arrival direction, the slope derived from a single LPDA yields an energy resolution of about 50\%. This is due to the statistical uncertainties of the fit and the intrinsic limitation due to the degeneracy between distance to the shower axis, height of the shower maximum and energy. Adding polarization information will improve the method. In addition, the contribution of noise has to be accounted for. Adding thermal noise to the simulations tends to flatten the reconstructed slope by about 30\%, and therefore also the uncertainty on the intercept increases correspondingly. This dominates the systematic uncertainty together with the uncertainties discussed in Section \ref{sec:abs}. 
Including noise effects, we estimate the energy of the five-fold coincident event to be between $8\times 10^{17}~ \text{eV}$ and $3\times 10^{18} ~\text{eV}$ from the slope of a single antenna. 

\begin{figure}
\includegraphics[width=0.5\textwidth]{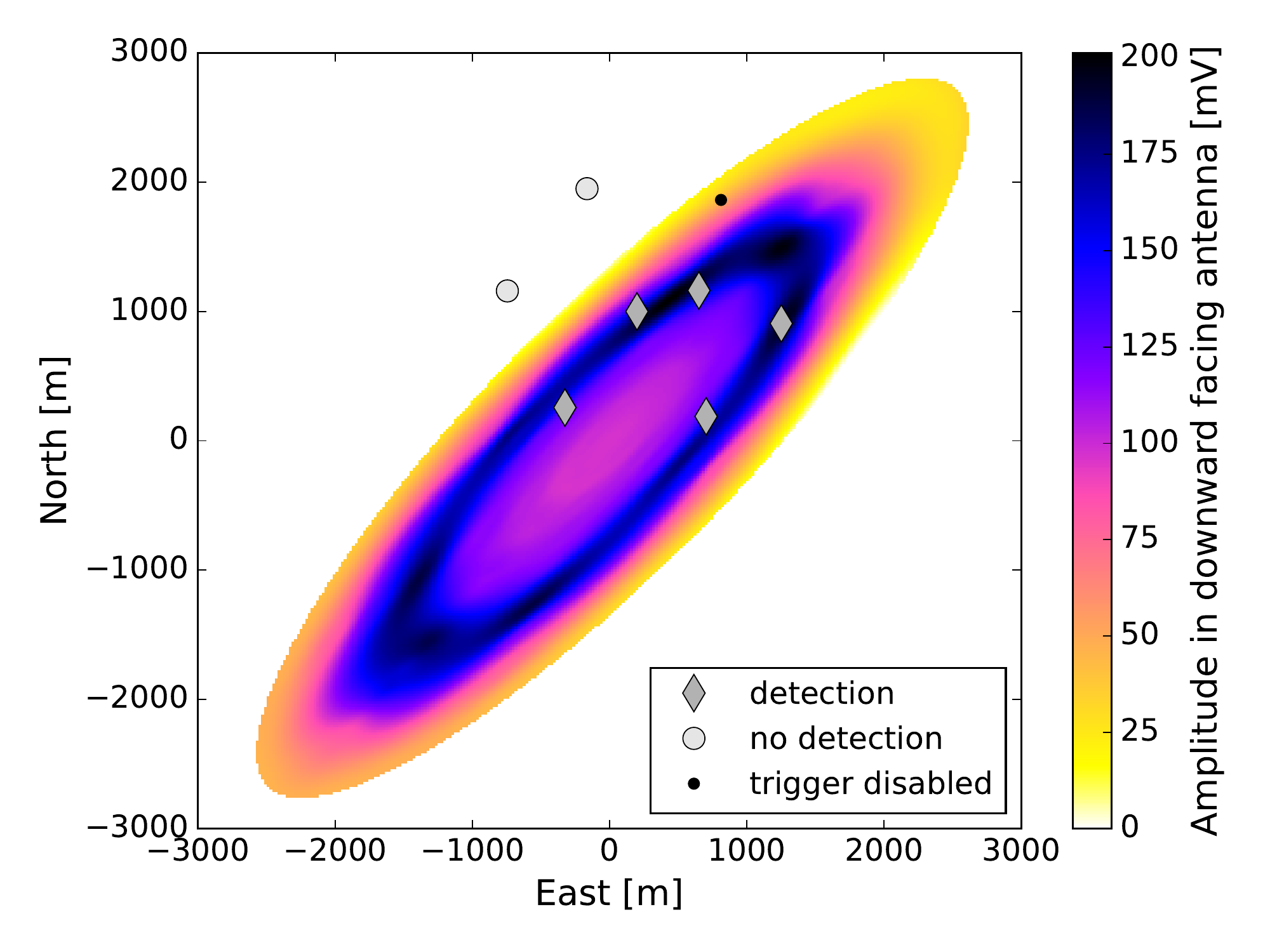}
\caption{One of the best fitting simulations for the five-fold coincidence event. The markers indicate the HRA station positions. The background map shows a simulated event of $75^{\circ}$ zenith angle, an energy $3\times10^{18}$eV and an $\text{X}_{\text{max}}$ of \unit[812]{g/cm$^2$}. The signal strengths are given for the downward-facing antennas. The pattern in the upward-facing antennas is slightly different and about a factor of 5 stronger for this particular arrival direction.}
\label{fig:golden}
\end{figure}

We cross-check this result with an adapted method similar to \cite{2014PhRvD..90h2003B}. We use all 30 showers that have been simulated with the reconstructed arrival direction. For each of these showers, we generate an amplitude map of the shower footprint and shift the core position of the shower within the array until the best agreement between measured and simulated amplitudes is found for every station position. There are only very few combinations of shower core, energy and $\text{X}_{\text{max}}$ that provide a good fit; one of them is shown in Figure \ref{fig:golden}. The energy of the best fitting event falls into the interval determined by the frequency slope method. 

To asses the full utility of the slope method and its energy resolution power, an HRA station will be upgraded. With four upward facing antennas, this station will allow us to better determine the uncertainties on the slope measurement and the arrival direction. In addition, a reconstruction of the polarization and the unfolding of the antenna response is foreseen, which will allow us to test hardware independent para\-me\-teri\-zations.  For the complete ARIANNA with 1296 stations, roughly 20\% of the detected events will be coincidences of three or more stations, so a combination of several slope measurements will yield additional improvements. 

\subsection{Cosmic ray flux calculation}
The HRA has detected 38 air showers in station X between December 2015 and May 2016. As mentioned, the energy of the individual events cannot be obtained from the data from station X alone due to the reduced capabilities to reconstruct the incoming direction. However, the most probable energy of an ensemble of events was obtained from simulations, which allows us to derive a flux measurement. 

Using the simulation chain described in Section \ref{chap:emiss}, the efficiency of Station X to detect air showers is established. As will be elaborated, this leads both to a most probable energy $\langle E \rangle$ and the time-integrated aperture or exposure $\mathcal{E}$ for station X in the season of 2015/16. The flux $J$ is then calculated as
\begin{equation}
J(E) = \frac{N_{\text{CR}}}{ \langle E \rangle \cdot \mathcal{E} \cdot \eta},
\label{eq:flux}
\end{equation}
where $\eta=0.98$ is the efficiency of having identified all cosmic rays in the measured sample (see Section \ref{sec:correl}).

In the analysis procedure, all showers are binned in energy and zenith angle, as the detection efficiency depends strongly on both and only secondarily on the azimuth angle since the magnetic field points almost vertically upwards at the ARIANNA site. For every simulated shower, 10,000 core positions are randomly thrown on a circular area $A$ with a radius of \unit[$20$]{km} around station X. This area is larger than the longest major axis of the longest detectable footprint obtained in a simulation and was chosen to avoid saturation effects in the efficiency due to geometrically large showers. 

\begin{figure}
\includegraphics[width=0.5\textwidth]{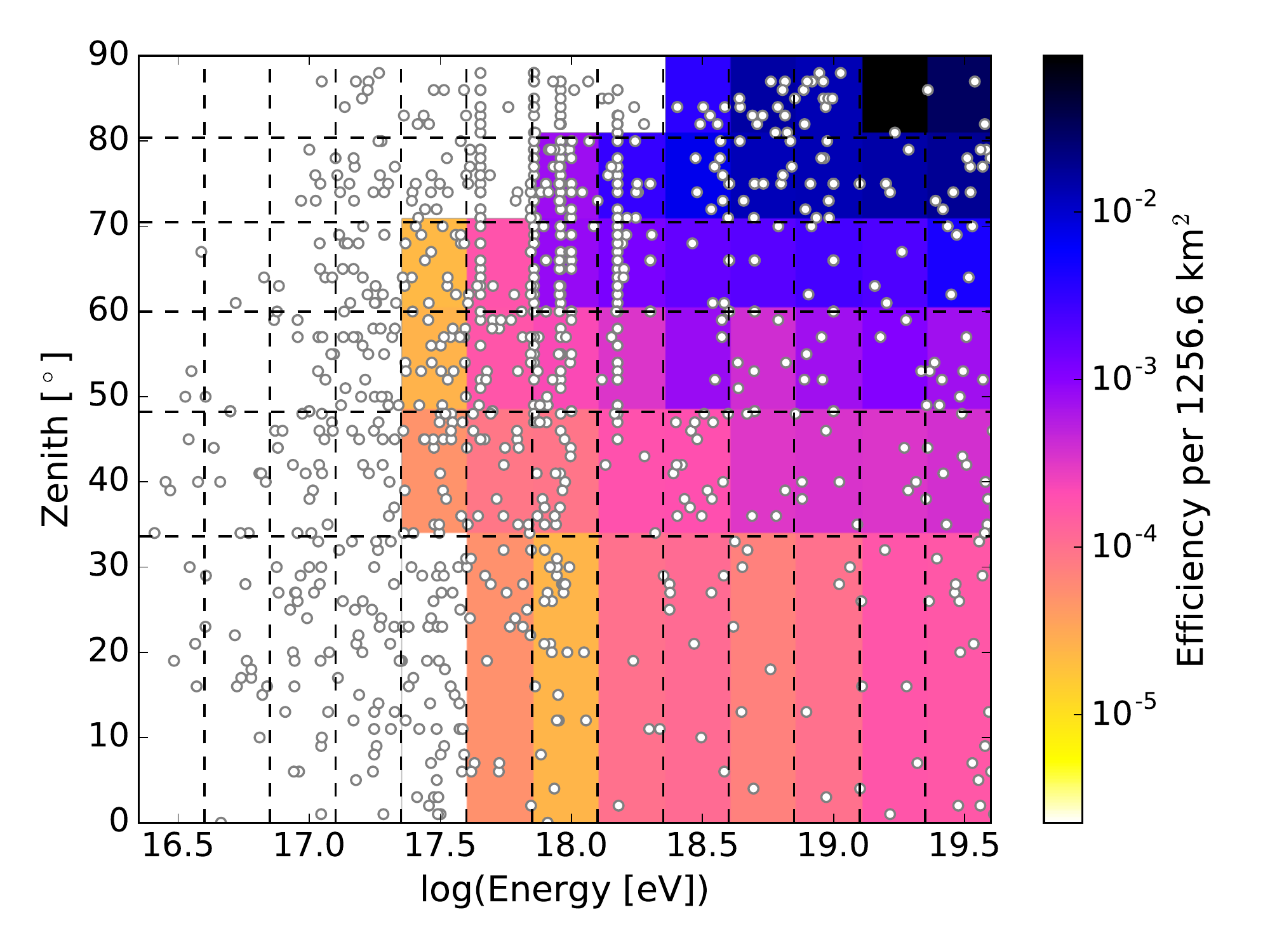}
\includegraphics[width=0.5\textwidth]{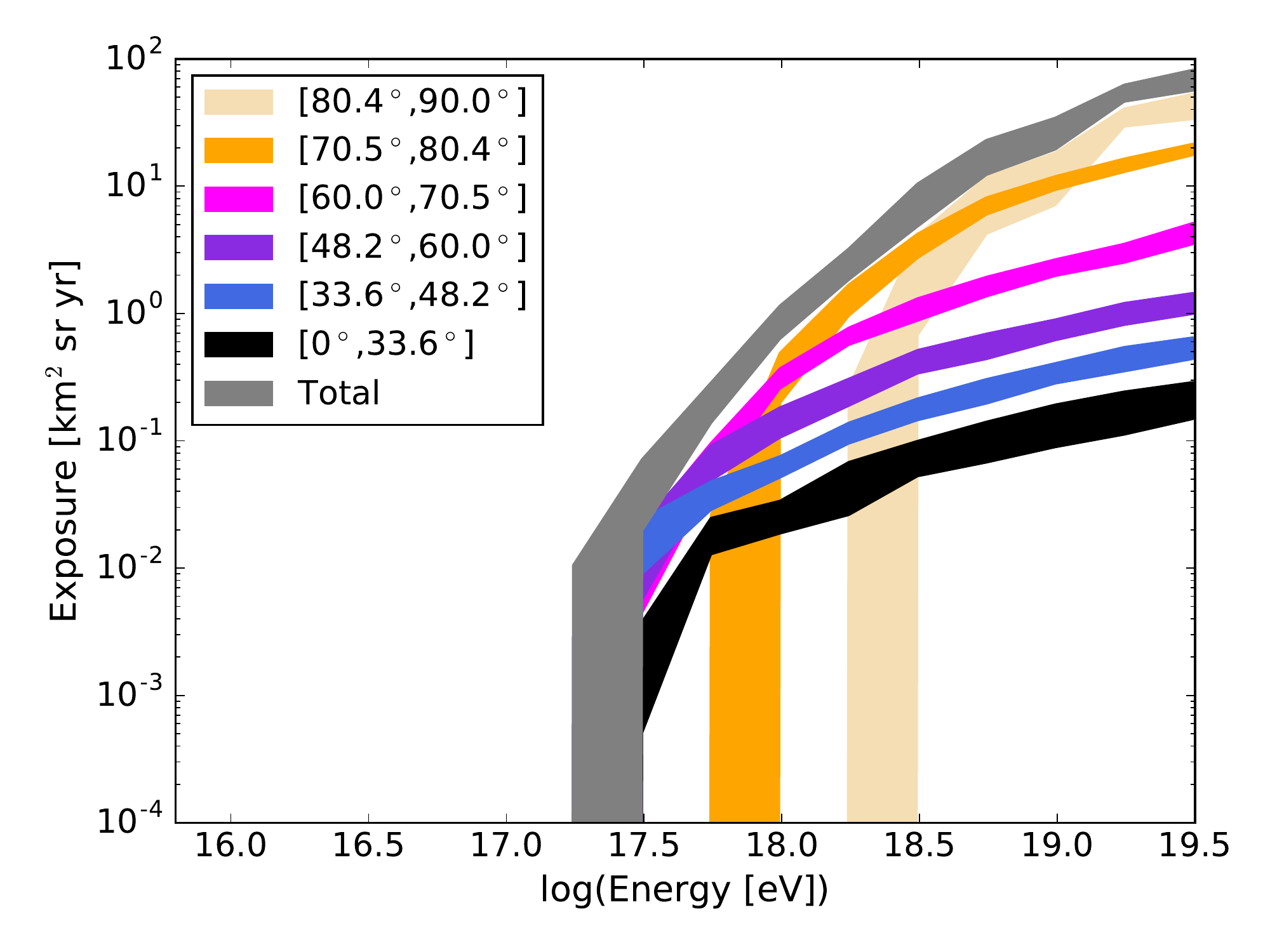}
\caption{Left: Detection efficiency $\varepsilon$ as a function of energy and zenith angle. The background colors show the binned efficiency, the markers show the number of simulated showers per bin. Left: Exposure $\mathcal{E}$ of station X in the 2015/16 season as a function of energy for different zenith angle bins.}
\label{fig:efficiency}
\end{figure}

For every one of the 10,000 realizations of the same shower, the relevant signal in the antenna is extracted from the simulation and a trigger decision is made. The fraction of showers that fulfill the trigger condition for the 10,000 shower realizations gives the efficiency $\varepsilon$ of Station X to detect this shower. The binned efficiency $\varepsilon$ for the ensemble of all simulations is shown on the left of Figure \ref{fig:efficiency}. The efficiency increases with zenith angle and energy, with the threshold energy increasing with inclination. For events above $80^\circ$ the Fresnel reflection starts to limit the efficiency as very little emission enters the ice.

Multiplying the efficiency $\varepsilon$ by the live-time of station X, the area $A$ over which the cores were simulated and the relevant steradian space angle, results in the time-integrated aperture, the exposure $\mathcal{E}$. This quantity is shown in bins of zenith angle on the right side of Figure \ref{fig:efficiency}. The bands in the figure provide a measure of the uncertainty and stem from two calculations of the exposure; the most optimistic and most pessimistic case with one sigma uncertainties for all relevant parameters. Binning the simulations adds an uncertainty, which is, however, small in comparison to the calibration uncertainties. 

As discussed already in Section \ref{chap:emiss}, only proton primaries were simulated. For the radio emission of air showers, the energy reconstruction is not affected by the type of primary particle, so this choice has no direct effect on the energy resolution \cite{2015APh....60...13N}. However, the choice of primary could affect the exposure calculation. Protons cover a wider range of values for the height of the shower maximum, than for example iron primaries. The exposure increases with zenith angle, owing to the increase in footprint size and the lack of attenuation of the radio emission in the atmosphere. If one had used only iron primaries, the propagation distance would have been larger on average, the efficiency better and therefore the exposure also larger. This effect can be simulated excluding showers with large values of $\text{X}_\text{max}$. The change introduced by this is smaller than the other uncertainties that factor into the current analysis. For future analyses, a more realistic composition will have to be studied. 

\begin{figure}
\includegraphics[width=0.5\textwidth]{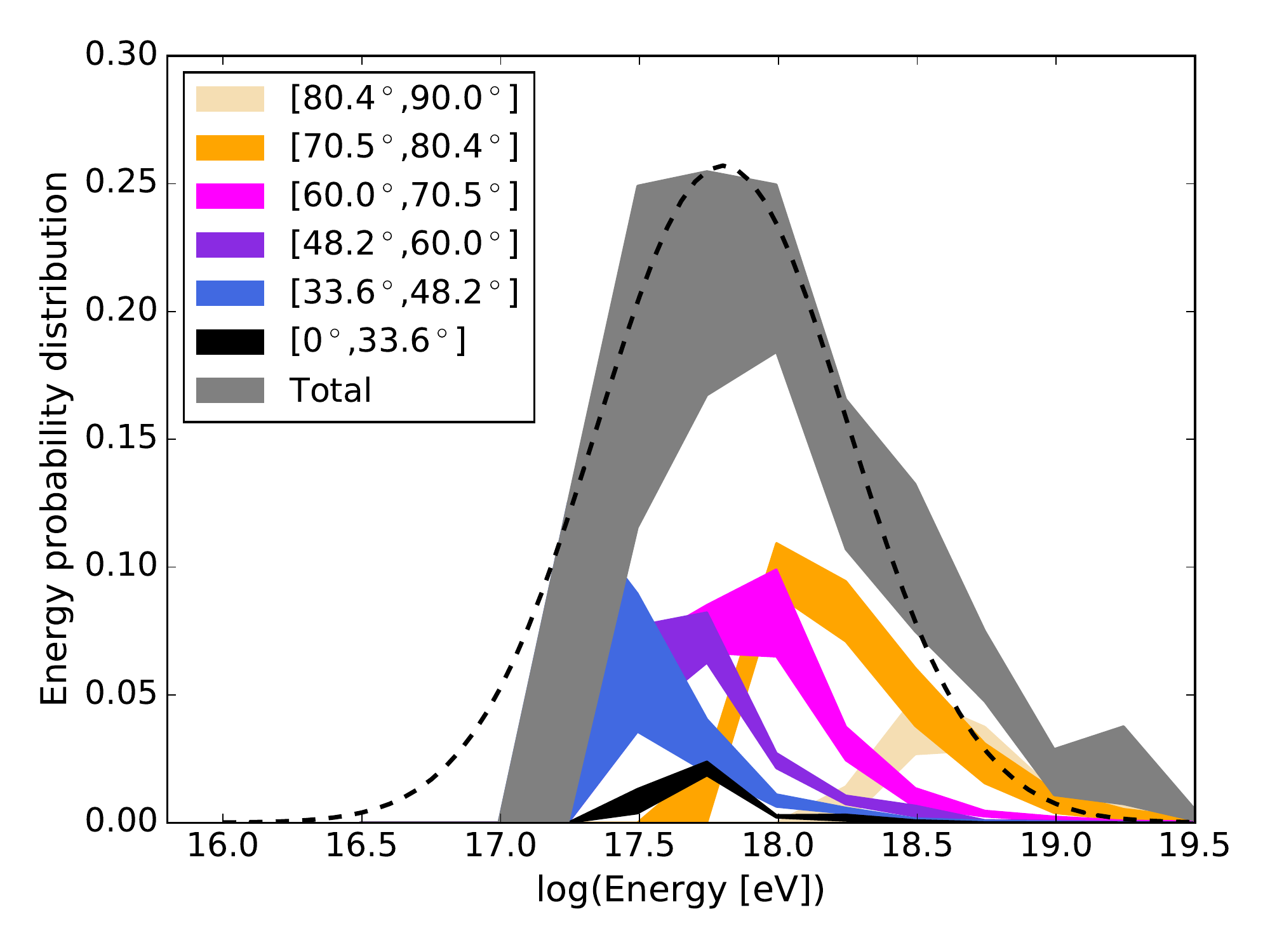}
\includegraphics[width=0.5\textwidth]{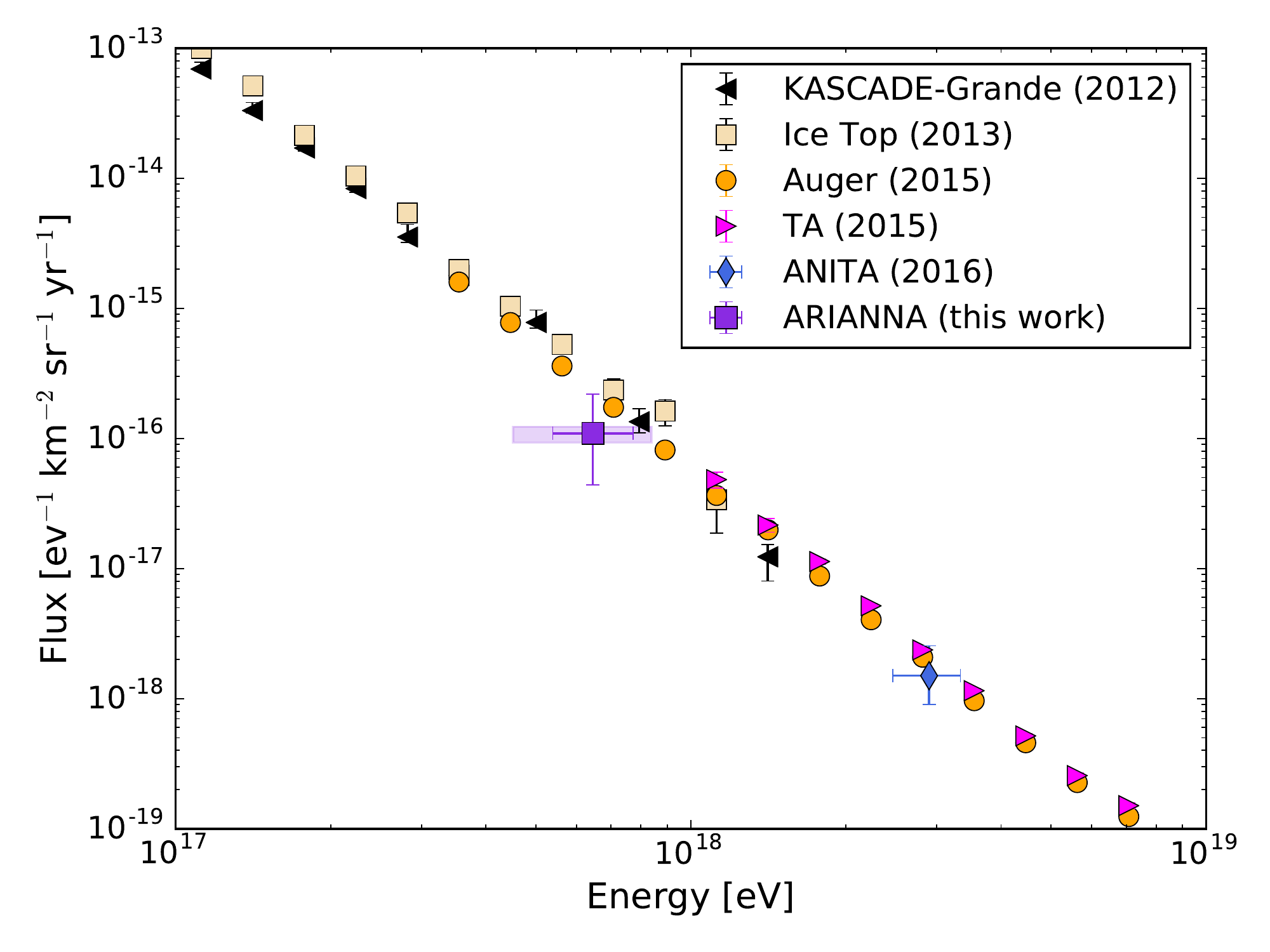}
\caption{Left: Probability of a detected air shower of a certain arrival direction to be of a certain energy. Curves including uncertainties are shown for different zenith angle bins. A Gaussian fit to the total distribution is also shown. Right: Flux of cosmic rays as function of energy. The data from the HRA as derived in this work is shown together with published flux data from KASCADE-Grande \cite{2012APh....36..183A}, the Pierre Auger Observatory \cite{2015arXiv150903732T}, Telescope Array \cite{2015APh....61...93A}, IceTop \cite{2013PhRvD..88d2004A} and ANITA \cite{2016APh....77...32S}. The errorbars indicate the uncertainty on the mean energy as well as the exposure. The uncertainty on the scale of the energy is indicated by the shaded area.}
\label{fig:energy}
\end{figure}

With a given spectral index of the flux of cosmic rays and the efficiency $\varepsilon$, the most probable energy can be obtained. From published measurements such as \cite{2013PhRvD..88d2004A,2015arXiv150903732T}, the spectral index below \unit[$10^{18}$]{eV} is known to be $-3.3\pm0.04$. Since the most probable energy will be close to the threshold, which is also below \unit[$10^{18}$]{eV}, a change in the spectral index above the ankle is less relevant for this calculation, but was considered for the uncertainty calculation. By multiplying the efficiency $\varepsilon$ binned in zenith angle by an arbitrary absolute flux with a defined spectral index and normalizing the sum of the distribution to one, we obtain the result shown on the left of Figure \ref{fig:energy}. The energy distribution can be approximated by a Gaussian as most showers will be measured very close to the threshold energy. The best fit is centered around $\log(E) = 17.80$ (corresponding to \unit[$6.5\times10^{17}$]{eV}) with $\sigma=0.45$. The broad shoulder towards higher energies is driven by horizontal showers for which the efficiency and energy threshold are higher. The mean of the Gaussian is the most probable energy $\langle E \rangle$ and $\sigma$ determines the energy bin, where most of the air showers are expected. The statistical uncertainty on $\langle E \rangle$ is given by $\sigma/\sqrt{N}$, where $N$ are the 38 detected cosmic rays. 

Using equation \ref{eq:flux} with the most probable energy $\langle E \rangle$ leads to a flux measurement for the HRA as shown in Figure \ref{fig:energy}. The flux at $\langle E \rangle = \unit[6.5^{+1.2}_{-1.0}\times10^{17}$]{eV} has been measured to $ J = \unit[1.1^{+1.0}_{-0.7}\times10^{-16}$]{eV$^{-1}$km$^{-2}$sr$^{-1}$ yr$^{-1}$}, which is in agreement with values from the literature. The statistical uncertainties include all propagated uncertainties in the efficiency calculation, as well as the uncertainty of the spectral index of the flux. The band in Figure \ref{fig:energy} illustrates the systematic uncertainty on the absolute energy.

\section{Consequences for neutrino detection and the ARIANNA array}
From the studies of the air shower signals in the HRA a number of conclusions can be drawn for both neutrino and cosmic-ray detection with ARIANNA. Overall, no external background that mimics the predicted neutrino or cosmic ray signals has been found. This is both a statement of quality for the ARIANNA site on the Ross ice shelf and an indication that the current stations have reached the design goal of having a very low intrinsic noise level.

The full detector simulation of radio emission of air showers has produced signal waveform templates that show a high correlation with the measured waveforms. Average correlations of 0.8 are common and indicate that a good understanding of the detector has been reached. While there is room for improvement, as the data does not perfectly match the predictions, the quality of the predictions is at a stage where 98\% analysis efficiency for cosmic rays is reached. As the neutrino capability of ARIANNA is based on the same type of template matching, we expect high analysis efficiencies for the neutrino search. 

Since the radio signals of neutrinos and cosmic rays are similar, the more abundant cosmic rays will be a confusion background for neutrinos. Equipping every station with upward facing antennas will eliminate the confusion of direct signals with the help of the directivity of the antennas. In all cosmic ray simulations, no signal has been found that is stronger in the downward facing antennas than in the upward-facing antennas, which predicts a false positive rate of $\ll1$\% for cosmic rays in the neutrino analysis. 

It has been suggested that the transition radiation of air showers hitting the ice could produce a confusion signal \cite{2016APh....74...96D}. It is still unclear whether this phenomenon will still be a strong background for a detector at sea-level, where few showers still contain a significant amount of particles, and if realistic ice conditions are taken into account.  Strictly following the calculations in \cite{2016APh....74...96D}, we have extrapolated that there is an, albeit small, chance for the HRA to detect the transition radiation within a couple of months of run time. So far, no evidence was found that the measured rate of air showers is significantly higher than predicted from direct signals. This means either that there is no evidence yet for strong signals of transition radiation or that the signals no longer resemble the original air shower signals due to surface scattering or other propagation effects. Nonetheless, it is too early too exclude the phenomenon and more statistics need to be gathered. 

\begin{figure}
\includegraphics[width=0.5\textwidth]{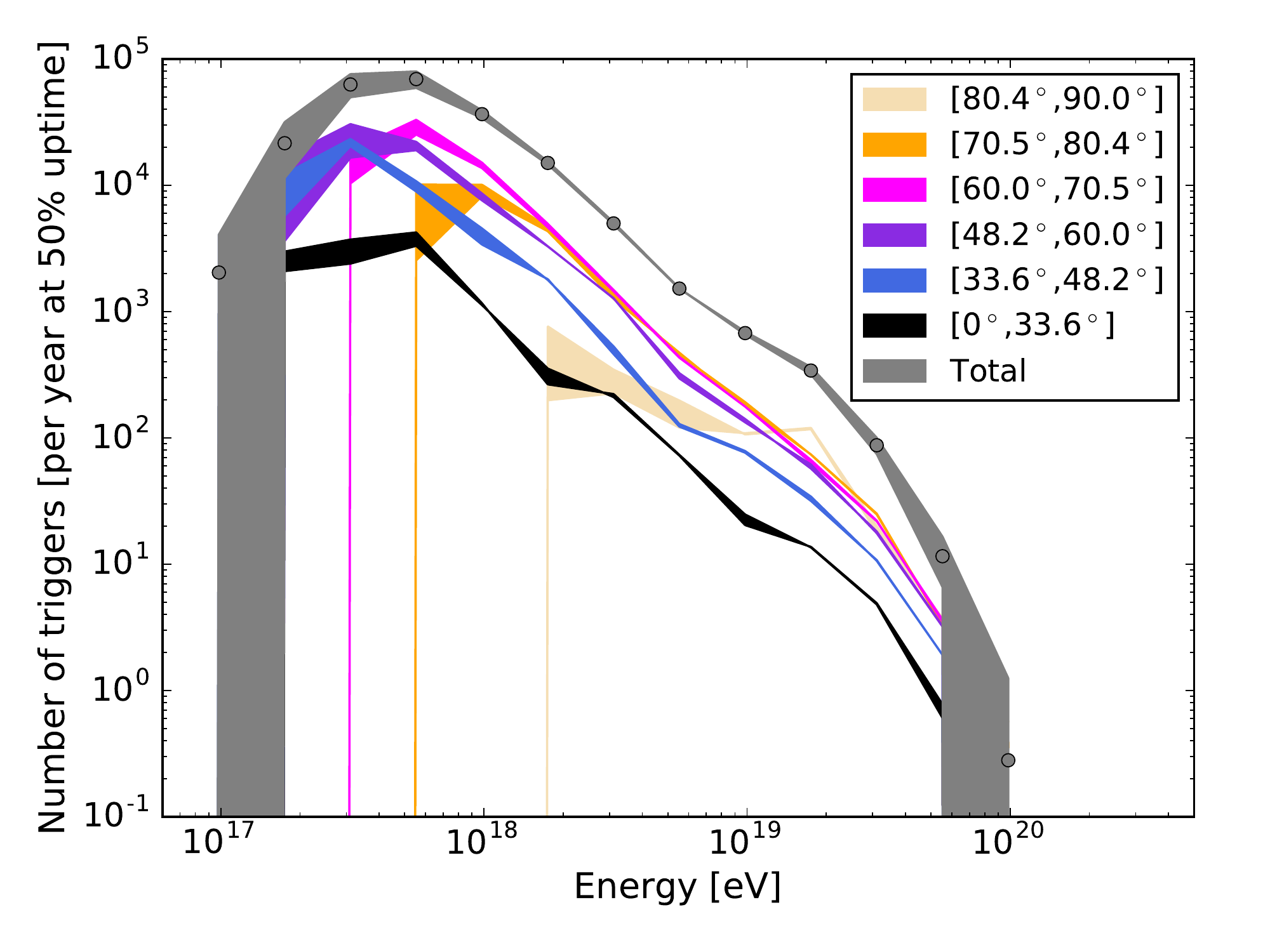}
\caption{Expected air shower measurements with ARIANNA with 1269 station with four upward facing antennas each, deployed over a hexagonal-grid of \unit[1]{km} baseline. Shown are the number of events per quarter decade of energy and per year, assuming a 50\% uptime. The calculations are based on the published flux of the Pierre Auger Observatory \cite{2015arXiv150903732T}. The bands depict the uncertainty on the numbers derived from simulations. No systematic uncertainty on the flux is included. }
\label{fig:arianna_spec}
\end{figure}

ARIANNA with 1296 stations will detect a significant number of cosmic rays. The exact numbers will depend on features such as the antenna orientation (straight up, down, tilted), station configuration (how many antennas, trigger on the coincidence of two or four channels) and trigger configuration (trigger on up- or downward antennas). As shown in Figure \ref{fig:arianna_spec} in an array triggering on two upward facing antennas about 300,000 air showers will be registered per half-year of live-time. These cosmic rays will have a threshold energy of \unit[$8\times10^{16}$]{eV} and event detections of up to \unit[$10^{20}$]{eV} are likely. If triggering only on the coincidence of two downward facing antennas, the threshold will rise to \unit[$8\times10^{17}$]{eV}, which reduces the number of total triggers to about 14,500 per half year of live-time, but does not affect the exposure at the highest energies. 

For efficient rejection of cosmic rays from the neutrino data-set, at least two upward facing antennas per station are needed, covering two perpendicular polarizations. Adding two more antennas, will convert ARIANNA into a fully functional cosmic ray detector, while retaining full efficiency for neutrino detection. Such a detector can provide arrival directions for all measured cosmic rays ($\Delta \theta = 1^{\circ})$, as well as an energy estimate. The precise energy resolution will depend on methods still to be developed. For horizontal air showers, which are likely to trigger more than one station, an energy-resolution of 20\% should be achievable, extrapolating from other radio detectors that employ multiple station detections \cite{2016PhRvL.116x1101A}.  At \unit[1296]{km$^2$} ARIANNA will be the largest stand-alone radio cosmic ray detector, being about half the size of the Pierre Auger Observatory and slightly larger than the current surface array of Telescope Array. It is anticipated that such a detector would be in a good position to measure both the energy spectrum, as well as a possible anisotropy of cosmic rays between \unit[$8\times10^{17}$]{eV} and \unit[$10^{20}$]{eV}.
 
Irrespective of potential independent cosmic ray science, the cosmic ray signals will act both as a calibration signal and as training set for neutrino reconstruction algorithms. Extended flux measurements of cosmic rays will lead to an energy calibration of the detector, as well as algorithms for the energy reconstruction of neutrinos. The polarization of the neutrino signal is needed to solve for the ambiguity between signal arrival direction and direction of the neutrino. Since the polarization structure of air showers is well understood, the data will determine the accuracy of the polarization measurements and thereby the angular resolution for neutrinos more precisely. 

\section{Conclusions}
The first measurements of cosmic ray induced air showers with the HRA of ARIANNA have been presented. Between December 2015 and April 2016, 38 air showers have been detected in one station optimized for cosmic ray detection. The air shower signals are identified with an efficiency of better than 98\%, with a trigger efficiency solely determined by live-time and signal amplitude and thereby the energy of the air showers. All air showers have been detected with a radio-self trigger without additional particle detectors. 

The cosmic ray flux measured with the HRA of ARIANNA at \unit[$\langle E \rangle = 6.5^{+1.2}_{-1.0}\times10^{17}$]{eV} has been measured to be \unit[$J = 1.1^{+1.0}_{-0.7}\times10^{-16}$]{eV$^{-1}$km$^{-2}$sr$^{-1}$yr$^{-1}$} and is in agreement with measurements by other experiments. With increasing statistics a more detailed measurement of the cosmic ray flux and the slope of the energy spectrum will be possible. The future ARIANNA array will not only be sensitive to neutrinos above \unit[$10^{16}$]{eV}, but also to cosmic rays above \unit[$8\times10^{16}$]{eV} with detections likely up to \unit[$10^{20}$]{eV} at an area of \unit[1296]{km$^2$}. 

There is currently no evidence for backgrounds that mimic cosmic ray or neutrino radio signals. Impulsive radio signals recorded during periods of high winds increase the trigger rates, but are cleanly removed during data-analysis. The trigger-rate of the stations is dominated by thermal noise fluctuations and a simple threshold trigger is sufficient to detect cosmic ray and neutrino signals. 

Measurements in the upcoming season will help to quantify both the energy and angular resolution of the ARIANNA stations, by measuring cosmic rays in a dedicated station with four upward facing antennas. This data will be used to further train reconstruction algorithms for energy and polarization.

\section{Acknowledgements}
The authors wish to thank the staff of the Antarctic Support Contractors and the entire crew at McMurdo Station for their continued support. 

This work was supported in part by U.S. National Science Foundation under grants PLR-08339133, PHY-0970175, and PLR-1126672, as well as PLR-1413661, PHY-1607719, and PHY-1607199. We acknowledge support from the U.S. Department of Energy under contract number DE-AC-76SF00098, from the German Research Foundation (DFG), grant NE 2031/1-1, a Uppsala University Vice-Chancellor's travel grant from the Knut and Alice Wallenberg foundation, and a Liljewalchs travel scholarship. 

\appendix

\section{Antenna simulations and measurements}
\label{sec:app_sim}
\begin{figure}
\includegraphics[width=0.5\textwidth]{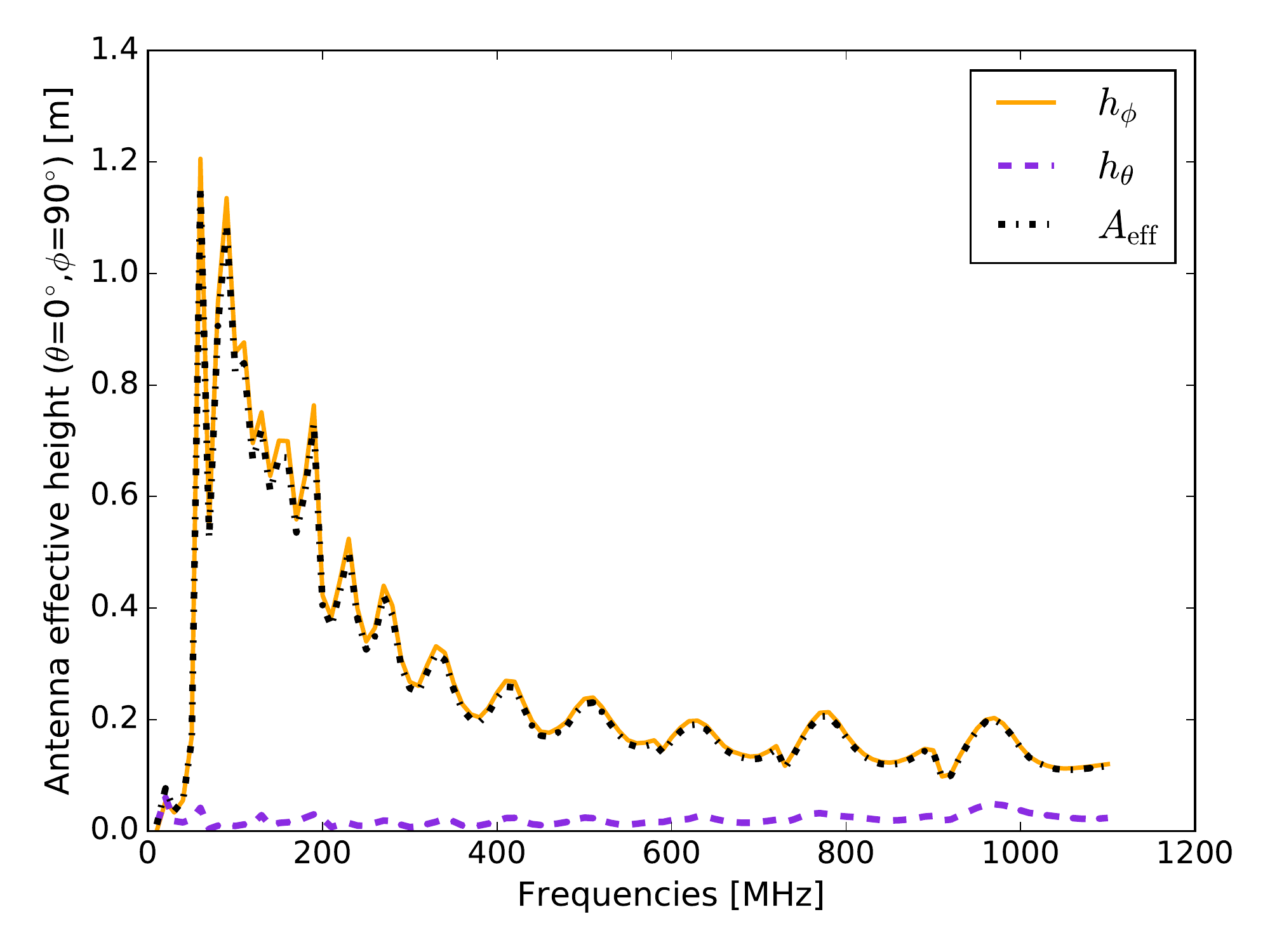}
\includegraphics[width=0.5\textwidth]{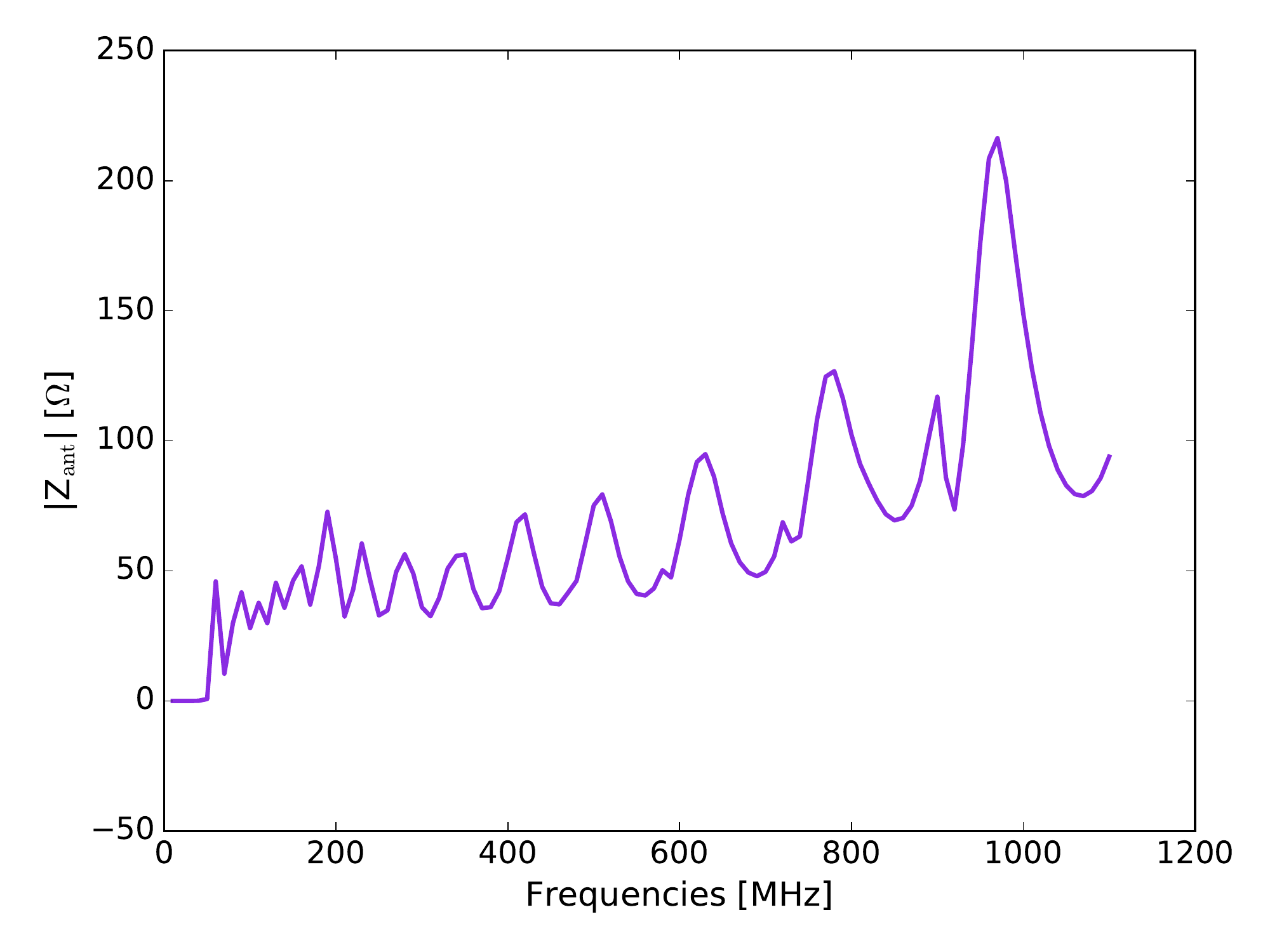}
\includegraphics[width=0.5\textwidth]{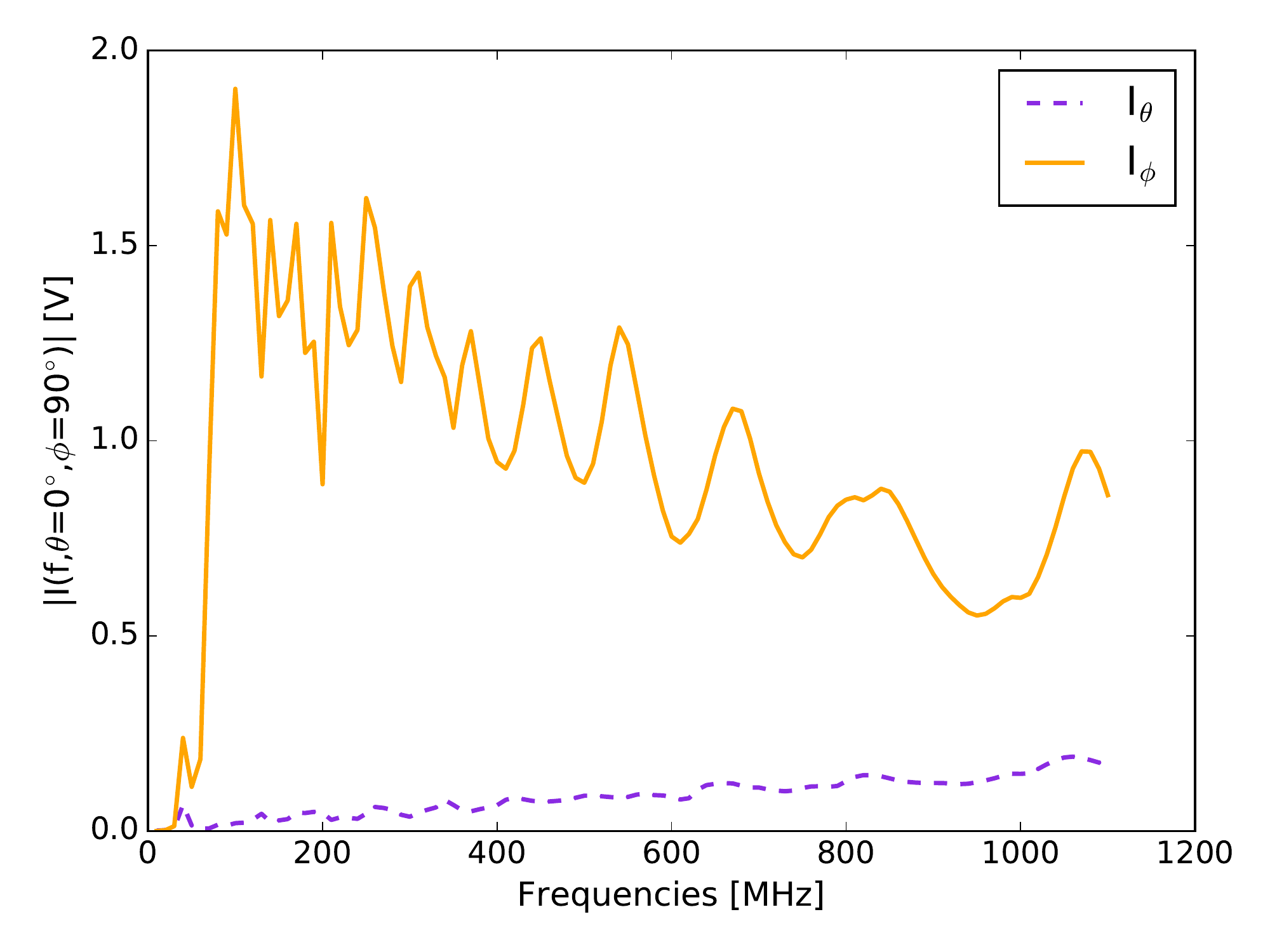}
\includegraphics[width=0.5\textwidth]{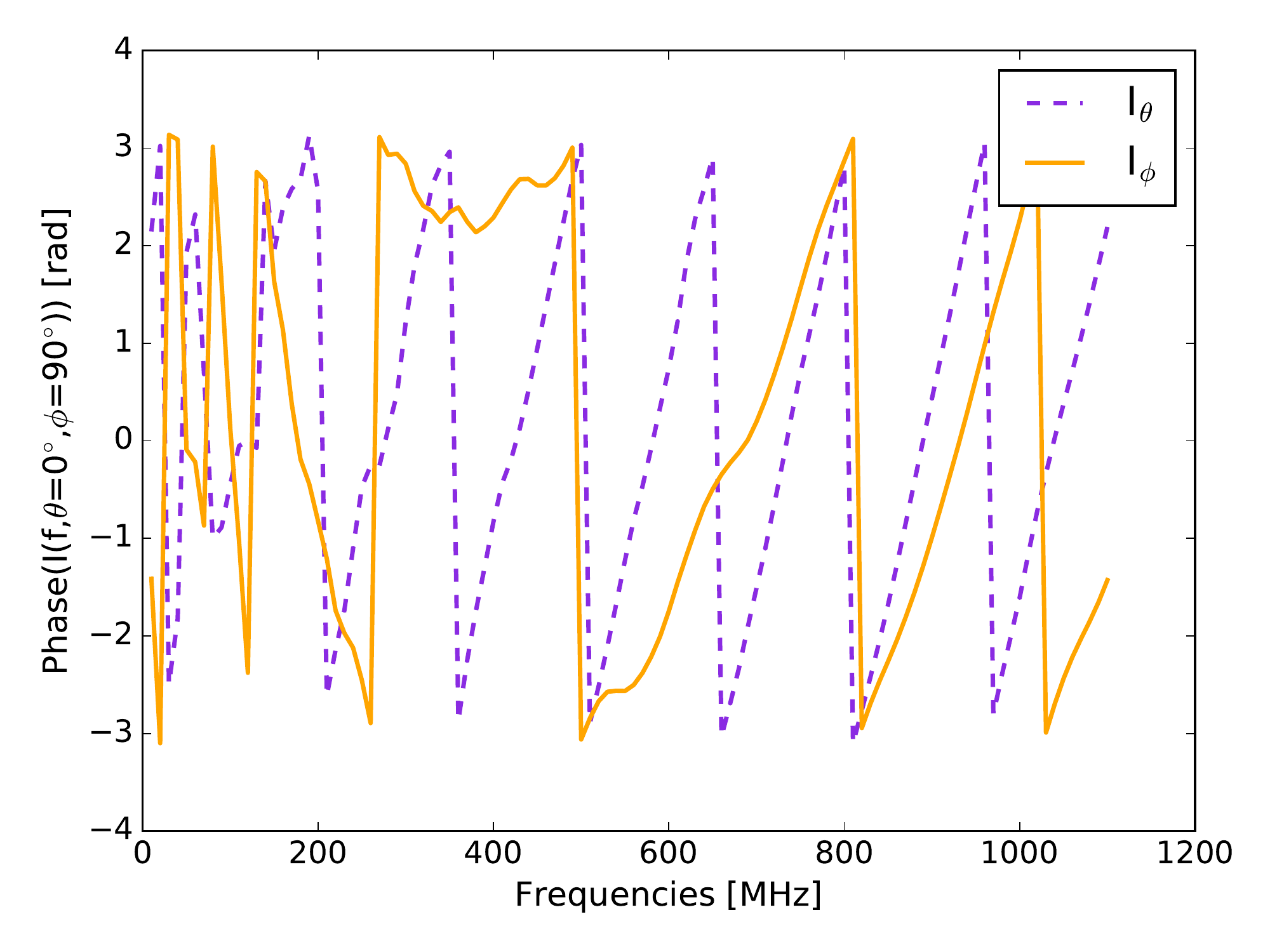}
\caption{Antenna characteristics as simulated with WIPL-D. The top left figure shows the antenna effective height as function of frequency. The antenna effective height is calculated using directly the complex currents and the gain of the simulations (see discussion in Section \ref{chap:galaxy}). The top right figure shows the absolute value of the antenna impedance as function of frequency. In the two bottom figures examples of the complex antenna voltages for an electric field arriving at the antenna front-lobe are shown. Shown are both the absolute component (left) and the complex argument (right). }
\label{fig:wipl-d}
\end{figure}

The ARIANNA LPDAs were simulated in detail using the software WIPL-D (WIPL-D Pro v13). The simulations were performed in an infinite firn medium using 1157 unknowns at $\epsilon=1.78$, $\mu=1$ and $\sigma(S/m)=10^{-6}$. Examples of the output of the simulation of the complex antenna response are shown in Figure \ref{fig:wipl-d}. 

\begin{figure}
\includegraphics[width=0.65\textwidth]{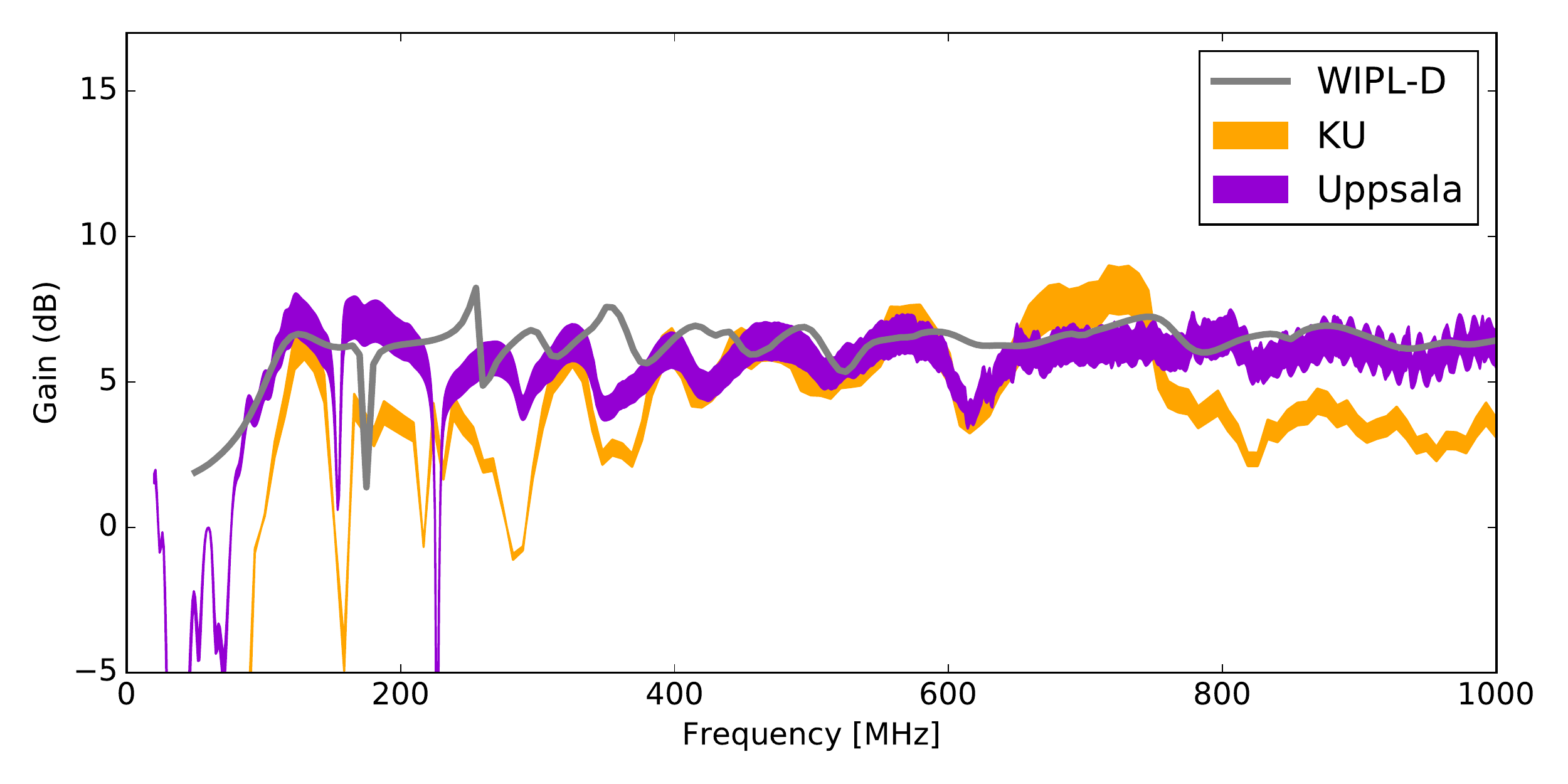}
\caption{Absolute gain of the ARIANNA LPDAs as a function of frequency. Shown are both measurements with uncertainties (colored bands), as well as the WIPL-D simulation (line).}
\label{fig:measure}
\end{figure}

\begin{figure}
\includegraphics[width=0.5\textwidth]{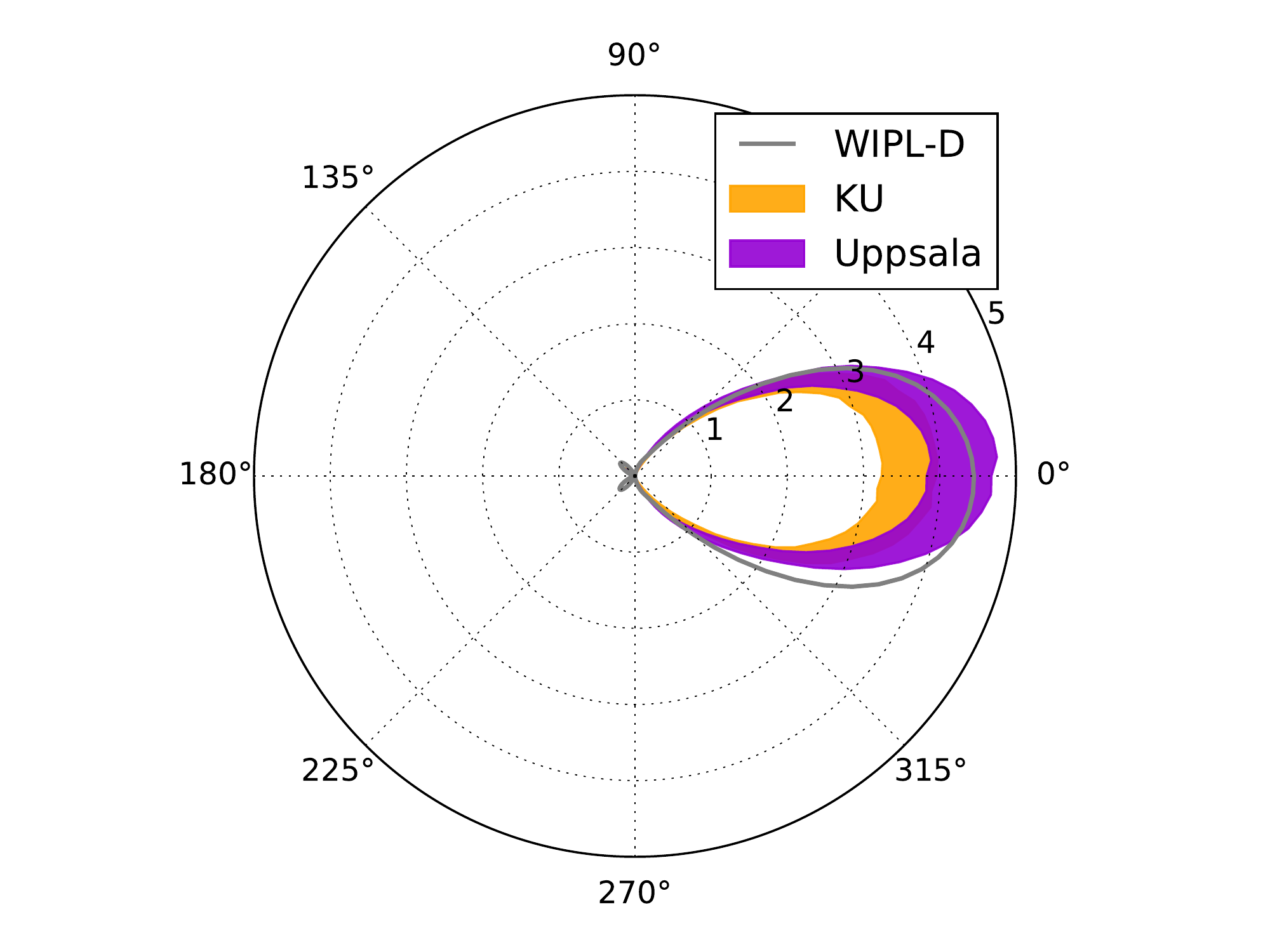}
\includegraphics[width=0.5\textwidth]{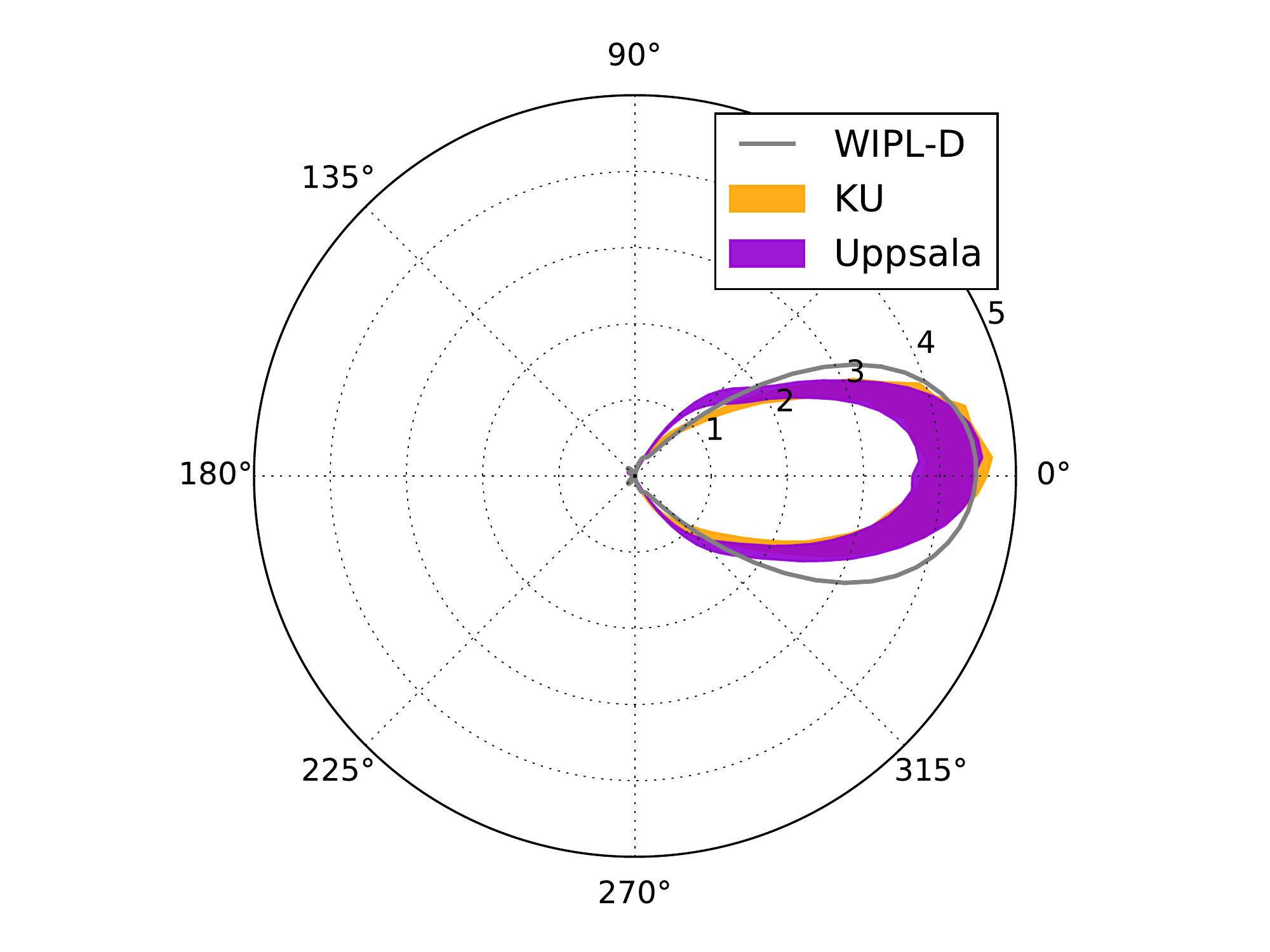}
\caption{Gain as function of direction in the plane of the antenna tines at 330 MHz (left) and 400 MHz (right). The highest gain is found along the axis in the direction of the smallest tines (here at $0^{\circ}$, front-lobe). Shown are both measurements (colored bands) and the simulated response of WIPL-D.}
\label{fig:measure2}
\end{figure}

The simulated antenna characteristics were compared to dedicated measurements both at the University of Kansas (KU) and Uppsala University in Sweden. Examples of these measurements and the corresponding simulations can be found in Figures \ref{fig:measure} and \ref{fig:measure2}. As discussed in Section \ref{sec:ant-mod}, the simulation and measurements show differences between 10\% and 50\% in linear gain, where a large fraction of this uncertainty stems from small details in the antenna simulations (differences in length of mm) and systematic differences between the two measurement set-ups. Measurements of two antennas of the same type reveal significantly smaller differences than between two measurements at different locations. The width of the uncertainty bands given in the Figure has been established based on the measurements of two antennas at the same site. 

\section{The spectral slope and its dependencies}
\label{sec:app_slope}
As was discussed in Section \ref{sec:slope}, the slope of the frequency spectrum of a detected air shower pulse can be used to reconstruct the energy of the shower. The presented analysis is detector-dependent, as the quantities used are not related to the frequency spectrum of the incoming electric field, but to the one measured in an HRA station, therefore containing also the frequency response of antenna and amplifier. A forthcoming publication will discuss a more general approach of using the frequency spectrum in a system with a large band-width. For this work, however, some additional insight in the dependencies might be interesting to the reader. 

\begin{figure}
\includegraphics[width=0.5\textwidth]{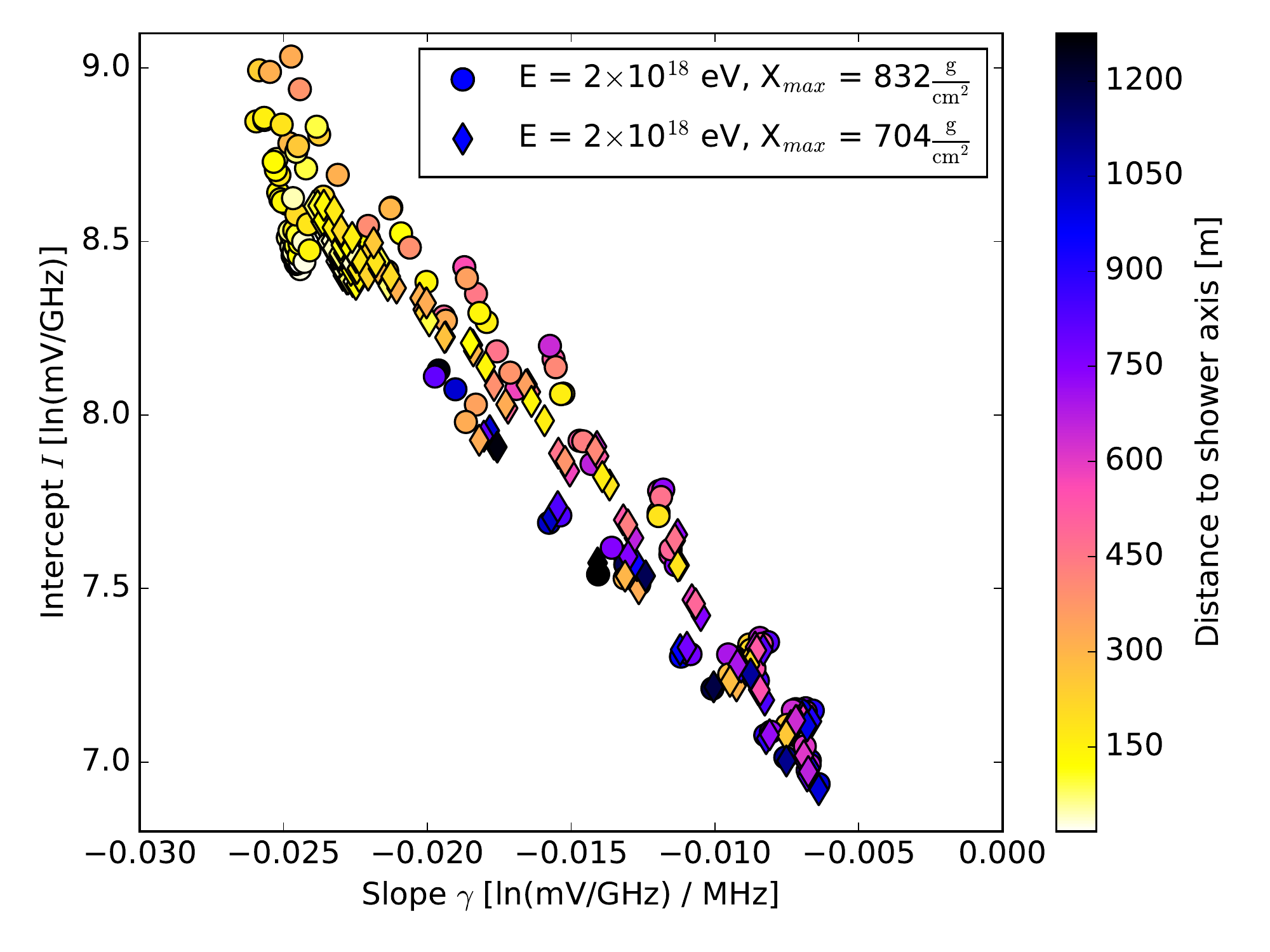}
\caption{A subset of the simulated air showers shown in Figure \ref{fig:spec} are depicted. Here, two showers of the same energy are contrasted. The colors describe the distance to the shower axis and the markers distinguish between the two showers of different values of the height of the shower maximum. }
\label{fig:slope2}
\end{figure}

Figure \ref{fig:slope2} shows the comparison of two air shower simulations with the same energy and arrival direction, but different heights of shower maximum. The Figure shows in detail how the frequency spectrum changes as function of distance to the shower axis. Near the shower axis the slope of the spectrum is steepest, while it is close to flat at the Cherenkov angle. It never reaches the theoretically expected fully flat spectrum in an HRA station, due to the response of the system and the distances between the sampling points, which not necessarily provide a position directly at the Cherenkov angle. It should be noted that the shape of the depicted distribution (\emph{"the slope of slopes"}) is strongly zenith angle dependent and when including the antenna response also azimuth dependent. 

The height of the shower maximum causes only a secondary effect in this way of presenting the data. The inherent spread is mostly driven by different observer angles around the shower axis. In a future analysis, the full electric-field will be reconstructed, which will avoid projection effects of the signal polarization into the antenna sensitivties, which introduces additional spread. Such a reconstruction of the full electric-field is foreseen for ARIANNA.

\bibliographystyle{elsarticle-num} 
\bibliography{CR-analysis}

\end{document}